\documentclass[longauth]{aa} 
\usepackage{graphicx}
\usepackage{subfig}
\usepackage{hyperref}
\usepackage{mathtools}
\DeclarePairedDelimiter\abs{\lvert}{\rvert}%
\usepackage{txfonts}
\usepackage{natbib}
\usepackage[colorinlistoftodos]{todonotes}
\def\lesssim{\mathrel{\hbox{\rlap{\hbox{\lower4pt\hbox{$\sim$}}}\hbox{$<$}}}}
\def\gtrsim{\mathrel{\hbox{\rlap{\hbox{\lower4pt\hbox{$\sim$}}}\hbox{$>$}}}}
\newcommand{\mincir}{\raise
-2.truept\hbox{\rlap{\hbox{$\sim$}}\raise5.truept
\hbox{$<$}\ }}
\newcommand{\magcir}{\raise
-2.truept\hbox{\rlap{\hbox{$\sim$}}\raise5.truept
\hbox{$>$}\ }}
\newcommand{\be}{\begin{equation}}
\newcommand{\ee}{\end{equation}}
\newcommand{\ba}{\begin{eqnarray}}
\newcommand{\ea}{\end{eqnarray}}

\newcommand {\hh} {$\mathrm{h^{-1}}$ Mpc}
\newcommand {\ks} {km~s$^{-1} \;$}
\newcommand {\kss} {km~s$^{-1}$}

\hypersetup{draft}
\begin{document} 


\institute{
INAF-Osservatorio Astronomico di Capodimonte, Via Moiariello 16, 80131 Napoli, Italy (e-mail: amata.mercurio@inaf.it)
\label{oac}  \and
Dipartimento di Fisica e Scienze della Terra, Universit\`a di Ferrara, Via Saragat 1, 44122 Ferrara, Italy
\label{unife}  \and
INAF - OAS, Osservatorio di Astrofisica e Scienza dello Spazio di Bologna, via Gobetti 93/3, I-40129 Bologna, Italy
\label{inafbo} \and
INAF-Osservatorio Astronomico di Trieste,  via G. B. Tiepolo 11, I-34143, 
Trieste, Italy
\label{oats}  \and
IFPU - Institute for Fundamental Physics of the Universe,  via Beirut 2, I-34014, Trieste, Italy 
\label{ifpu}  \and
Centro de Astrobiología, Instituto Nacional de Técnica Aeroespacial, Ctra de Torrejón a Ajalvir, km 4, 28850 Torrejón de Ardoz, Madrid, Spain
\label{madrid} \and
Dipartimento di Fisica, Univ. degli Studi di Trieste, via Tiepolo 11, I-34143, Trieste, Italy
\label{units}  \and
Dipartimento di Fisica, Universit\`a  degli Studi di Milano, via Celoria 16, I-20133 Milano, Italy
\label{unimilano} \and
INAF - IASF Milano, via A. Corti 12, I-20133 Milano, Italy 
\label{iasf-milano}\and
OmegaLambdaTec GmbH, Lichtenbergstrasse 8, 85748 Garching bei Munchen, Germany
\label{monaco} \and
Max-Planck-Institut f\"ur Astrophysik, Karl-Schwarzschild-Str. 1,
D-85748 Garching, Germany
\label{maxplank1} \and
Instituto de Física, Pontificia Universidad Católica de Valparaíso, Casilla 4059, Valparaíso, Chile
\label{unicile}  \and
Max Planck Institute for Extraterrestrial Physics, Giessenbachstrasse, D-85748 Garching, Germany 
\label{maxplank} \and
Universitäts-Sternwarte, Fakultät für Physik, Ludwig-Maximilians Universität München, Scheinerstr. 1, D-81679 München, Germany
\label{sterwarte} \and
INAF - Osservatorio Astrofisico di Arcetri, Largo E. Fermi, I-50125 Firenze, Italy
\label{oaa}  \and
INFN–Sezione di Trieste, Trieste, Italy
\label{infnts} \and
Departamento de Astronom\'ia, Facultad de Ciencias F\'isicas y Matem\'aticas, Universidad de Concepci\'on, Concepci\'on, Chile
\label{uniconception} \and
INAF - Osservatorio Astronomico di Brera, via Brera 28, I-20121, Milano, Italy
\label{oabre}  \and
Institute for Particle Physics and Astrophysics, ETH Z{\"u}rich, Wolfgang-Pauli-Str. 27, 8093 Z{\"u}rich, Switzerland
\label{zurigo}\and
Academia Sinica Institute of Astronomy and Astrophysics (ASIAA), No. 1, Section 4, Roosevelt Road, Taipei 10617, Taiwan
\label{asiaa} \and
Carnegie Observatories, 813 Santa Barbara Street, Pasadena, California, 91101 USA\label{canargieobs}\and
University of Vienna, Department of Astrophysics, Tuerkenschanzstrasse 17, 1180 Vienna, Austria\label{univienna}\and
Space Telescope Science Institute: 3700 San Martin Dr., Baltimore, MD 21218. US.
\label{stsi} \and
Dipartimento di Fisica e Astronomia, Università degli Studi di Padova, Italy \label{unipd}
}

\title{CLASH-VLT: Abell~S1063}
\subtitle{Cluster assembly history and spectroscopic catalogue}
\titlerunning{CLASH-VLT: Abell~S1063}

\author{
A.~Mercurio\inst{\ref{oac}}\fnmsep\thanks{ESO Prog.~ID 186.A-0798}\and 
P.~Rosati\inst{\ref{unife},\ref{inafbo}}\and 
A.~Biviano\inst{\ref{oats},\ref{ifpu}}\and
M.~Annunziatella\inst{\ref{madrid}}\and
M.~Girardi \inst{\ref{units}} \and 
B.~Sartoris\inst{\ref{oats},\ref{ifpu}}\and
M.~Nonino\inst{\ref{oats}}\and
M.~Brescia\inst{\ref{oac}}\and
G.~Riccio\inst{\ref{oac}}\and
C.~Grillo\inst{\ref{unimilano},\ref{iasf-milano}}\and
I.~Balestra\inst{\ref{monaco}}\and
G.~B.~Caminha\inst{\ref{maxplank1}}\and
G.~De~Lucia\inst{\ref{oats}}\and
R.~Gobat\inst{\ref{unicile},\ref{oac}}\and
S.~Seitz\inst{\ref{maxplank},\ref{sterwarte}}\and
P.~Tozzi\inst{\ref{oaa}}\and
M.~Scodeggio\inst{\ref{iasf-milano}}\and
E.~Vanzella\inst{\ref{inafbo}}\and
G.~Angora\inst{\ref{unife},\ref{oac}}\and
P.~Bergamini\inst{\ref{inafbo}}\and
S.~Borgani\inst{\ref{units},\ref{oats},\ref{ifpu},\ref{infnts}}\and
R.~Demarco\inst{\ref{uniconception}} \and
M.~Meneghetti\inst{\ref{inafbo}}\and
V.~Strazzullo\inst{\ref{oats},\ref{oabre}} \and
L.~Tortorelli\inst{\ref{zurigo}} \and
K.~Umetsu\inst{\ref{asiaa}} \and
A.~Fritz\inst{\ref{monaco}}\and
D.~Gruen\inst{\ref{sterwarte}}\and
D.~Kelson\inst{\ref{canargieobs}}\and
M.~Lombardi\inst{\ref{unimilano}}\and
C.~Maier\inst{\ref{univienna}}\and
M.~Postman\inst{\ref{stsi}}\and
G.~Rodighiero\inst{\ref{unipd}}\and
B.~Ziegler\inst{\ref{univienna}}
}
\authorrunning{Mercurio et al.}

   

   \abstract
   {
Understanding the processes responsible for galaxy evolution in different environments as a function of galaxy mass remains heavily debated. Rich galaxy clusters are ideal laboratories for disentangling the role of environmental versus mass quenching, hosting a full range of galaxies and environments.}
   {Using the CLASH-VLT survey, we assembled an unprecedented sample of 1234 spectroscopically confirmed members in Abell~S1063, finding a dynamically complex structure at $\left<z_{\rm cl}\right>=0.3457$ with a velocity dispersion $\sigma_\mathrm{v}=1380_{-32}^{+26}$ km s$^{-1}$. We investigate cluster environmental and dynamical effects by analysing the projected phase-space diagram and the orbits as a function of galaxy spectral properties. 
   }
   {We classify cluster galaxies according to the presence and strength of the [OII] emission line, the strength of the H$\delta$ absorption line, and colours. We investigate the relationship between the spectral classes of galaxies and their position in the projected phase-space diagram. We analyse separately red and blue galaxy orbits. By correlating the observed positions and velocities with the projected phase-space constructed from simulations, we constrain the accretion redshift of galaxies with different spectral types.}
   {Passive galaxies are mainly located in the virialised region, while emission-line galaxies are outside r$_{200}$, and are accreted later into the cluster. Emission-lines and post-starbursts show an asymmetric distribution in projected phase-space within r$_{200}$, with the first being prominent at $\Delta \mathrm{v}/\sigma \lesssim-1.5$, and the second at $\Delta \mathrm{v}/\sigma \gtrsim$ 1.5, suggesting that backsplash galaxies lie at large positive velocities.
   We find that low-mass passive galaxies are accreted in the cluster before the high-mass ones. This suggests that we observe as passives only the low-mass galaxies accreted early in the cluster as blue galaxies, that had the time to quench their star formation. We also find that red galaxies move on more radial orbits than blue galaxies. This can be explained if infalling galaxies can remain blue moving on tangential orbits.}
  {}


\maketitle

%

\section{Introduction}
Massive galaxy clusters at intermediate redshifts represent ideal test-beds for studying the impact of hierarchical cluster assembly on galaxy evolution. The galaxy population in clusters has evolved rapidly over the last 5 Gyr \citep{but78,but84} with the star-forming spiral galaxies found at z$\sim$0.2-0.4 replaced mainly by the S0 galaxies in local clusters \citep{dre97,tre03}. This suggests that clusters accrete blue gas-rich star-forming spirals at $z\gtrsim 0.5-1$. Then these galaxies are somehow transformed into the passive S0s found in the local clusters. This transformation is due to the depletion of their gas reservoir by one or more cluster-related mechanisms, such as ram-pressure (see \citealt{sheen17,Foltz18} for an example) and/or strangulation, tidal stripping, harassment, mergers, and group-cluster collisions (see \citealt{bos06}). 

Observational evidence reveals that, at least at $z<$1, the current properties and past evolution of galaxies are strongly dependent on the environment (e.g., \citealt{bla05, tan05, pen10, nan16, lem19}). Moreover, the fraction of quenched galaxies strongly depends on galaxy mass (e.g., \citealt{vdb20}). For massive galaxies, the star-formation histories and morphologies seem to be determined by their build-up through mergers and the probable consequent feedback from SN and AGN (e.g., \citealt{gom03,tan04,hai06,Fritz2009}). Dwarf galaxies, instead, are more strongly affected by environmental effects. Passive dEs are found as satellites within massive halos, whether that be a cluster, group or massive galaxy \citep{hai06,hai07,bos14, Fritz2014,rob19}. 

Despite decades of work (e.g., \citealt{dre80,kau04,bal04,pos05,smi05,mer10,mor10,kov14,jus15,ann14,Jaffe15,ann16,bos16,rhee2017,oem17,owe19,rhee20}), the evolutionary pathways and the relative importance of processes responsible for galaxy transformations as a function of mass (mass-quenching, e.g., AGN/SN feedback) and environment (environmental-quenching, e.g., ram-pressure and/or tidal stripping, harassment, group-cluster collisions and “starvation”) remain heavily debated. To properly address this issue, it is necessary to observe a large sample of infalling galaxies, all the way, from the cluster centre out to the field, in halos of different masses and at epochs when the galaxy population is still rapidly evolving. 

Spectroscopic information is needed to identify the infalling population from the background and foreground galaxies. In this case, a useful tool to investigate the quenching of star-formation in cluster galaxies is the study of their velocities and positions in the projected phase-space diagram. Cosmological simulations have confirmed that cluster galaxies tend to follow a common path in the 3D phase-space diagram (see Fig.~1 in \citealt{rhee2017}). Thus, it is possible to trace back the accretion histories of galaxies in clusters, associating different populations to different phase-space locations (virialised, infalling, backsplash; e.g. \citealt{Pasquali19}). 

In this context, we present a detailed study of the galaxy population of the galaxy cluster Abell~S1063 (hereafter A~S1063, \citealt{abe89}), a very massive cluster at $z=0.348$ \citep{kar15} with total mass of $(2.9 \pm 0.3)\times10^{15}\ \mathrm{M}_{\odot}$ \citep{sar20}. The cluster was also catalogued as RXJ2248.7$-$4431, as it was detected in the ROSAT All-Sky Survey \citep{deg99,guz99}. It has a high X-ray luminosity ($L_X\approx 8\times 10^{45}$ erg s$^{-1}$) and high X-ray temperature, $T_X\approx 13$~KeV \citep{gom12}. A SZ signal is also detected with high significance in the Planck data \citep{plk11,pla10}.  

\citet{gom12} proposed that A~S1063 hosts a recent merger event close to the plane of sky along the NE-SW direction. They derived their conclusion as based on the spectroscopic GMOS data for 51 members and X-ray Chandra data. This NE-SW elongation is also visible in X-ray and in the Dark-Matter (DM) distribution as reconstructed by the strong lensing analysis (see Fig.2 in Bonamigo et al. 2018). However, the DM and hot-gas mass distributions have different shapes and centres, due to their intrinsically different physical properties. While the DM component is roughly centred on the of the brightest cluster galaxy (BCG), the hot-gas mass distribution is skewed toward the northeast and is rounder than the DM one. Moreover, \citet{Xie20} found a giant radio halo with a size of $\sim$1.2~Mpc and an integrated spectral index that steepens between 1.5 and 3.0 GHz. 
 
\cite{sar20}  performed a full dynamical reconstruction of the mass density profile from the very centre ($\sim 1$~Kpc) out to the virial radius. They disentangled the DM profile from the total mass profile and showed the different contributions of the stellar mass profile of cluster members, BCG, and of the intra-cluster gas mass profile. They found the inner slope of the DM density profile modelled as a gNFW, $\gamma_{\rm DM}=0.99 \pm 0.04$, in agreement with the predictions from the $\Lambda$CDM model.
 
Other previous studies of this cluster include the weak lensing (WL) analysis presented in \cite{gru13}, the detection of ultra-diffuse galaxies in \cite{lee17}, the analysis of the Kormendy relation in \cite{tor18}, the analysis of the enhancement in (O/H) \cite{ciocan2020}, and the analysis of Chandra X-ray observations and 325 MHz Giant Metre Radio Telescope observations in \cite{ram21}. 

A~S1063 was part of the HST treasury program CLASH \citep{pos12} and the Frontier Field initiative\footnote{\footnotesize{http://www.stsci.edu/hst/campaigns/frontier-fields/HST-Survey}} (\citealt{lot17}, hereafter FF). In this paper, we exploit and make public the extensive CLASH-VLT spectroscopic campaign (\citealt{ros14}, Rosati et al. 2021, in prep.) of the A~S1063 field with the VIMOS spectrograph, augmented with MUSE integral field spectroscopy in the cluster core \citep{cam17}. 

This paper specifically presents the analysis of low and medium-resolution VIMOS spectroscopic data, combined with multi-band photometry from HST/FF and the WFI at the ESO 2.2m telescope. The purpose of this study is to explore the dominant quenching processes by focusing on the accretion histories of cluster galaxies through the analysis of the projected phase-space diagram enabled by the unique spectroscopic sample available. The structure of this paper is the following. Observations and catalogues are described in Sect.~\ref{sec:2} and Appendix~\ref{app:A}. The member selection, the dynamical analysis and the substructures are discussed in Sects.~\ref{sec:3}, \ref{sec:4}, and \ref{sec:5} and Appendix~\ref{app:B}, respectively. We describe the spectroscopic classification of galaxies in Sect.~\ref{sec:6}. In Sect.~\ref{sec:7}, we discuss the accretion history of galaxies in the cluster, using the location in the phase space diagram as a diagnostic of the accretion redshift, the comparison with simulations, and the analysis of the orbits. Finally, in Sects.~\ref{sec:8}, we discuss and summarise our results.

Throughout the paper we adopt a cosmology with $\Omega_{\rm m}$ = 0.3, $\Omega_\Lambda$ = 0.7, and $H_0$ = 70 \ks Mpc$^{-1}$. According to this cosmology, 1 arcmin corresponds to 0.294 Mpc at $z$ = 0.3457. Unless otherwise specified, figures are oriented with North at the top and East to the left, magnitudes are in the AB system, and stellar masses are obtained by using a Salpeter IMF \citep{sal55}.

\section{Observations and catalogues}
\label{sec:2}

\begin{figure*}
\centering
\includegraphics[scale=0.6]{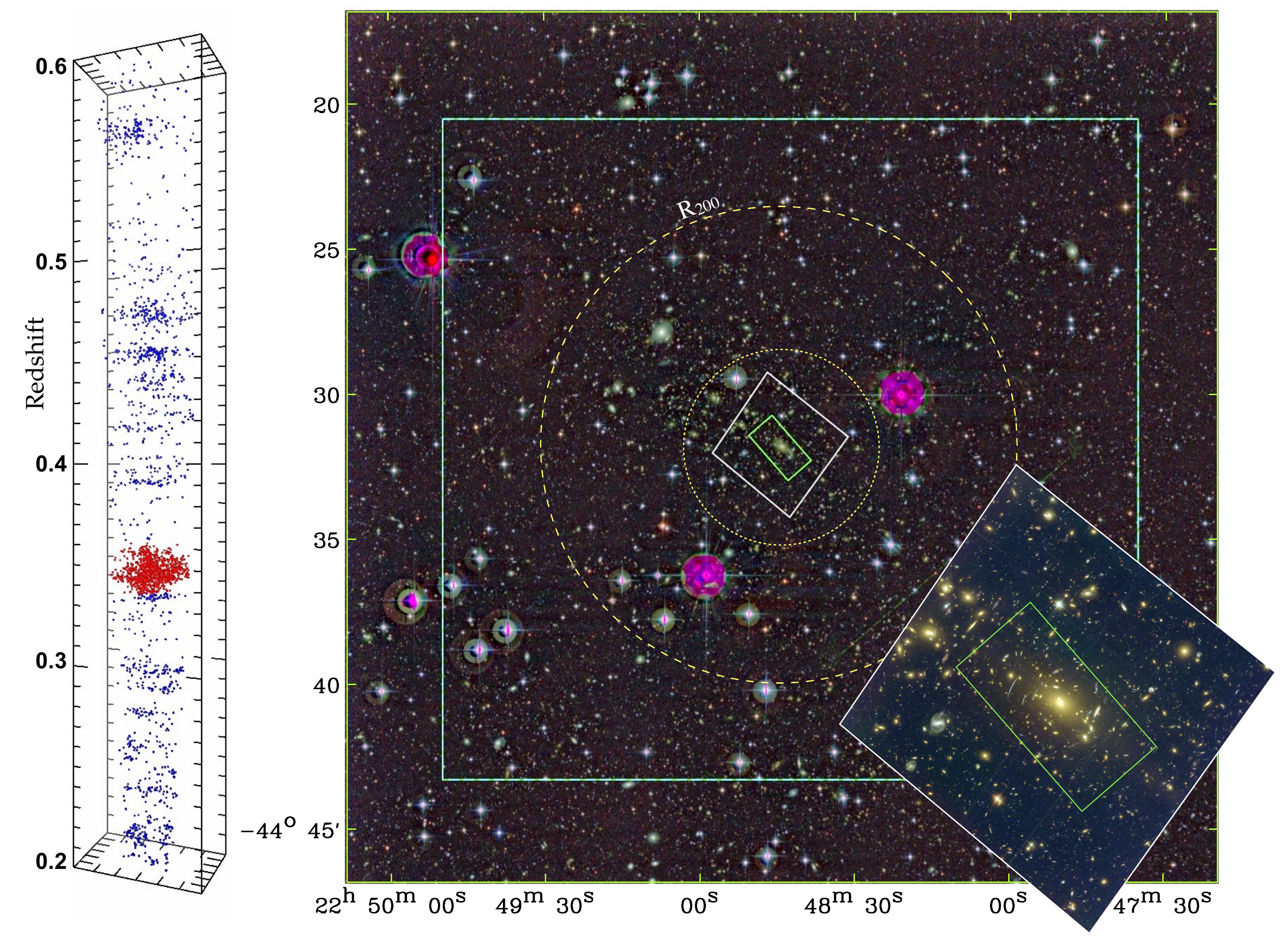}
\caption{{\it Right panel}: WFI colour composite image of A~S1063 (UBVRIz, $30\times 30$ arcmin$^2$). The cyan square indicates the VIMOS spectroscopic survey area ($28.9\times 22.7$ arcmin$^2$), while the white and green polygons indicate the HST FF colour composite image (F435W, F606W, and F814W) and the footprint of MUSE observations (2 pointings, $2\times 1 $ arcmin$^2$) blown up in the inset. The dotted and dashed circles have radii R = 1~Mpc and R = r$_{200}$, respectively. {\it Left panel}: 3D spatial distribution of all measured redshifts at $0.2<z<0.6$, where cluster members are marked in red.}
\label{fig1}
\end{figure*}

In this paper, we examine the spectro-photometric dataset of A~S1063. It provides a unique combination of photometric coverage over a wide wavelength range [0.2-1.6] $\mu$m, from HST, ground-based WFI observations, and wide-field VLT-VIMOS spectroscopy, further complemented with the integral-field VLT-MUSE spectroscopy in the central 0.3 Mpc region. The photometric data and catalogues are presented in Appendix~\ref{app:A}, while, in this section, we describe the spectroscopy. Figure~\ref{fig1} summarises the data used in this paper.

\subsection{Spectroscopy}
\label{sec:21}

The analysis presented here is based on low and medium-resolution VIMOS spectroscopic data obtained in 25$\times$25~arcmin$^2$. Moreover, we use MUSE integral-field spectroscopy in the cluster core that we published in \citet{cam17} and additional redshifts from other sources (see Sect.~\ref{sec:22}).

\subsubsection{VIMOS Data}
\label{sec:211}

The cluster A~S1063 was observed with VIMOS as part of the ESO Large Programme 186.A-0798 "Dark Matter Mass Distributions of Hubble Treasury Clusters and the Foundations of $\Lambda$CDM Structure Formation Models" (PI: P. Rosati, \citealt{ros14}, hereafter CLASH-VLT), which performed a panoramic spectroscopic survey of the 13 CLASH clusters visible from ESO-Paranal. 

\begin{figure}[ht]
   \centering
   \includegraphics[width=0.5\textwidth]{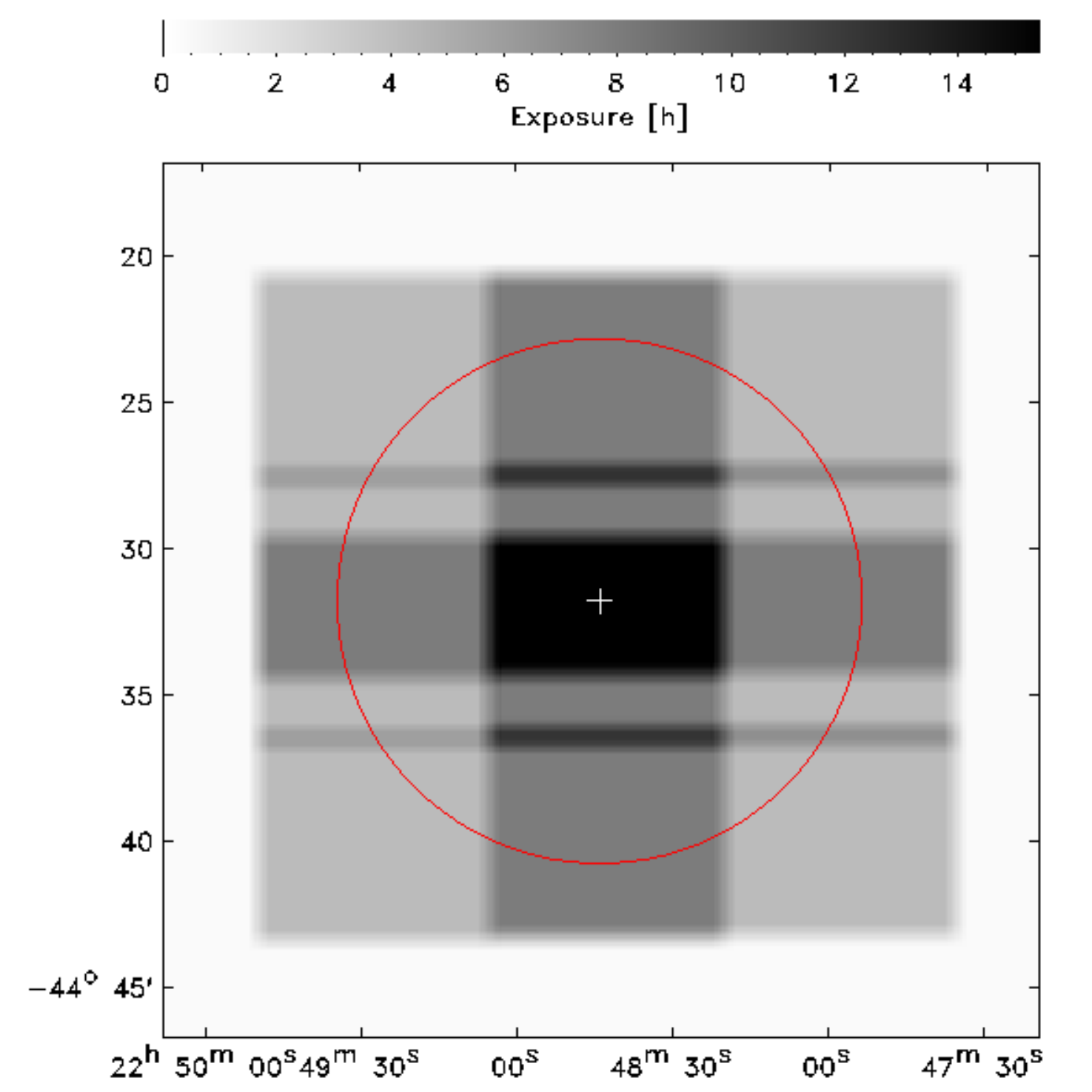}
   \caption{Exposure map of the 16 VIMOS pointings in the field of A~S1063. The red circles are centred on the BCG and have radii of r$_{200}=2.63$ Mpc (see Sect.~\ref{sec:4}), the cross marks the BCG position.
   }
          \label{fig2}%
    \end{figure}

The VIMOS observations were designed in sets of four separate pointings, each with a different quadrant centred on the cluster core (see Fig.~\ref{fig2}), for a total of 16 masks. We used the low-resolution blue grism (LRb) for twelve of those, covering the spectral range 3700-6700 $\AA$ with a resolution of R = 180. For the remaining four, we used the medium resolution grism (MR) in the range 4800-10000 $\AA$ with a resolution of R = 580. The pointings overlap on the cluster centre to achieve longer integration times on faint arcs and other interesting strong lensing (SL) features. They have the largest possible number of slits on candidate cluster members in the cluster crowded central region. 
The resulting exposure map of the 16 VIMOS pointings footprint is shown in Fig.~\ref{fig2} and the number of exposures and integration time are reported in Table~\ref{tab:1}.

\begin{table}[]
    \caption{\label{tab:1} List of VIMOS Observations of A~S1063.}
    \centering
    \begin{tabular}{|c|c|c|}
\hline
Mask ID & Date & Exp. Time (s)\\	
\hline	
\multicolumn{3}{|l|}{Low-resolution masks} \\
\hline
MOS\_R2248\_LRb\_1\_M1 & Jun 2013 & 3$\times$1200 \\
MOS\_R2248\_LRb\_2\_M1 & Jun 2013 & 3$\times$1200 \\
MOS\_R2248\_LRb\_3\_M1 & Jul 2013 & 3$\times$1200 \\
MOS\_R2248\_LRb\_4\_M1 & Jul 2013 & 3$\times$1200 \\
MOS\_R2248\_LRb\_1\_M2 & Oct 2013 & 3$\times$900 \\
MOS\_R2248\_LRb\_2\_M2 & Oct 2013 & 3$\times$900 \\
MOS\_R2248\_LRb\_3\_M2 & Oct 2013 & 3$\times$900 \\
MOS\_R2248\_LRb\_4\_M2 & Oct 2013 & 3$\times$900 \\
MOS\_R2248\_LRb\_1\_M3 & Aug 2014 & 3$\times$1200 \\
MOS\_R2248\_LRb\_4\_M3 & Aug 2014 & 3$\times$1200 \\
MOS\_R2248\_LRb\_2\_M4 & May 2015 & 3$\times$1200 \\
MOS\_R2248\_LRb\_3\_M4 & Sep 2014 & 3$\times$1200 \\
\hline	
\multicolumn{3}{|l|}{Medium-resolution masks}\\
\hline
MOS\_R22248\_MR\_1\_M1 & Jul 2013 & 3$\times$1200 \\
MOS\_R22248\_MR\_2\_M1 & Jul 2013 & 3$\times$1200 \\
MOS\_R22248\_MR\_3\_M1 & Jul 2013 & 3$\times$1200 \\
MOS\_R22248\_MR\_4\_M1 & Jul 2013 & 3$\times$1200 \\
\hline
\end{tabular}
\tablefoot{In col.~1, the mask identification number is reported. In col.~2, the date of the observations. In col.~3, the number of exposures and integration time of single exposures.}
\end{table}

\begin{figure}[ht]
   \centering
   \includegraphics[width=0.5\textwidth]{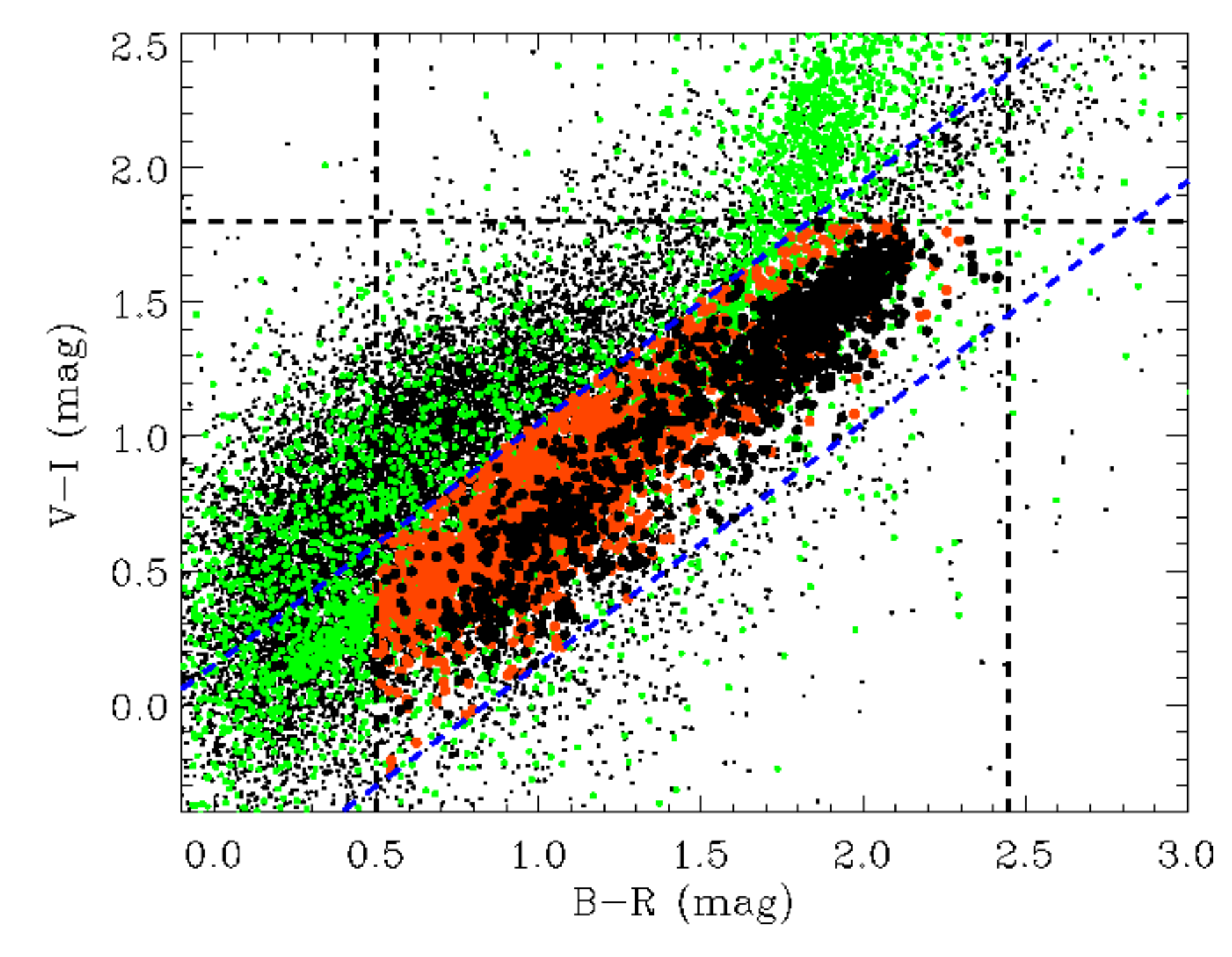}
   \caption{WFI ${\rm V-I}$ vs. ${\rm B-R}$ colours for all of the extracted sources with ${\rm R}$ magnitudes $\le$ 24. The blue and black dashed lines show the colour cuts defining the box used for target selection. Small black and green dots are the sources classified as galaxies and stars in the WFI photometric catalogue, respectively. Larger data points mark cluster member galaxies in black and non-members in red, belonging to our CLASH-VLT spectroscopic catalogue.}
          \label{fig3}%
    \end{figure}

We selected spectroscopic targets through specifically defined cuts in the colour-to-colour space using the WFI photometry. These cuts were set to include both blue and red galaxies at the cluster redshift, according to the expected colours. Figure~\ref{fig3} shows the selection box in the V–I versus B–R diagram.

We reduced the data using the VIMOS Interactive Pipeline Graphical Interface (VIPGI, \citealt{sco05}) pipeline, which performs bias subtraction, flat-field correction, bad-pixel cleaning, sky subtraction, fringing correction, and wavelength calibration. 

The redshift determination follows the procedure described in \cite{bal16}. First, we run the EZ software \citep{gar10} for cross-correlation with template spectra. Then, we visually inspect redshift solutions obtained in the first step. During the visual check, we also assigne a Quality Flag (QF) to each redshift, according to four classes: "Secure" (QF = 3), several emission lines and/or strong absorption features are identified, 100\% reliability; "Emission-line" (QF = 9), redshift based on one or more emission lines, $>$90\% reliability; "Likely"(QF = 2), intermediate-quality spectra with at least two spectral features well-identified, $\sim$80\% reliability; "Insecure" (QF = 1), low signal-to-noise ratio spectra, i.e., with spectral features less clearly identified, 20-40\% reliability.

We extract 6477 spectra and measure 4199 redshifts with QF $\ge 1$, corresponding to a success rate of $\sim$ 65\%. When considering repeated observations of the same objects, we obtaine 3607 redshifts of different objects. We use repeated observations to check the redshift uncertainties as a function of the spectral resolution and the assigned quality flag. We found that the uncertainties vary between 75 and 150 \kss for MR and LRb observations, respectively \citep[consistent with our previous estimates, see][]{biv13}. 

\subsubsection{MUSE Data}
\label{sec:212}

A~S1063 was observed with the MUSE integral field spectrograph \citep{kar15,cam16,kar17} as part of ESO Programme IDs 60.A-9345(A) (P.I. Caputi \& Grillo) and 095.A-0653(A) (P.I. Caputi). Two pointings, covering the North-East (NE) and the South-West (SW) sides of the cluster core, were observed (see the bottom right panel in Fig.~\ref{fig1}). The SW pointing has a total exposure time of 3.1 h, with a seeing of $\sim 1.1\arcsec$, while the NE pointing has a coadded exposure time of 4.8 h with $0.9\arcsec$ seeing. Data reduction and redshift measurements are fully described in \cite{kar15,kar17}, and the catalogues of cluster members and multiply lensed images, together with the first SL model, in the $\sim\! 2\times 1$ arcmin$^2$ central region, were presented in \citet{cam16}. 
The MUSE data provided 175 additional redshifts. 

   \begin{figure}[ht]
   \centering   
   \includegraphics[width=0.5\textwidth]{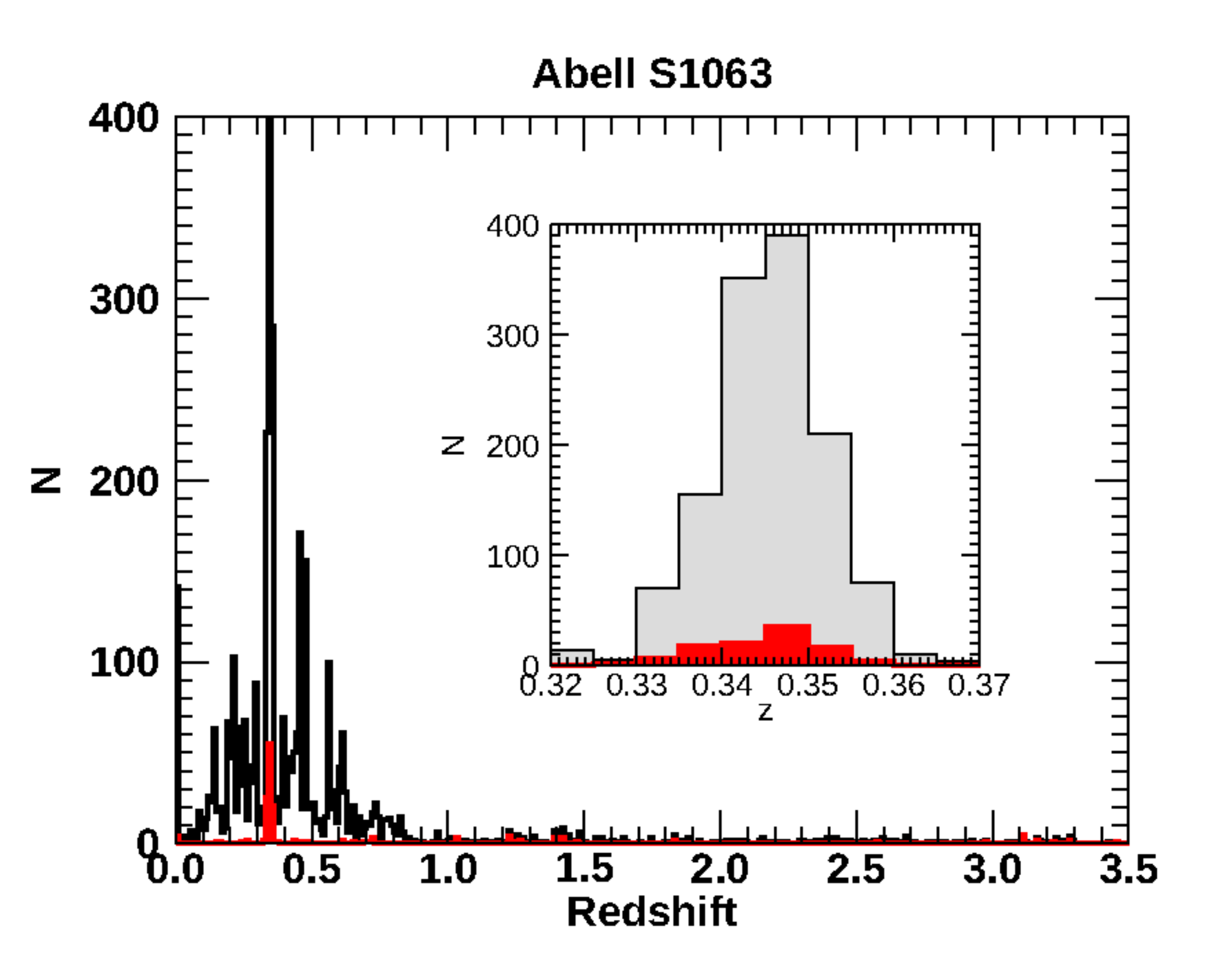}
   \caption{Spectroscopic redshift distribution of 3850 sources in A~S1063 from the CLASH-VLT catalogue, including 175 MUSE redshifts. The inset shows a zoom-in around the mean cluster redshift $\left<z_{\rm cl}\right>$ = 0.3457 (red histogram for MUSE galaxies) of the 1234 cluster members found from the kinematic analysis.}
          \label{fig4}%
    \end{figure}

\subsection{Spectroscopic catalogue}
\label{sec:22}

\begin{figure}[ht]
   \centering
   {\includegraphics[width=0.4\textwidth]{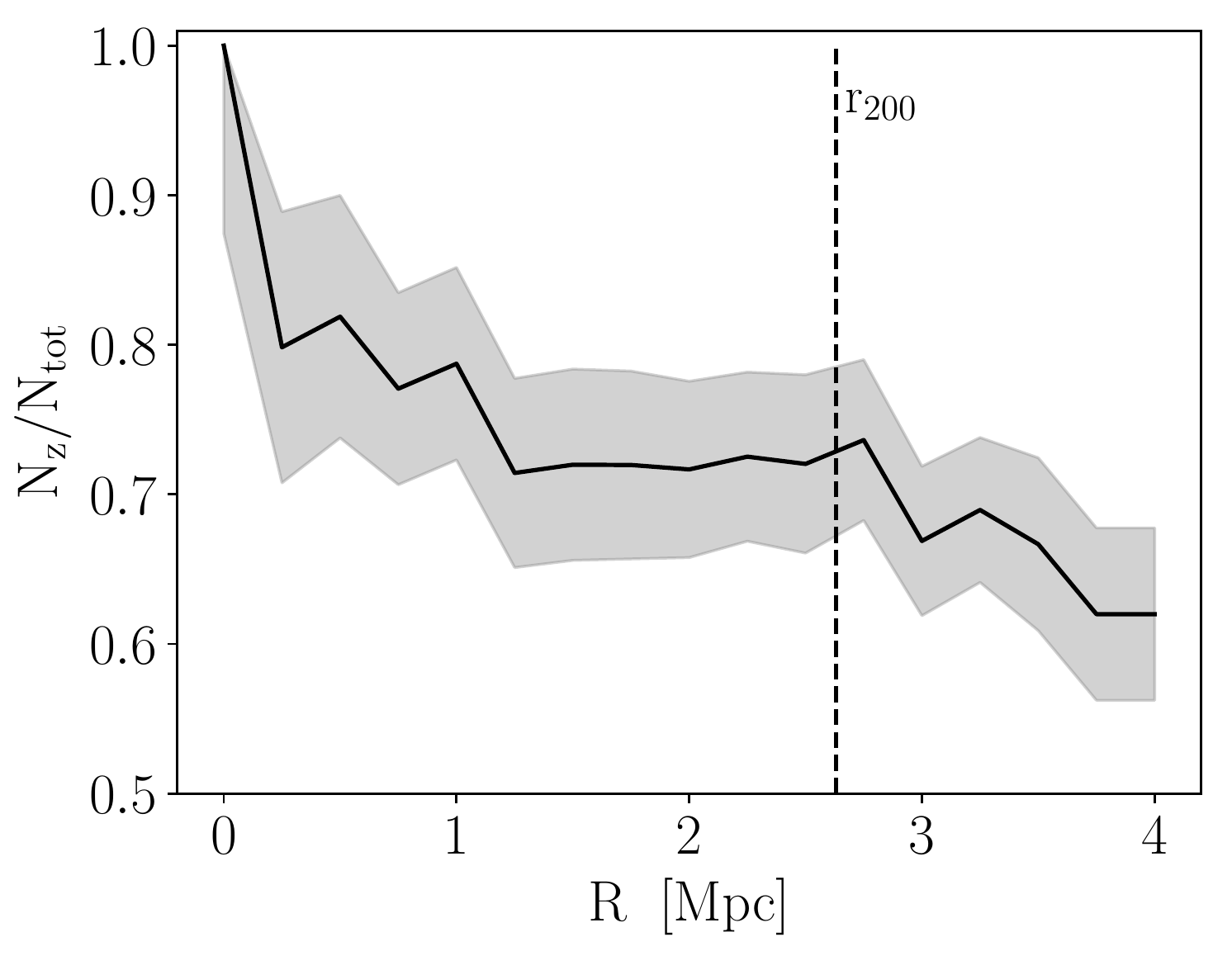}}
   \caption{Completeness of the spectroscopic sample as a function of the custercentric distance. The shaded area indicates the 68\% confidence regions, according the  algorithm  of  Gehrels (\citealt{geh86}). The dashed line indicates of r$_{200}$ =2.63 Mpc (see sect~\ref{sec:4}).}
   \label{fig5}%
    \end{figure}

The final spectroscopic catalogue contains 3850 redshifts, with QF $\ge$ 2, where a single entry\footnote{\footnotesize The redshift of the single entry is the mean of the redshifts of the spectra with QF = 3 or 9 if available or, if only observations with QF~=~2 are available, the mean of the spectra with QF = 2.} is preserved in duplicate observations of the same object. 3607 redshifts are measured from VIMOS spectra, 175 are from MUSE data, 21 from \cite{gom12}, 30 from the Grism Lens-Amplified Survey from Space (GLASS, GO-13459, PI: Treu; \citealt{tre16}), and 17 unpublished redshifts from Magellan observations (D. Kelson, private communication). We measure and visually check the redshifts for VIMOS, MUSE, and GLASS data and assign the QF, as explained in Sect.~\ref{sec:211}. Thus, for these 3812 sources, we obtain 3005 Secure, 183 Emission-line, and 624 Likely redshifts. Figure~\ref{fig4} shows the redshift distribution of all the 3850 sources. The 175 MUSE objects are highlighted in red. The inset shows a zoom-in around the mean cluster redshift ($\left<z_{\rm cl}\right>$ = 0.3457, see below). 
The redshift catalogue is publicly available at the CLASH-VLT website\footnote{Currently located at \url{https://sites.google.com/site/vltclashpublic/}}.
We check the spectroscopic sample completeness as a function of position on the sky and magnitude. We define completeness as the ratio between the galaxies with measured redshifts and the number of galaxies in the colour-colour box we use to select targets (see Fig.~\ref{fig3}), down to the fiducial limiting magnitude of ${\rm R}=24$. 
Figure~\ref{fig5} shows the spectroscopic completeness as a function of the distance from the BCG ([R.A.=$22^{\mathrm{h}}48^{\mathrm{m}}44.0^{\mathrm{s}}$, Dec.=$-44^{\mathrm{d}} 31^{\mathrm{m}}51^{\mathrm{s}}$ (J2000.0)]), adopted as the cluster centre. As it can be inferred from this plot, the completeness is approximately 1 in the central 0.25 Mpc, thanks to the MUSE coverage, then it decreases to 0.8 out to 1.25 Mpc, down to 0.7 from 1.25 Mpc to 2.75 Mpc.

\section{Selection of cluster members}
\label{sec:3}

\begin{figure}[ht]
   \centering   
   \includegraphics[width=0.5\textwidth]{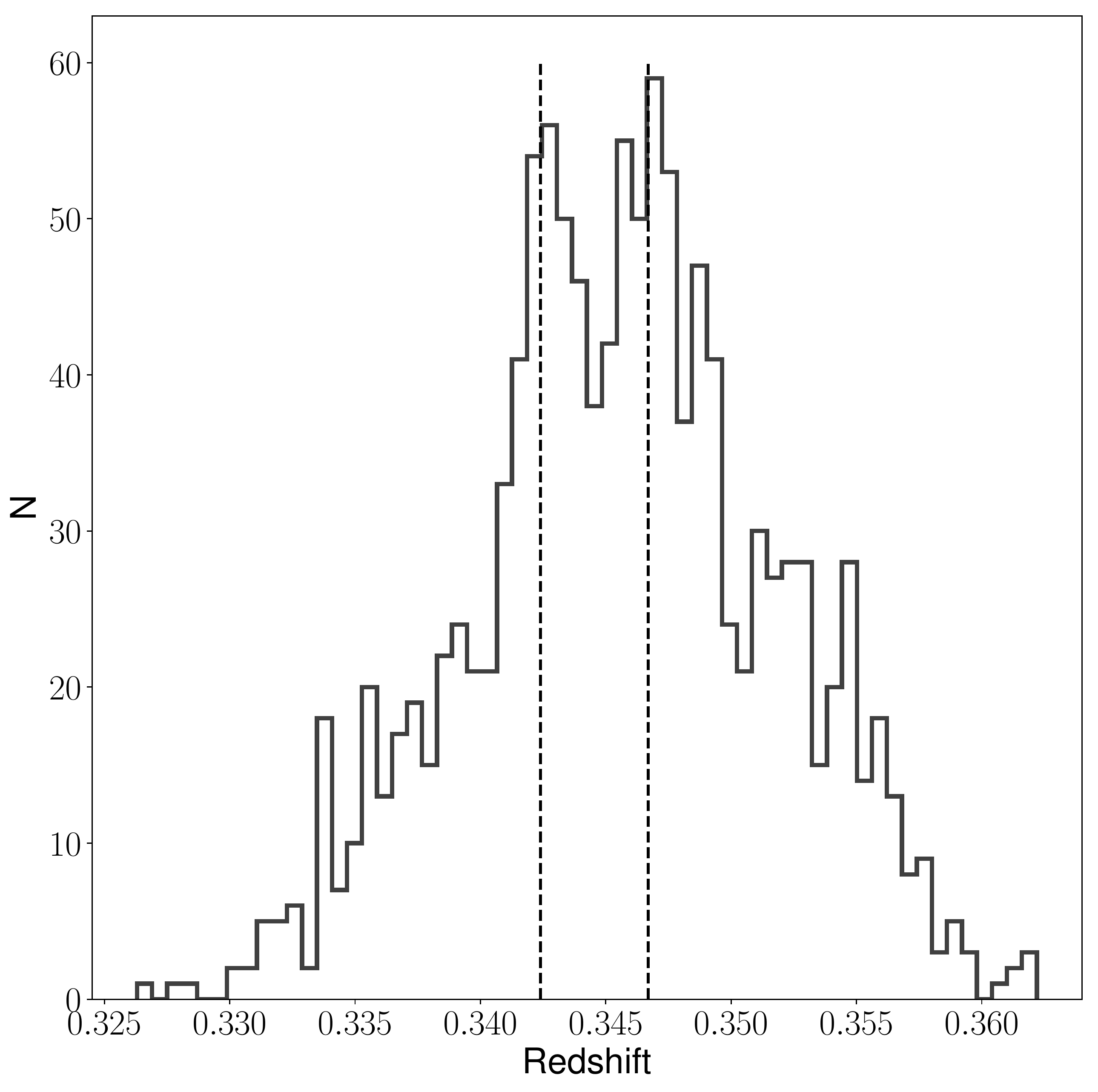}
   \caption{Redshift distribution of the 1234 cluster members found from the kinematical analysis of the members in A~S1063. The two dashed lines show the two overlapping peaks of 590 and 715 galaxies at $z=0.3424$ and $z=0.3467$ detected with 1D-DEDICA.}
          \label{fig6}%
    \end{figure}

In order to select cluster members, we apply the two-step method called ``peak+gap'' (P+G) already applied in \citet[][see also refs. therein]{girardi2015}. The method is a combination of the 1D adaptive-kernel method DEDICA \citep{pisani1993} and the ``shifting gapper'', that uses both position and velocity information \citep{fadda1996,girardi1996}. In the first step, the 1D-DEDICA method detects A~S1063 as two overlapping peaks of 590 and 715 galaxies at $z=0.3424$ and $z=0.3467$, respectively, in the range $0.30905\leq z \leq 0.3716$ (see Fig.~\ref{fig6}), for a total of 1305 candidate members.

\begin{figure}[ht]
\centering
\includegraphics[width=0.5\textwidth]{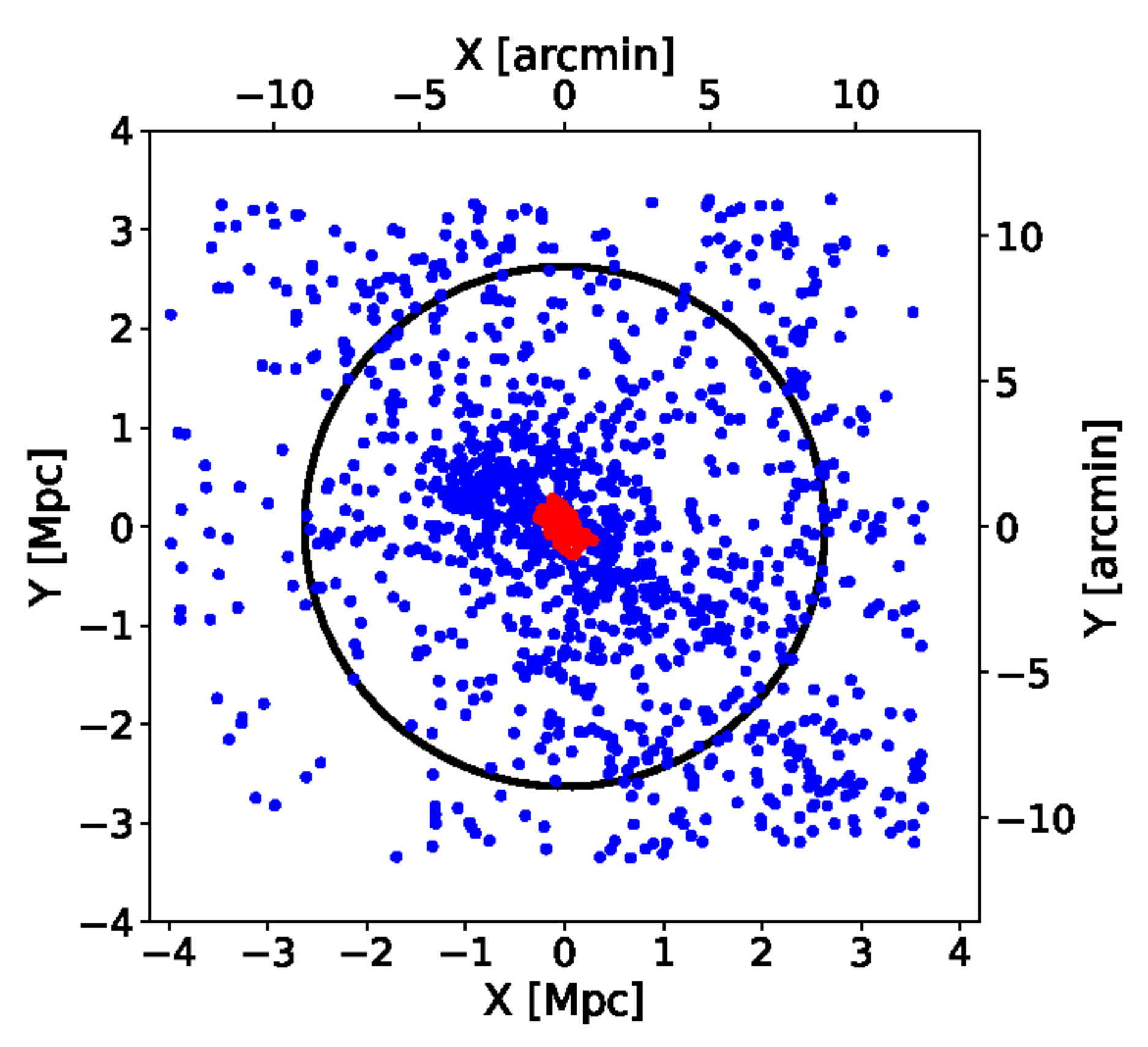}
\caption{2D spatial distribution of VIMOS (blue circles) and MUSE (red circles) spectroscopically confirmed members. The large black circle has a radius equal to r$_{200}$ = 2.63 Mpc (see Sect.~\ref{sec:4}).
}
\label{fig7}
\end{figure}

The second step in the member selection procedure combines galaxy positions and velocities to reject sources that are too far in velocity from the main body of galaxies within a fixed radial bin that is shifted along with the clustercentric distance.  The procedure is iterated until the number of cluster members converges to a stable value.  We use a velocity gap of $800$ \ks -- in the cluster rest-frame -- and a bin of 0.6 \hh, or large enough to include 15 galaxies.  As for the centre of A~S1063, we adopt the position of the BCG. The ``shifting gapper'' procedure rejects other 71 interlopers that survived to the first step of our member selection procedure. Thus, we obtain a sample of 1234 fiducial members (see Fig.~\ref{fig6}), whose spatial distribution is shown in Fig.~\ref{fig7}, with MUSE members highlighted in red.

By applying the biweight estimator \citep[ROSTAT software]{bee90} to the 1234 cluster members, we compute a mean cluster line-of-sight (LOS) velocity $\left<V\right>=\left<cz\right> =(103\,640\pm39$) \kss, corresponding to a mean cluster redshift $\left<z_{\rm cl}\right>=0.3457\pm0.0001$.

\begin{figure}[ht]
   \centering
   \includegraphics[width=0.5\textwidth]{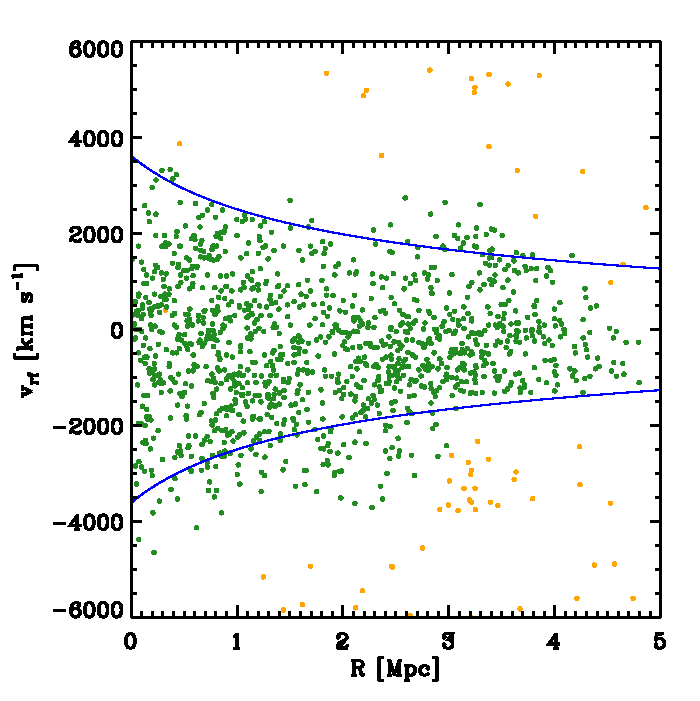}
   \caption{Phase-space diagram including 1305 galaxies from the spectroscopic sample, where rest-frame velocities are plotted vs. the clustercentric distance.  Green circles highlight the 1234 cluster members. Yellow circles indicate galaxies which lie in the selected cluster redshift range but are rejected in the second step of our member selection procedure. The escape velocity curves are also shown (blue lines).}
          \label{fig8}%
    \end{figure}

The positions of cluster members in the (projected) phase-space are shown in Fig.~\ref{fig8}. The inspection of Fig.~\ref{fig8} suggests that escape velocities (blue lines, refer to the following section for how these are computed) are adequate to describe the position of the A~S1063 galaxies in the phase-space. This can also be considered as a posteriori validation of our member selection procedure (see Sect.~\ref{sec:4}).  

\section{Dynamical analysis}
\label{sec:4}

In this paper, we use the mass model derived from \citet{sar20} to compute the escape
velocity profile, and we briefly summarise the results presented in \citet{sar20} to compute the total mass profile of A~S1063, using the same spectroscopic dataset presented here. Such analysis is based on the combined constraints obtained from the velocity dispersion profile of the stellar component of the BCG and the velocity distribution of cluster member galaxies. \citet{sar20} used an extension of the MAMPOSSt technique \citep{mamon13} to solve the Jeans equation of dynamical equilibrium. \citet{sar20} determined a value $\gamma_{\rm DM} =0.99 \pm 0.04$ for the inner slope of the gNFW profile, in agreement with the expectation from the CDM model. They also derived the contributions of the stellar mass component of cluster members and of the intra-cluster gas and of the DM profile to the cluster total mass profile. Their study also showed an excellent agreement among the projected total mass profiles obtained from a combined strong plus weak lensing analysis, hydrostatic X-ray analysis and their dynamical analysis, thus indicating a negligible hydrostatic mass bias.

In this paper, we use the total mass profile of A~S1063 derived by  \citet{sar20} adopting a simple NFW parametrization, whose best fit parameters are $\mathrm{r_s = 0.84 \pm 0.18}\, {\rm Mpc}$, ${\rm r_{200,c}} = 2.63 \pm 0.09\, {\rm Mpc}$, and ${\rm M = 2.9 \pm 0.3 \times 10^{15}\, M_\odot}$.

This mass model is used to compute the projected escape velocity profile in Fig.~\ref{fig8}, which is directly related to the potential well as $v_{\rm esc} = \sqrt{2\phi}$ and, thus, to the cluster total mass. To derive the $v_{\rm esc}$ profile, we follow the procedure described in \citet[and references therein]{stark16}. According to \citet{nandra12} and \citet{stark16}, a massive particle in the vicinity of a galaxy cluster with gravitational potential $\tilde{\phi}(r)$ experiences an effective potential, $\phi(r)$, that is the sum of the cluster potential and a second term related to the expansion of the universe, that can be thought as a repulsive force that opposes the inward pull of the cluster mass distribution: $\nabla \phi(r) = \nabla \tilde{\phi}(r)+q(t)H^2(t) r$. We integrate this equation between a given $r$ and the equivalent radius ($r_{\rm eq}$), that is where the acceleration due to the cluster gravitational potential and the acceleration of the expanding universe are equivalent \citep{behroozi2013}. 
Following \citet{lokas01}, we calculate the cluster potential $\tilde{\phi}$ for a NFW mass density profile. This potential is a function of the NFW mass profile parameters: r$_{200}$, and r$_{\mathrm{s}}$, that we obtained from the MAMPOSSt analysis. 
Finally, we project the 3D escape velocity profile rescaling it with a function of the velocity anisotropy, according to the formulation presented in \citet{diaferio97, diaferio99}. The values of the velocity anisotropy profile parameters were also measured in the MAMPOSSt analysis.
%
\section{Substructures}
\label{sec:5}

We estimate the cluster LOS velocity dispersion, $\sigma_V$, using the biweight estimator and applying the cosmological correction and the standard correction for velocity errors \citep{dan80}. We obtain $\sigma_V=1380_{-32}^{+26}$ \kss, where errors are estimated through a bootstrap technique.    

The BCG velocity is $V_{\rm BCG}=(104\,088\pm40$) \ks 
and its rest-frame relative velocity with respect to the mean velocity is $\Delta V=(V_{\rm BCG}-\left<V\right>)/(1+z)=333$ \kss. The corresponding value of $\abs{\Delta V}/\sigma_V=0.24$ is a rather typical velocity offset for BCGs \citep{lauer2014}, as also confirmed by \cite{geb91}. 

\begin{figure*}
\centering
\includegraphics[width=0.45\textwidth, bb= 80 220 530 570,clip]{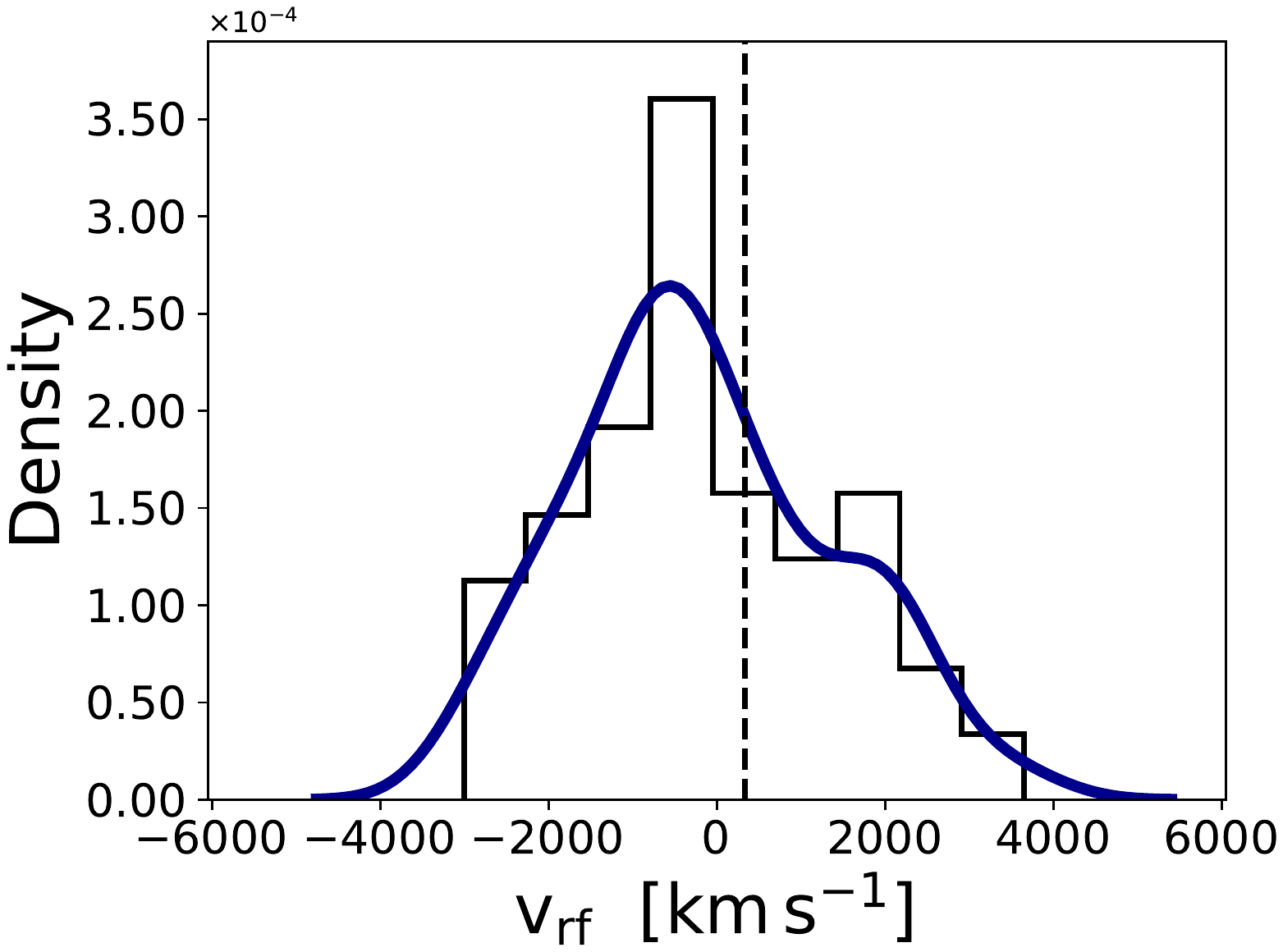}
\includegraphics[width=0.45\textwidth, bb= 80 220 530 570,clip]{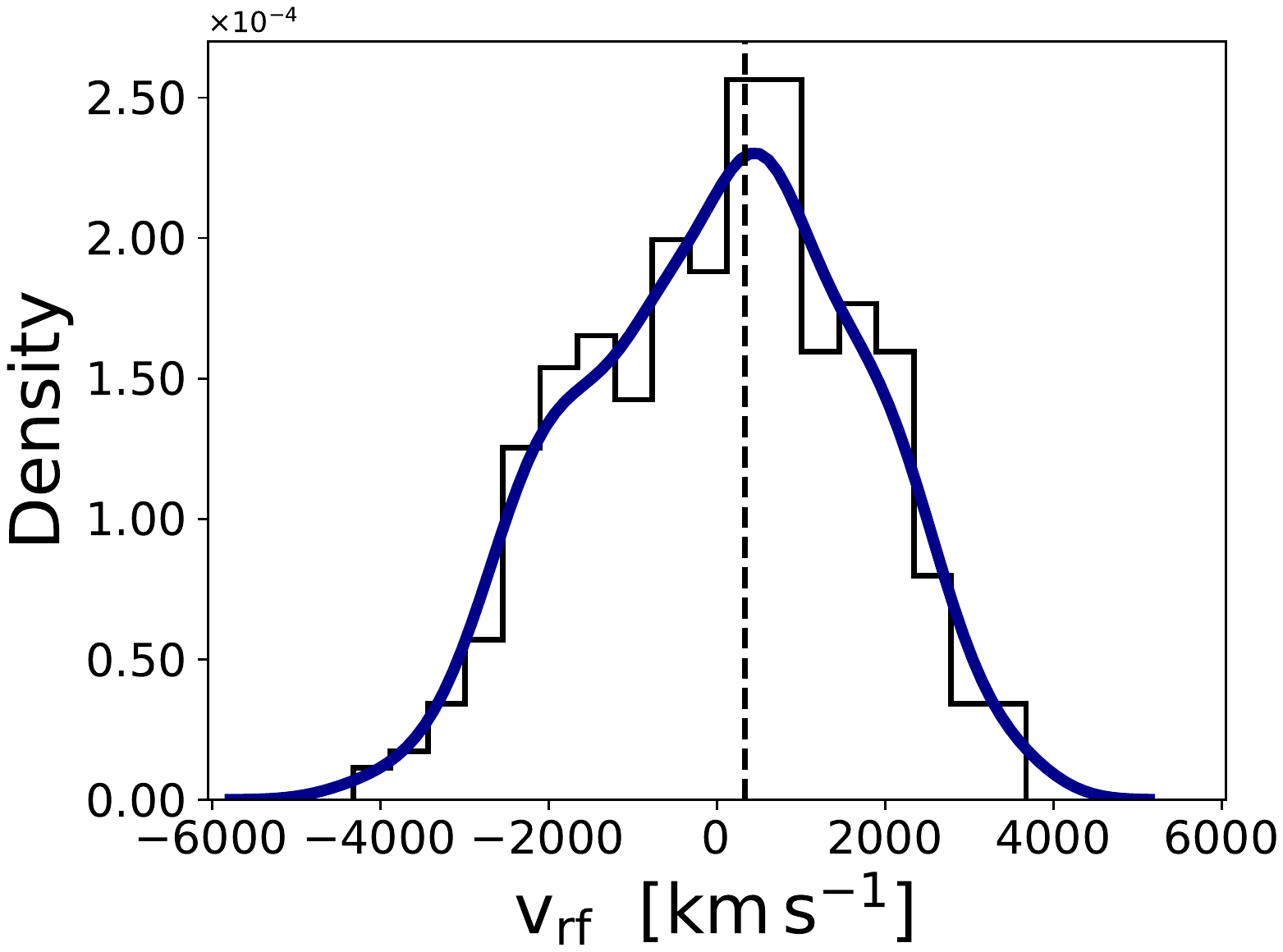}
\caption
{Rest-frame velocity distribution relative to the mean cluster velocity of the 120 galaxies having R $<$ 20 ({\it Left panel}), and of the 395 galaxies within 1~Mpc from the cluster centre ({\it Right panel}). The smooth curve is the probability density function obtained with a kernel density estimator. In both panels, the dashed line indicates the rest-frame  velocity of the BCG: $(V_{\rm BCG}-\left<V\right>)/(1+z)=333$ \kss.}
\label{fig9}
\end{figure*}

When selecting galaxies with R $<$ 20 (120 galaxies, hereafter indicated as "bright galaxies"), they show a bimodality in the velocity distribution (left panel of Fig.~\ref{fig9}), with the BCG in the middle (the arrow in the figure).  The secondary peak is at velocities higher than that of the BCG.
We apply the 1d KMM algorithm (cf. \citealt{ashman1994, girardi2008}) to the velocity distribution of the R $<20$ galaxy subsample, and we find that the superposition of two Gaussians with n$_1$ = 96 and n$_2$ = 24 members, at mean redshifts z$_1$=0.3413 (102617 \kss) and z$_2$=0.3555 (106565 \kss), is a better description of the velocity distribution than a single Gaussian, with a probability of 94.9\%. 

An asymmetry in the velocity distribution is also evident by selecting the 395 member galaxies in the cluster inner 1~Mpc (right panel of Fig.~\ref{fig9}). In this subsample, the secondary peak is at lower velocities than that of the BCG, with the BCG lying in the main peak. Also, in this case, by applying the 1d KMM algorithm, we find a very high probability (99.3\%) that a mixture of two Gaussians, with n$_1$ = 135 and n$_2$ = 260 members, at mean redshifts z$_1$=0.3377 (101245 \ks) and z$_2$=0.3501 (105955 \ks), is a better description of the velocity distribution than a single Gaussian.

We further test for the presence of substructures by considering the velocity and spatial distributions of cluster galaxies at the same time by using the 3D-DEDICA method described in Appendix~\ref{app:B} and the test of \cite{dressler1988}, discussed below. 
The test of \cite{dressler1988} compares the local mean velocity and velocity dispersion as computed around each galaxy with the global cluster values. We use the modification of the method introduced by \cite{girardi1997a,girardi2010}, which considers only the more useful and immediate kinematical indicator based on the local mean. Following the methodology of \citet[hereafter DSv-test]{girardi2010}, the kinematical indicator is based on the deviation of the local mean, $\delta_{{\rm v},i}^2= [(N_{\rm nn}+1)/\sigma_{\rm v}^2][(\overline {\rm v_l} - \overline {\rm v})^2]$, where the subscript "l" denotes the local quantity computed over the group formed by the galaxy {\it i} itself and its $N_{\rm{nn}}=10$ neighbors. The value $\Delta$ (i.e., the sum of the $\delta_{{\rm v},i}$ of the individual $N$ galaxies) gives the cumulative deviation of the local mean velocities from the global cluster mean velocity. The significance of $\Delta$, that is, how far is the system from dynamical equilibrium, is checked by running 1000 Monte Carlo simulations, randomly shuffling the galaxy velocities.

\begin{figure}[ht]
\centering
\includegraphics[width=0.5\textwidth]{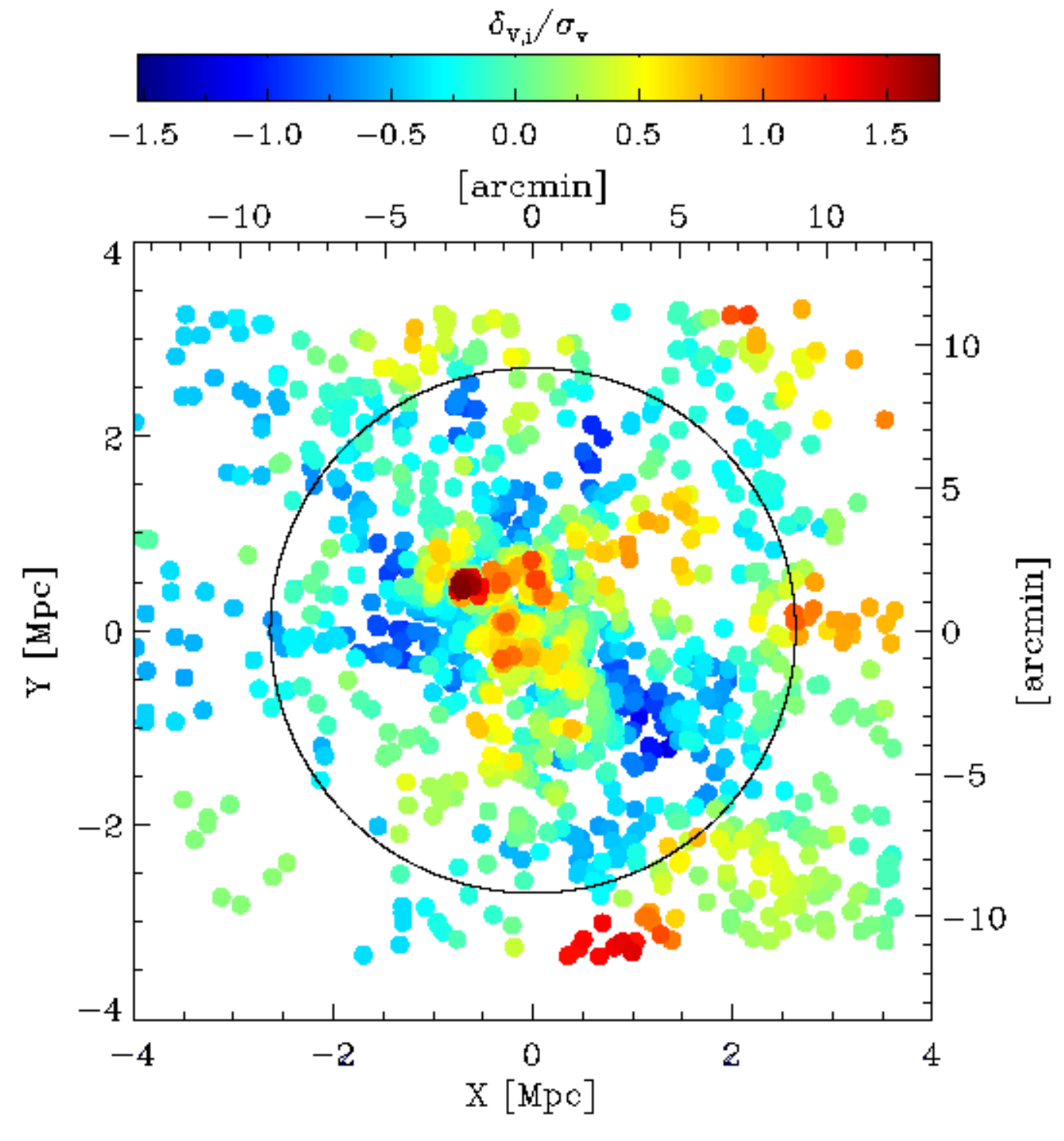}
\caption{2D Spatial distribution of the 1234 member galaxies colour-coded according to the local mean velocity deviations $\delta_{{\rm v},i}$. The circle has a radius equal to r$_{200}$ = 2.63 Mpc.}
\label{fig10}
\end{figure}

The DSv-test reveals that A~S1063 is not relaxed, at the $>99.9\%$ c.l., as also suggested by \citet{gom12}, from the analysis of the X-ray emission, the optical imaging and the spectroscopy of 51 cluster members. 
In particular, within r$_{200}$, there is a high-velocity region at the NE (X=$-0.5$~Mpc Y=0.5~Mpc in Fig.~\ref{fig10}), and a low-velocity region at the SW (X=1.0~Mpc Y=$-1.0$~Mpc in Fig.~\ref{fig10}). This peak is similar to the NE peak reported in Fig.~15 (right panel) of \citep{gru13}, from the photometric redshift information and the WL analysis.
There is also a small group with high velocity at the South, outside r$_{200}$ (X=1.0~Mpc Y=$-3.2$~Mpc in Fig.~\ref{fig10}) and, considering the whole spatial distribution of the 1234 members (see Fig.~\ref{fig7} and Fig.~\ref{fig10}), there is also a NE-SW elongation. 

\begin{figure}[ht]
\centering
\includegraphics[width=0.5\textwidth]{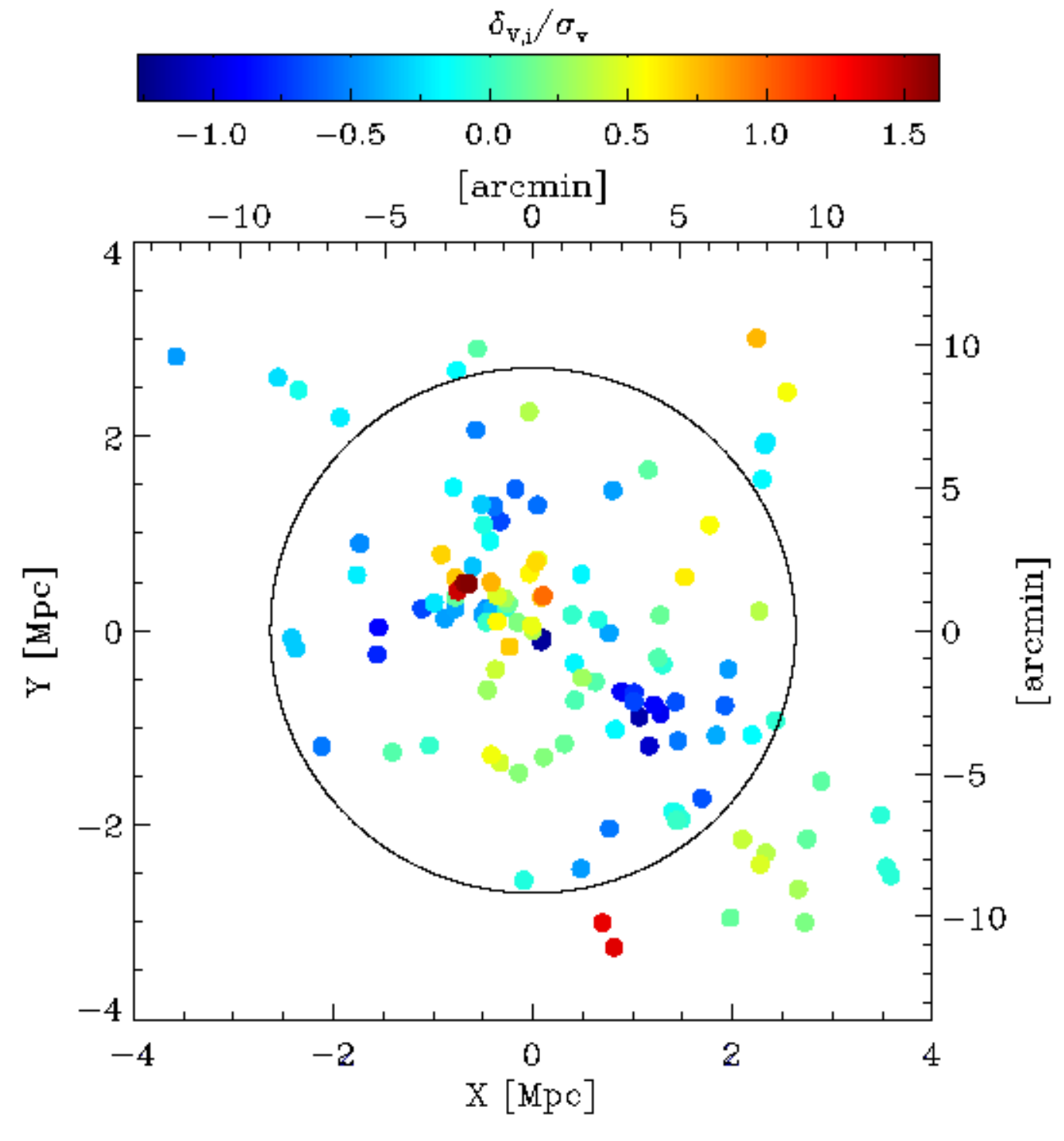}
\caption{2D spatial distribution of the 120 member galaxies with R $<$ 20, colour-coded as in Fig.~\ref{fig10}. The circle has a radius equal to r$_{200}$ = 2.63 Mpc.}
\label{fig11}
\end{figure}

When considering only bright galaxies (see Fig.~\ref{fig11}), the NE-SW elongation is more evident. The $\delta_{{\rm v},i}/\sigma_{\mathrm{V}}$ parameter shows that, within r$_{200}$, there are at least two groups populated by bright galaxies, one of which is a low-velocity region at the SW of the cluster centre (at $\sim$ X=$1.0$~Mpc Y=$-1.0$~Mpc) mentioned above. As expected, bright galaxies serve as signposts for the substructures (e.g., \citealt{girardi2008} and see also below).

These results are in agreement with \citet{ram21}, which identified the merger axis along the same direction as we obtained from the dynamical analysis by using Chandra X-ray observations and 325 MHz Giant Metre Radio Telescope observations.

\section{Galaxy classification}
\label{sec:6}

\begin{figure}
   \centering   
   \includegraphics[width=0.5\textwidth]{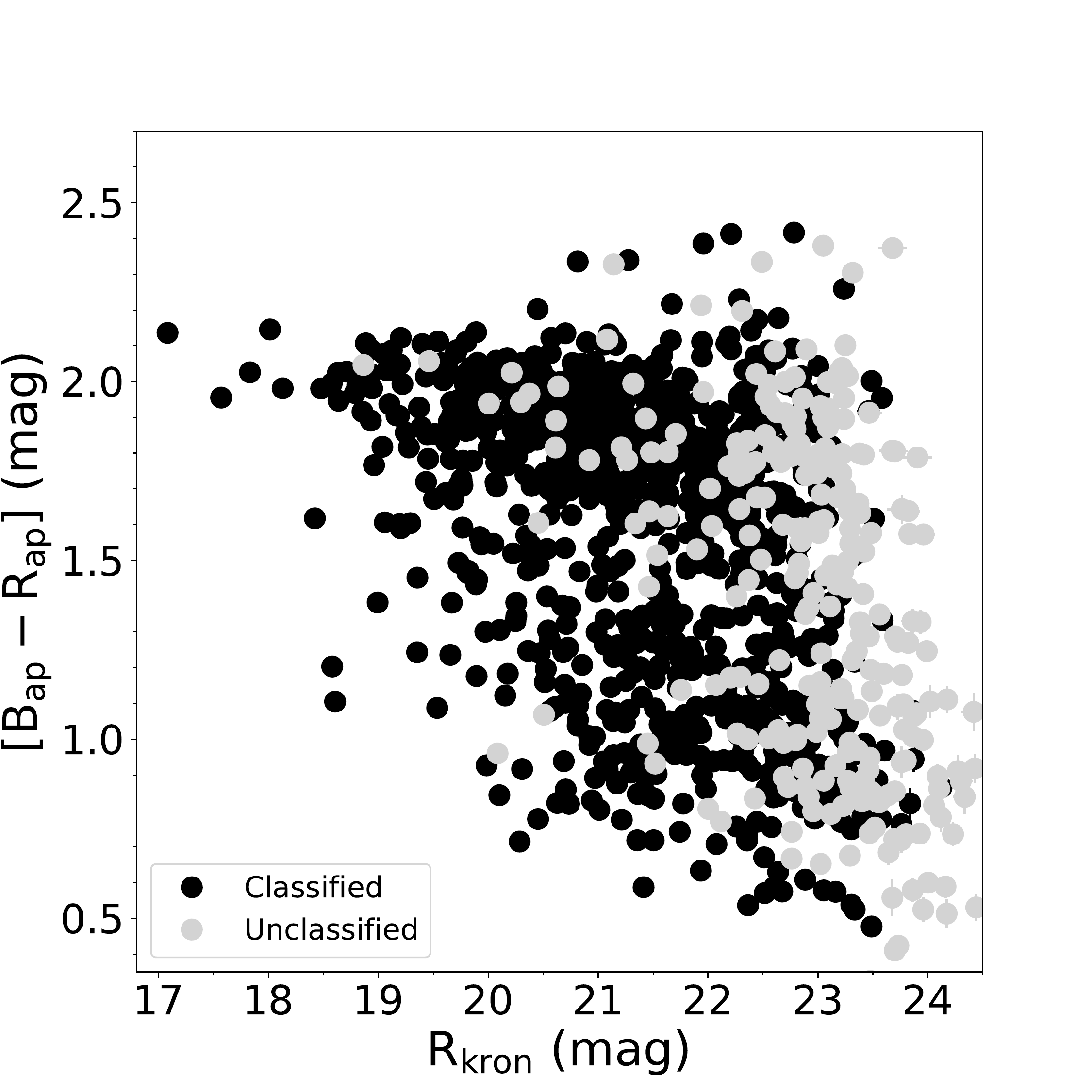}
   \caption{Colour-Magnitude diagram of the spectroscopic members in A~S1063. Black circles marks the 960 galaxies with robust spectral classification, while grey filled circles indicate galaxies with reliable colours and a spectral signal-to-noise ratio too low to be classified.}
          \label{fig12}%
    \end{figure}

We classify galaxies according to the presence of emission lines and the strength of the H$\delta$ absorption line. After a careful inspection of the spectra, out of a total of 1234 members, we classify 960 members, excluding 256 galaxies with too low signal-to-noise and 18 with unavailable colours. In Fig.~\ref{fig12}, we show that  unclassified galaxies are uniformly distributed at faint magnitudes in the colour-magnitude diagram.
Within r$_{200}$, we spectroscopically classify 700 members. We measure the equivalent widths (EWs) for the emission lines [OII], [OIII] and, when available, H$\alpha$, and the EW of the H${\delta}$ line (see \citealt{mercurio2004} for the definition of wavelength ranges).

\begin{figure}[ht]
   \centering
   {\includegraphics[width=0.4\textwidth]{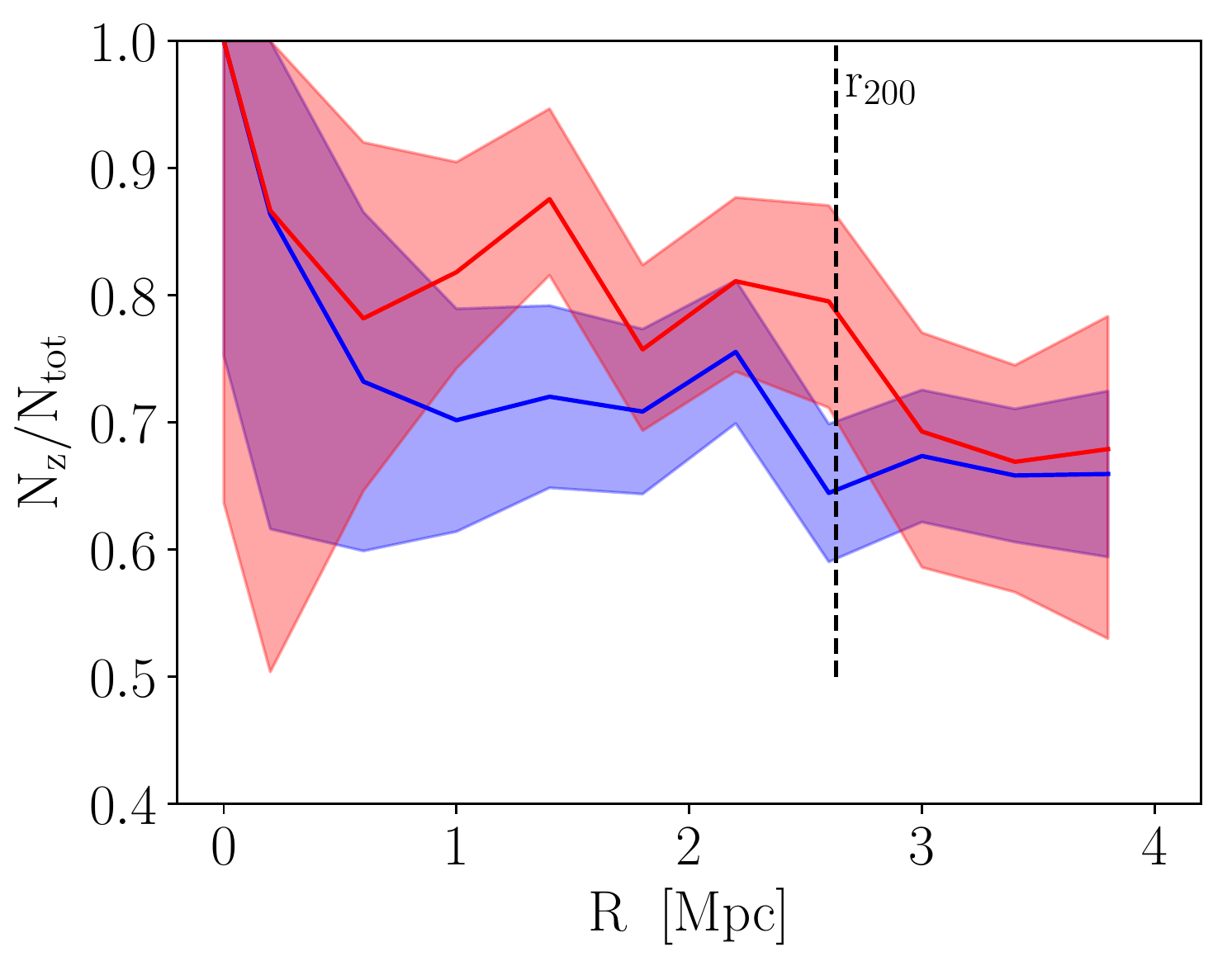}}
   \caption{Completeness of the spectroscopic sample of blue and red galaxies as a function of custercentric distance. The shaded regions indicate the 68\% confidence intervals, following the  algorithm  of  Gehrels (\citealt{geh86}). The dashed line indicates r$_{200}$ =2.63 Mpc.}
   \label{fig13}
    \end{figure}

\begin{figure*}
   \centering   
   \includegraphics[width=0.8\textwidth]{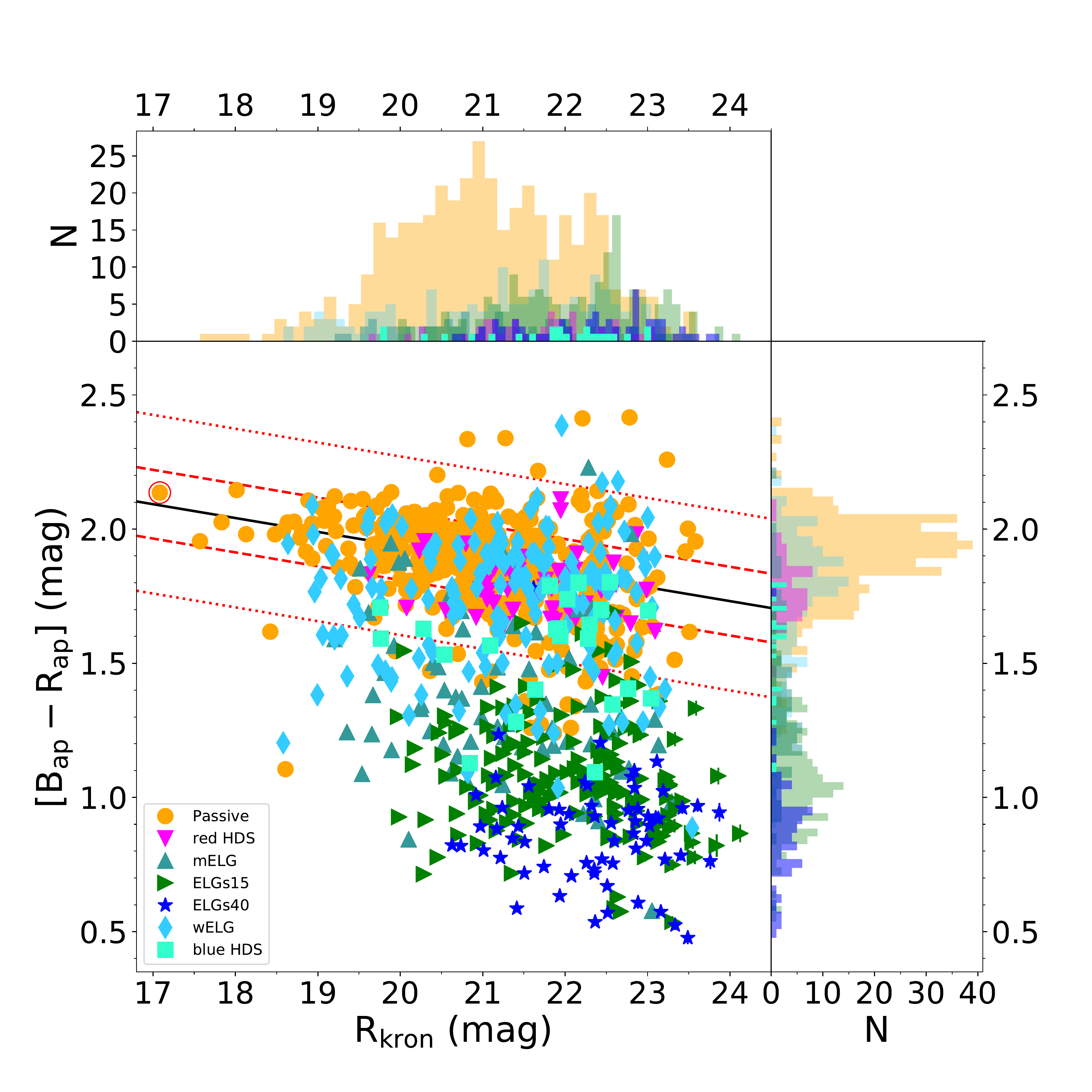}
   \caption{Colour-Magnitude relation of 960 spectroscopic members in A~S1063 with reliable spectral classification. The best-fit of the CM is indicated by the black line, while 68\% and 99\% limits are marked as red dashed and dotted lines, respectively. The red empty circle highlights the BCG.}
          \label{fig14}%
    \end{figure*}

We detect 459 galaxies with evidence of emission lines. These are divided in four subclasses depending on the strength of the [OII] emission line: faint (EW[OII]$>-7$\AA, hereafter wELG, 158 galaxies), medium ($-15$\AA $<$EW[OII]$\le-7$\AA, hereafter mELG, 69 galaxies), strong ($-40$\AA$<$ EW[OII]$\le-15$\AA, hereafter ELGs15, 172 galaxies) and very strong (EW[OII]$\le-40$\AA, hereafter ELGs40, 65 galaxies) emission-line galaxies (ELGs). Very strong emission lines (EW[OII]$<$-40\AA) are generally taken as a significant indication of starburst (e.g., \citealt{dre99}), while weaker lines can also be due to continuous star-formation (e.g., \citealt{oem09}). Among the non-emission line galaxies, the 70 galaxies with EW(H${\delta})>$3\AA\
are classified as strong H${\delta}$ absorption galaxies (HDS sample, or post-starburst), interpreted as post-starbursts or galaxies with truncated star formation (e.g., \citealt{oem09}).  However, bluer galaxies require larger EW(H$\delta$) to be identified as HDS galaxies.  Thus, a more precise way to classify strong H$\delta$ galaxies uses the diagram of EW(H$\delta$) versus the B$-$R colour or versus the strength of the 4000 \AA\ break, that correlates with colour \citep{cou87,bar96,bal99}. In this separation, we followed \cite{mercurio2004}. First, we fit the colour-magnitude relation of spectroscopic members (Fig.~\ref{fig14}):
\begin{equation}
\begin{split}
\mathrm{B-R =2.97(\pm 0.13)-0.052(\pm 0.006) \times R} \ \ , \\
\noindent \mathrm{Observed \ \ rms \ \ scatter \ \ \Delta=0.13  \,}
\end{split}
\end{equation}
where the $\mathrm{B-R}$ colour is obtained from aperture magnitudes, and $\mathrm{R}$ are Kron magnitudes. Then, we calculate the corrected colours as $\mathrm{(B-R)_{corr}= [(B-R)_{\rm obs}-(B-R)]}$ and we use the same threshold of \cite{mercurio2004}, $\mathrm{(B-R)_{corr}=-0.5}$. In their original study, \cite{mercurio2004} based their scheme directly on the strength of the 4000 \AA\ break, while we use the corresponding colour\footnote{\footnotesize We use the threshold $\mathrm{(B-R)_{corr}=-0.5}$ to separate also blue and red galaxies in the analysis of the orbits described in Sect.~\ref{sec:73}.}. In Fig.~\ref{fig13}, we check separately the spectroscopic completeness of red and blue galaxies as a function of their radial position and find that they are consistent within 1$\sigma$ uncertainties. As a result, we adopted the same completeness function to correct different galaxy types.

Thus, according to the criteria mentioned above, we find 47 strong H${\delta}$ galaxies having red colours and EW(H$\delta$)$>3$\AA\ (hereafter, rHDS or red HDS or red post-starburst) and 23 strong H${\delta}$ galaxies having blue colours and EW(H$\delta$)$>5$\AA\ (hereafter, bHDS or blue HDS or blue post-starburst). 
The remaining 431 galaxies are classified as passive (hereafter P). 

Finally, among the 960 members with spectroscopic classification, we find 44.9\% are P, 4.9\%/2.4\% rHDS/bHDS; 16.4\% wELG; 6.7\% mELG and 17.9\%/6.8\% ELGs15/s40. 
Moreover, the y-axis histograms in Fig.~\ref{fig14} show that the spectral sequence P, rHDS, bHDS, wELG, mELG, ELGs corresponds to a colour sequence from red to blue. Red HDS follow the colour distribution of P galaxies, lying on the red sequence, although at magnitudes fainter than R=19.5, while bHDS ones have intermediate colours, lying in a region between the red P galaxies and the blue ELGs. The spectral sequence also roughly corresponds to a magnitude sequence at the bright end, only P galaxies are brighter than R=18.5, and all ELGs are fainter than R=20.5 (see x-axis histograms in Fig.~\ref{fig14}).


\section{Accretion history of different galaxy populations in the cluster}
\label{sec:7}

This section investigates the accretion history of galaxies in the cluster and the possible differences among different galaxy populations, as classified above. More specifically, {\it i)} we analyse the 2D distribution as well as the position of galaxies of different spectral classes in the projected phase-space diagram (Sect.~\ref{sec:71}), {\it ii)} we use a semi-analytical model to investigate the relationship between the position of the different spectral classes and the accretion redshift (Sect.~\ref{sec:72}), and {\it iii)} we calculate the orbits of cluster galaxies (Sect.~\ref{sec:73}).

\subsection{The projected Phase-Space Diagram}
\label{sec:71}

    
\begin{figure*}
\centering
\hbox{\includegraphics[width=1.05\textwidth]{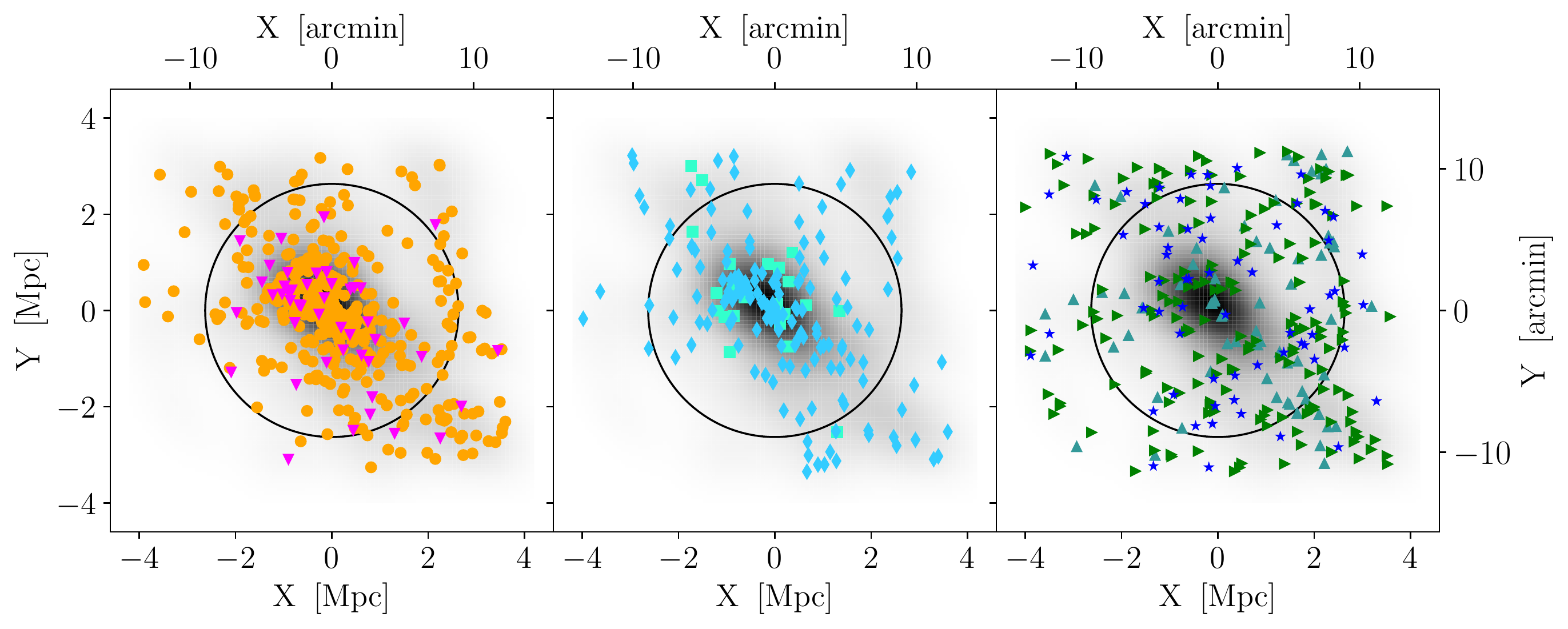}}
\hspace*{-1cm}
\hbox{\includegraphics[width=1.05\textwidth]{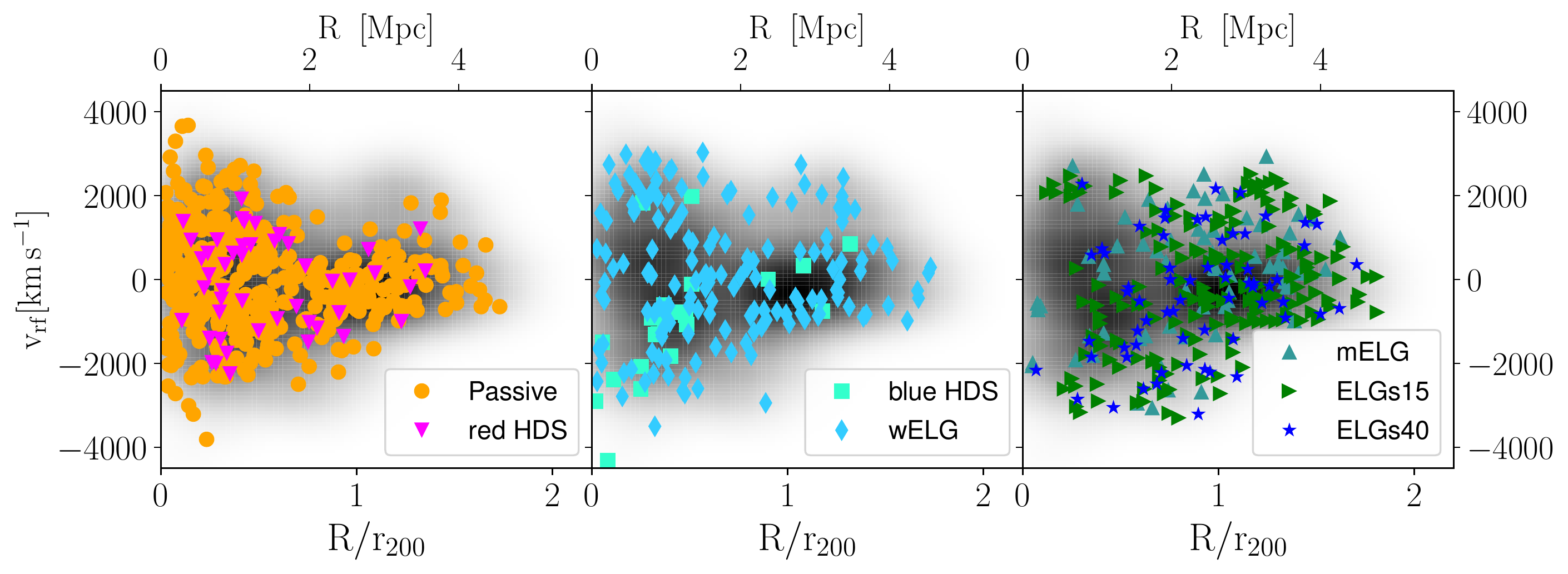}}
\caption{{\it Top panels:} 2D spatial distribution of the 960 spectroscopic member galaxies, classified according to their spectral properties. The circle has a radius equal to r$_{200}$ = 2.63~Mpc. {\it Bottom panels:} Phase-space diagram of the spectroscopic members, according to their spectral types. The underlying 
grayscale image in the top and bottom panels shows the galaxy number density distribution in 2D and in the phase-space, respectively, considering the whole population.}
\label{fig15}
\end{figure*}

    
\begin{figure}[ht]
\centering
\hbox{\includegraphics[width=0.5\textwidth]{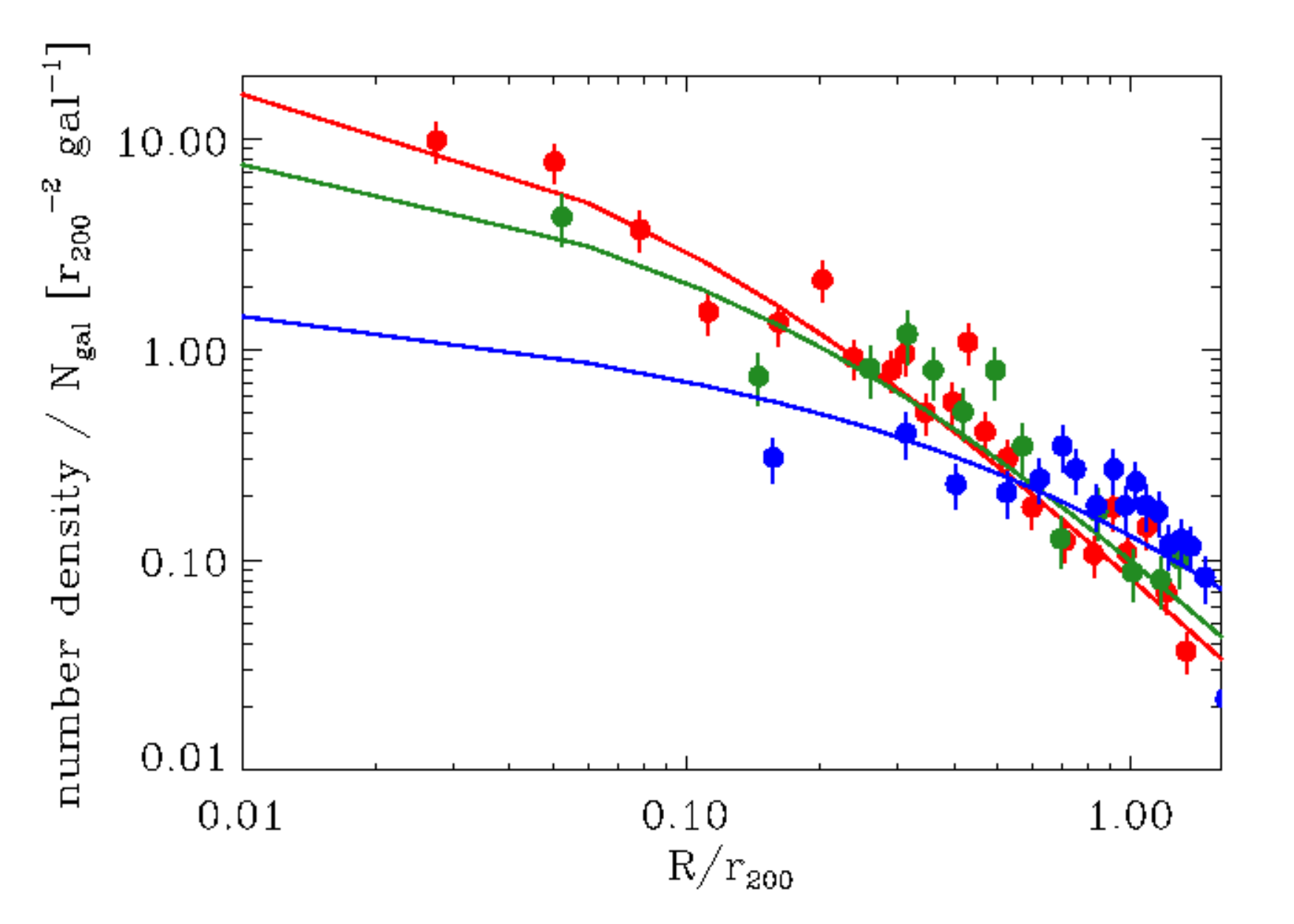}}
\caption{Number density profiles (normalized by the total number of galaxies in each sample, $\mathrm{N_{gal}}$) of the passive and red HDS (red points and lines), blue HDS, and wELG (green), and mELG, ELGs15, and ELGs40 (blue) populations. Error bars are 1 $\sigma$. Solid lines are best-fit NFW models for each population.}
\label{fig16}
\end{figure}

The top panels of Fig.~\ref{fig15} show the 2D distribution of galaxies with different spectral types. Passive and red HDS galaxies (top-left panel) trace the NE-SW elongation of the cluster. They are concentrated in the centre, with only 18\% of these galaxies outside r$_{200}$. A similar elongation is traced by blue HDS and wELG galaxies (top-central panel). However, these populations are less concentrated in the centre, and the fraction of galaxies outside r$_{200}$ is slightly higher (25\%). When considering the three populations of ELGs with larger EWs (mELG, ELGs15, ELGs40, top-right panel), the 2D distribution shows that these galaxies avoid the centre, and almost half (44\%) of these galaxies are located outside r$_{200}$.

The bottom panels in Fig.~\ref{fig15} show the projected phase-space diagrams as a function of galaxy spectral properties. In all the panels, the x-axis shows the projected distance from the cluster centre, normalised by r$_{200}$, while the y-axis shows the rest-frame line-of-sight velocity. By comparing the three panels, it is evident that: passive and red HDS populations are concentrated at low clustercentric distances, in a region that all the remaining spectral classes mostly avoid; blue HDS and wELGs trace the projected phase-space diagram at a greater distance from the centre; and, as expected, galaxies with stronger emission lines occupy the outer regions of the projected phase-space. 

In the following analyses, to increase the statistics, we merge some of the spectral classes defined in Sect.~\ref{sec:6}: {\it 1)} P and red HDS (hereafter P+red HDS or P+rHDS), {\it 2)} blue HDS and wELG (hereafter blue HDS+wELG or bHDS+wELG), and {\it 3)} mELG, ELGs15, and ELGs40 (hereafter mELG+ELGs15+ELGs40). We confirm, according to the two-dimensional Kolmogorov-Smirnov test \citep{FF87}, that the classes we merge have similar spatial distributions. 
Moreover, a 2-dimensional Kolmogorov-Smirnov test confirms the visual impression obtained bottom panels of Fig.~\ref{fig15}, namely the phase-space distributions of the three samples defined above are all different from one another with a probability $>0.99$. 

\begin{figure}[ht]
\centering
\hbox{\includegraphics[width=0.5\textwidth]{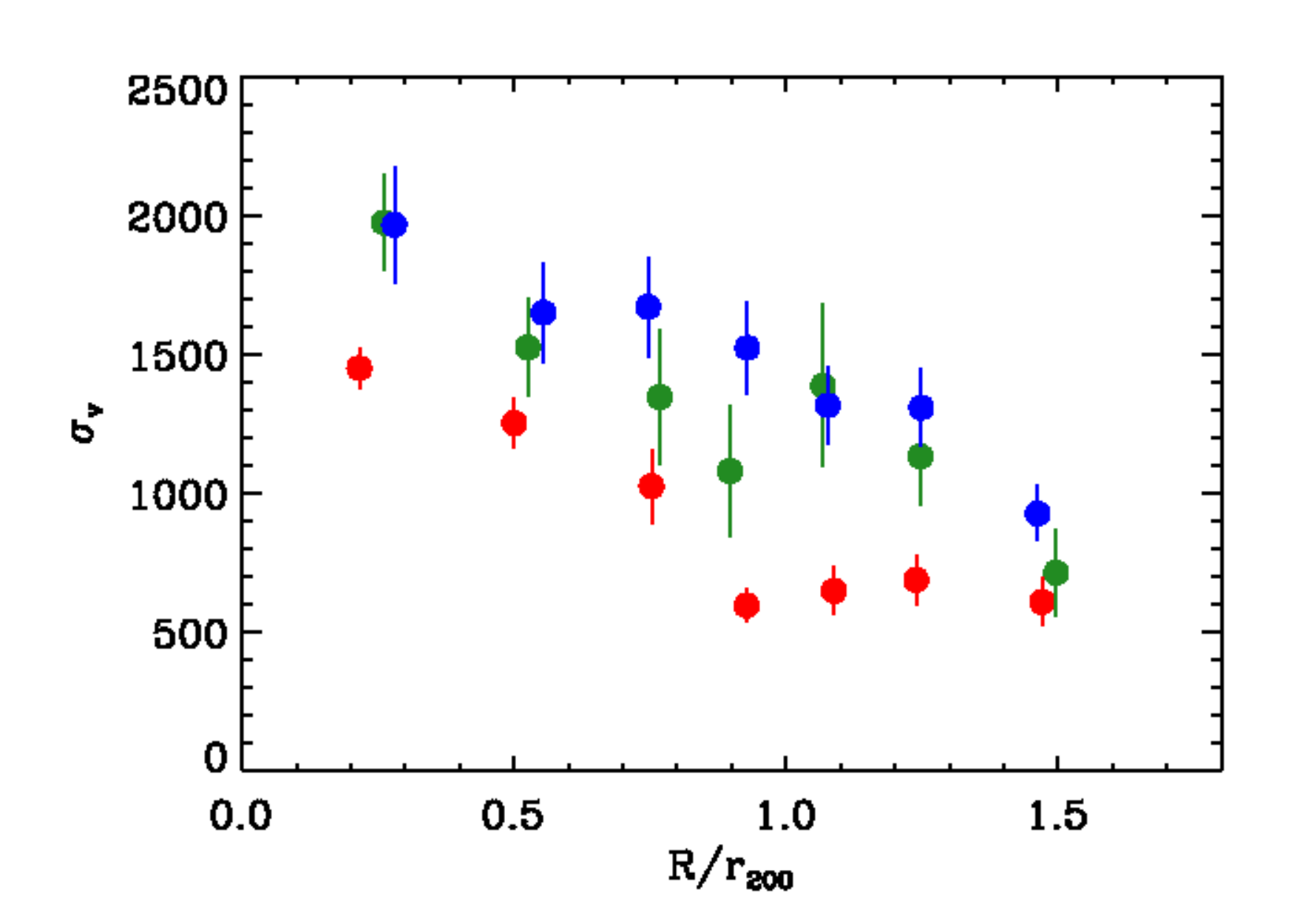}}
\caption{Velocity dispersion profiles of the passive and red HDS (red points), blue HDS and wELG (green), and mELG, ELGs15, and ELGs40 (blue) populations. Error bars correspond to 1 $\sigma$. Velocity dispersions are in \ks.}
\label{figsigv}
\end{figure}

To quantify this difference, we consider the spatial and velocity distributions of the different sub-populations separately. We fit the incompleteness-corrected number density profiles of the three samples with a NFW model \citep{navarro1997}. We used a universal completeness correction for different galaxy types, as  discussed Sect.~\ref{sec:6} (see Fig.~\ref{fig13}).

The number density profiles and their best fits are shown in Fig.~\ref{fig16}. We find that the best-fit concentrations ($c \equiv \mathrm{r_{200}/r_s}$ where $\mathrm{r_s}$ is the scale radius of the NFW model) are $c=9_{-2}^{+1}$ for Passive+red HDS, $c=5 \pm 1$ for blue HDS+wELG, and $c=0.8_{-0.1}^{+0.2}$ for mELG+ELGs15+ELGs40. Therefore, the Passive+red HDS and the blue HDS+wELG populations are only marginally different from one another (see red and green best-fit curves in Fig.~\ref{fig16}), whereas the profile of the blue population (mELG+ELGs15+ELGs40) appears to be significantly different from the other two samples. The velocity dispersion profiles of the three samples are shown in Fig.~\ref{figsigv}. We compare the three profiles two by two using the $\chi^2$ distribution,

\begin{equation}
    \chi^2 = \Sigma_i (\sigma_{i,j}-\sigma_{i,k})^2 / (\delta_{i,j}^2+\delta_{i,k}^2),
\end{equation}
where the sum runs over the seven radial bins, $j, k$ identify the two samples compared, $\sigma_{i,j}$ is the value of the velocity dispersion of sample $j$ at the radial bin $i$, and $\delta_{i,j}$ is its error. We find that the population of passive P+red HDS galaxies has a systematically lower velocity dispersion profile with respect to the other two populations of blue HDS+wELG and mELG+ELGs15+ELGs40, which show instead consistent velocity dispersion profiles. We then conclude that most of the difference we see in the projected phase-space distribution of the different populations is due to their different velocity distributions.

\begin{figure*}
   \centering
   \includegraphics[scale=0.6]{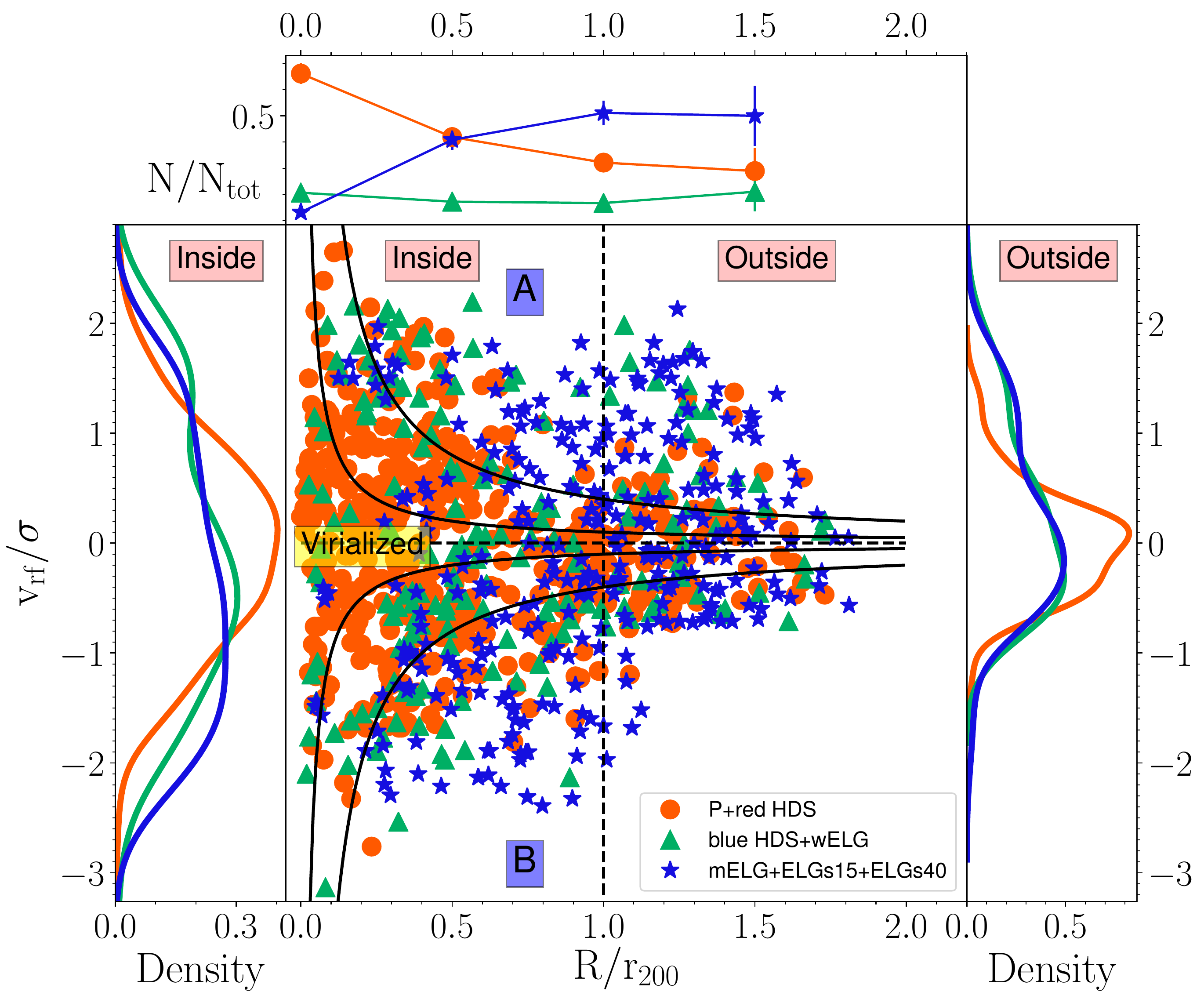}
   \caption{{\it Main panel}: Projected phase-space distribution of galaxies with different spectral types. The two lines of constant (${\rm v_{rf}}/\sigma)\times  ({\rm R/r_{200}})= 0.1,\, 0.4$ are plotted as in \citealt{Noble2013} to delineate the separation into regions of virialized (within the inner caustic), backsplash (between caustics), and infall galaxies (along and outside caustics). The black dashed line delimits the regions inside and outside r$_{200}$. We indicate as region A and B the two areas with ${\rm R<r_{200}}$, outside the virialized region. {\it Top panel}: the fraction of P+red HDS (orange circles), blue HDS+wELG (green triangles), and mELG+ELGs15+EGs40 (blue stars) as a function of the clustercentric distance. {\it Left and right panels} report the velocity distribution of the three corresponding samples, inside and outside r$_{200}$, respectively. The smoothed curves are the probability density functions obtained with a Kernel density estimator.}
          \label{fig18}%
    \end{figure*}

We explore the accretion histories of these three classes of galaxies by dividing the projected phase-space diagram into the regions inside and outside r$_{200}$ (Fig.~\ref{fig18}). We found that 75\% of the cluster members located inside r$_{200}$ are passive. This is in agreement with the results of \citet{bakels2020}, which, using a high-resolution cosmological dark matter-only simulation, found that 79\% of the galaxies, in projection, inside r$_{200}$ have passed their orbital pericenter at least once. 
Moreover, black lines in Fig.~\ref{fig18} show the lines corresponding to constant values of $({\rm R/r}_{200}) \times ({\rm v}_\mathrm{rf}/\sigma)$ in projected phase space, which correspond to caustic profiles \citep{Noble2013}, and can be used to identify regions in projected phase-space containing infalling, backsplash, and virialised galaxy populations. The relation between caustic and the observed spectral properties of member galaxies can shed light on the importance of a dynamically defined environment and the cluster accretion scenario.

\cite{Noble2013} defined {\it i)} as virialized the region within the caustic lines of constant (R/r$_{200}$) $\times$ (v$_\mathrm{rf}$/$\sigma$) equal to 0.1 (the inner caustic in Fig.~\ref{fig18}), {\it ii)} the backsplash region as the one between caustic equal to 0.1 and to 0.4, and  {\it iii)} the infall region along and outside the caustic equal to 0.4. Galaxies located at R$\le$r$_{200}$ and low rest-frame velocities ($\abs{\Delta \mathrm{v}}/\sigma \lesssim$ 1.5 \citep[i.e., within the virialized region]{Jaffe15,mahajan2011}, may have experienced many pericentric passages and were accreted when the cluster core was forming. At R$<$r$_{200}$ and intermediate value of (R/r$_{200}$) $\times$ (v$_\mathrm{rf}$/$\sigma$) one expects a mix of all galaxy types, but backsplash galaxies are most likely present \citep{mahajan2011,haines2012,Noble2013}. Galaxies falling into the cluster for the first time, can be found outside the caustic equal to 0.4 within r$_{200}$ and at R$>$ r$_{200}$ \citep{mamon04,dunner07,mahajan2011,haines2012,haines2015}.

Our data confirm this scenario. The top panel of Fig.~\ref{fig18} shows the fraction of the different spectral types as a function of the clustercentric distance. As discussed above, ELGs are mainly located outside r$_{200}$ and avoid the cluster centre. The fraction of blue post-starbursts and galaxies with weak emission lines is almost constant with the clustercentric distance (see also below about the discussion of post-starbursts only). The fraction of P+rHDS decreases as the clustercentric distance increases. 

We also indicate as A and B the two regions within r$_{200}$, with large radial velocities, which have uncertain accretion histories. \citet{Jaffe15} suggest that in the projected phase-space region R$\le$r$_{200}$ and $\abs{\Delta \mathrm{v}}/\sigma \gtrsim$ 1.5 ram-pressure stripping could play an important role in quenching galaxy star-formation because of the high density of the intra-cluster gas near the centre and the high galaxy velocities. \citet{ciocan2020} showed an enhancement of metallicities for galaxies with 9.2$<$Log($M^*$/M$_\odot$)$<$10.2 in A~S1063 compared to the field galaxies (see their Fig.~8). This was interpreted as a scenario in which only the hot halo gas of these cluster galaxies is removed due to strangulation, leading to an increase in their gas-phase metallicity (because no dilution of the interstellar-medium with pristine inflowing gas occurs anymore). However, the galaxies continue to form stars using the available cold gas in the disk, which is not stripped. Thus, strangulation could be another mechanism affecting infalling galaxies at R$\le$r$_{200}$. \citet{rhee2017}, using a schematic galaxy's trajectory after infalling into the cluster, show that after the first pericenter passage, the galaxy is found as "backsplash" also in this region. Then, the galaxy may settle into the virialised area. 

Our data suggest a different mix of post-starbursts and star-forming galaxies in the two regions region A (R$\le$r$_{200}$ and $({\rm v}_\mathrm{rf}/\sigma) \ge$0) and B (R$\le$r$_{200}$ and $({\rm v}_\mathrm{rf}/\sigma) \le$0), outside the virialized region. We find that post-starbursts are approximately 20\% ($\pm$ 3\%) of the total sample of galaxies both in region A and B, while star-forming galaxies are 27\% ($\pm$ 4\%) in region A and 33\% ($\pm$ 4\%) in Region B. By considering only the galaxies with $\abs{\Delta \mathrm{v}}/\sigma \lesssim 1.5$, we find that there are 19 post-starbursts out of a total of 61 (31\%) and 16 emission-lines out of a total of 64 with $\Delta \mathrm{v}/\sigma \ge 1.5$, while there are 18 post-starbursts out of a total of 74 (24\%) and 34 emission-lines out of a total of 107 (32\%) at $\Delta \mathrm{v}/\sigma \le -1.5$. This is shown also in the left panel of Fig.~\ref{fig18}, where we plot the smoothed velocity distribution of galaxies with different spectral types obtained with a Kernel density estimator. The green curve indicating the velocity distribution of post-starbursts is higher than that of star-forming galaxies in region A, while it is the contrary in region B. This suggests a different accretion scenario of galaxies in the two regions, with galaxies that have already experienced the first pericenter passage located mainly into region A, while galaxies falling into the cluster for the first time lie mainly in region B. As shown by the right panel of Fig.~\ref{fig18}, star-forming and post-starburst galaxies have similar distribution outside the virial radius.

\begin{figure}[ht]
   \centering
   \includegraphics[width=0.4\textwidth]{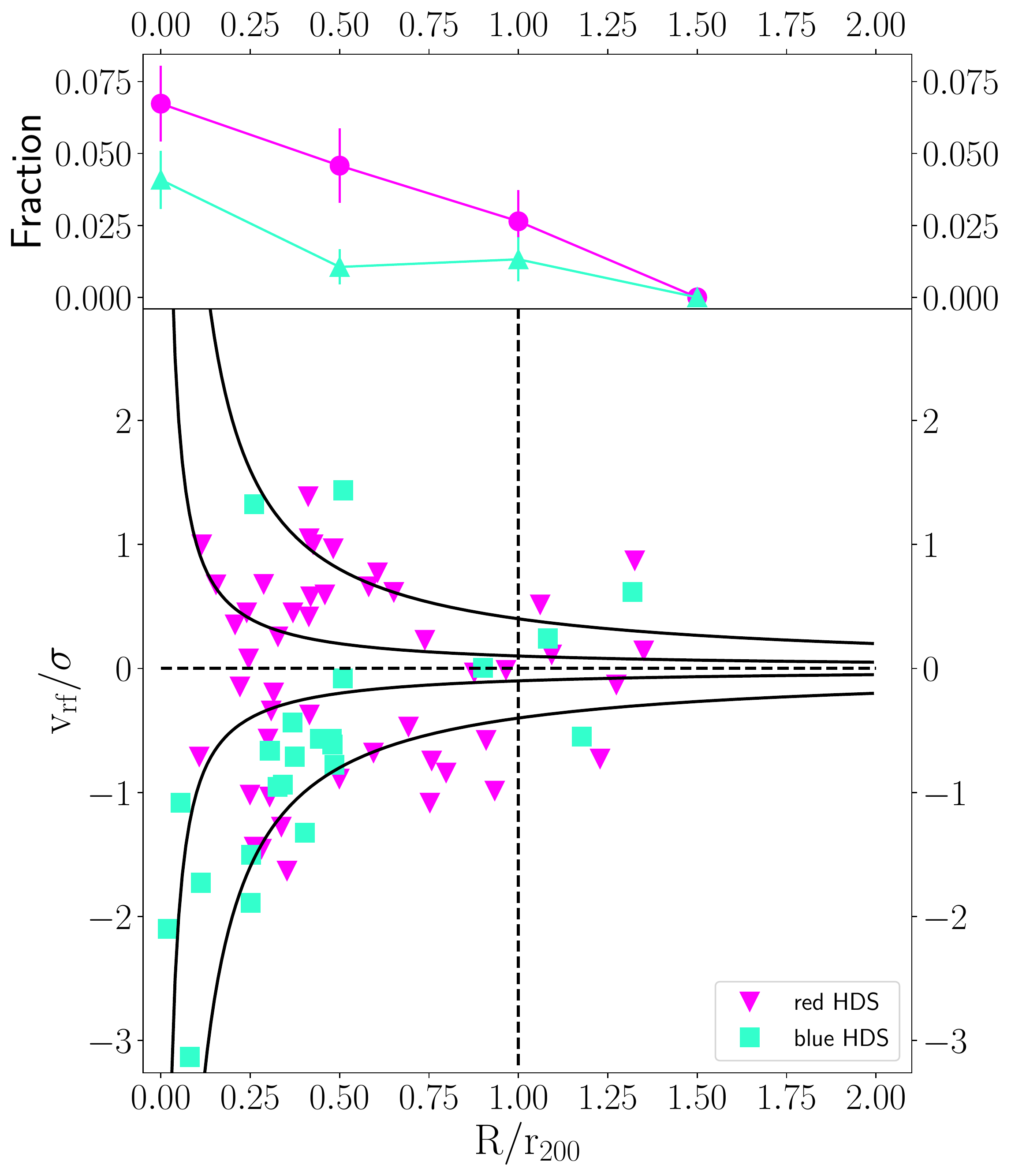}
   \caption{{\it Bottom panel}: Projected phase-space distribution of red (magenta) and blue (cyan) post-starburst galaxies. Lines are the same as those in Fig.~\ref{fig18}.  {\it Top panel}: the fraction of post-starbursts with respect to all classified members as a function of the clustercentric distance.} 
          \label{fig19}%
    \end{figure}

    It is also interesting to note that red and blue post-starburst galaxies lie mainly in the region between caustic parameter 0.1 and 0.4 (see Fig.~\ref{fig19}), or along the outer caustic. Blue post-starbursts seem to prefer regions where v$_\mathrm{rf}/\sigma \le 0$, where also Fig.~\ref{fig18} shows an excess of star-forming galaxies respect to the region where v$_\mathrm{rf}/\sigma \ge 0$. The top panel of Fig.~\ref{fig19} shows that the fraction of post-starbursts decreases as a function of the clustercentric distance. The asymmetry of the distribution of blue HDS in projected phase-space is striking. \citet{muzzin2014} found a similar asymmetry in the projected phase space distribution of post-starburst galaxies in a sample of z~1 clusters (see their Fig.~1), although Muzzin et al.'s asymmetry and ours are opposite in sign in velocity.


\subsection{Accretion redshift from simulations}
\label{sec:72}

To interpret the results obtained in the previous section in the context of the cluster assembly history, we compared the observed distribution of galaxies in the projected phase-space diagram with that predicted from a simulated cluster with a mass similar to that of A~S1063. The projected phase-space distribution of galaxies is an important tool to constrain the epoch of their accretion from their location in the diagram.  We use the semi-analytical model of \citep[hereafter, DLB07]{del07} to investigate the relation between the location of galaxies in the pseudo-phase-space diagram and the redshift at which they are accreted into the cluster. This model was run on the Millennium Simulation \citep{springel2005} which assumes a WMAP1 cosmology with $\mathrm{\Omega_{\Lambda}\, =\, 0.75,\, \Omega_m\, =\, 0.25,\, \Omega_b\, =\, 0.045, \sigma_8\, =\, 0.9,\, and \, h\, =\, 0.73}$. The semi-analytical model includes physical ingredients firstly introduced by \cite{white1991} and later refined by \citet{springel2001,del04,del07}. In particular, it includes prescriptions for gas accretion and cooling, star formation, feedback, galaxy mergers, the formation of supermassive black holes, and treatment of the "radio mode" feedback from galaxies located at the centres of groups or clusters of galaxies. The accretion redshift is defined as the first time when galaxies that reside in the cluster today are accreted onto its main progenitor.

To compare observations to the SAM predictions, we select the only halo available in the simulations with $M^{MS}_{200}$ similar to A~S1063 at $z=0.3$. $M^{MS}_{200}$ masses are computed from the N-body simulation as the mass enclosed within $R^{MS}_{200}$, the radius of a sphere that is centred on the most bound particle of the group and has an overdensity of 200 with respect to the critical density of the universe at the redshift of interest. We consider galaxies within a cylinder with a radius of 2 r$_{200}$ and height of $\mathrm{2r_{200}}$, in such a way to approximately match the data coverage, and with masses greater than $\mathrm{Log(M^*/M_\odot)=9.5}$, which is also the mass limit used for the analysis of the orbits described in Sect.~\ref{sec:73}. 


\begin{figure*}[ht]
\hbox{\includegraphics[width=0.50\textwidth]{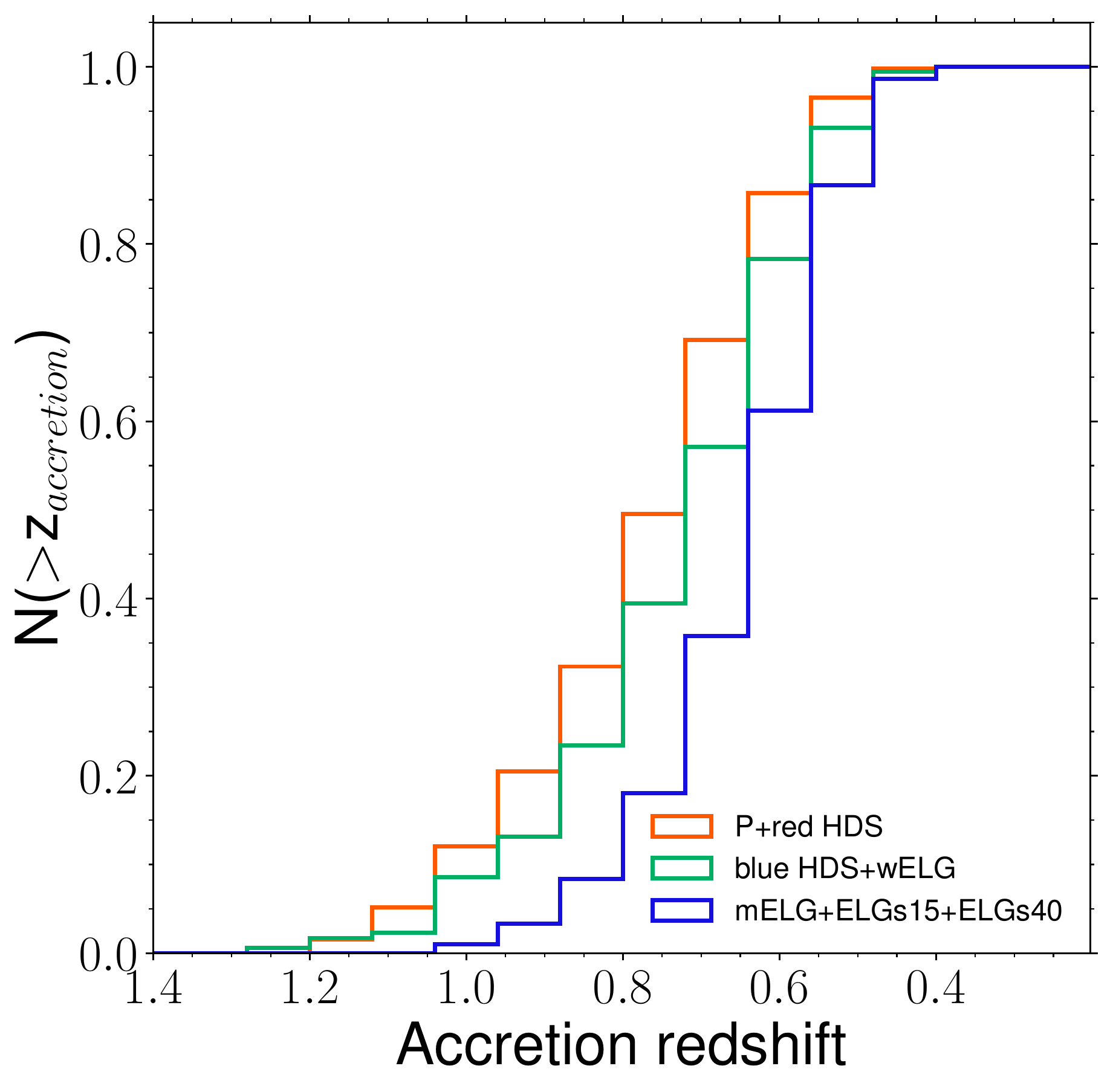}
{\includegraphics[width=0.50\textwidth]{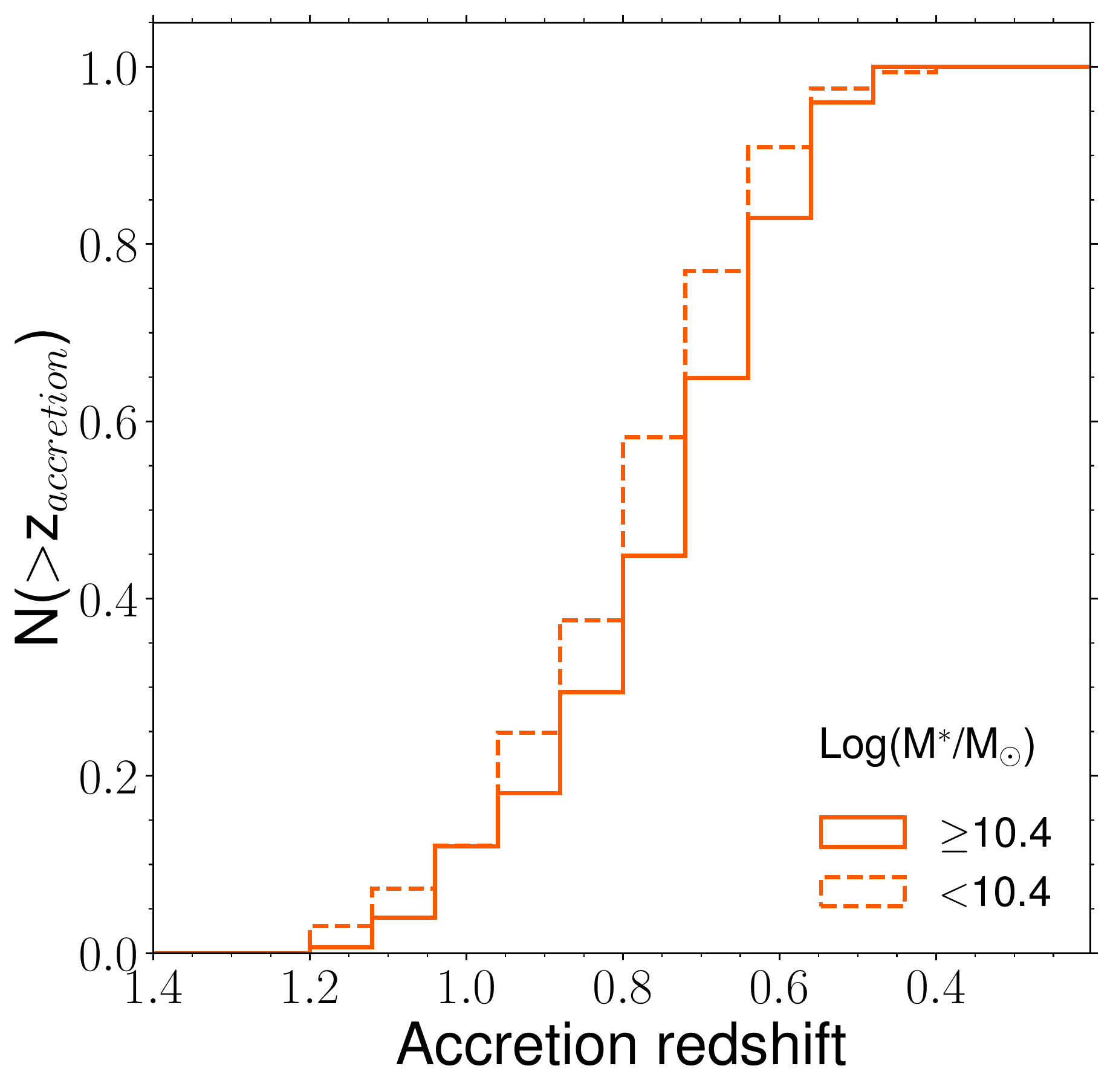}}
}
\hbox{\includegraphics[width=0.5\textwidth]{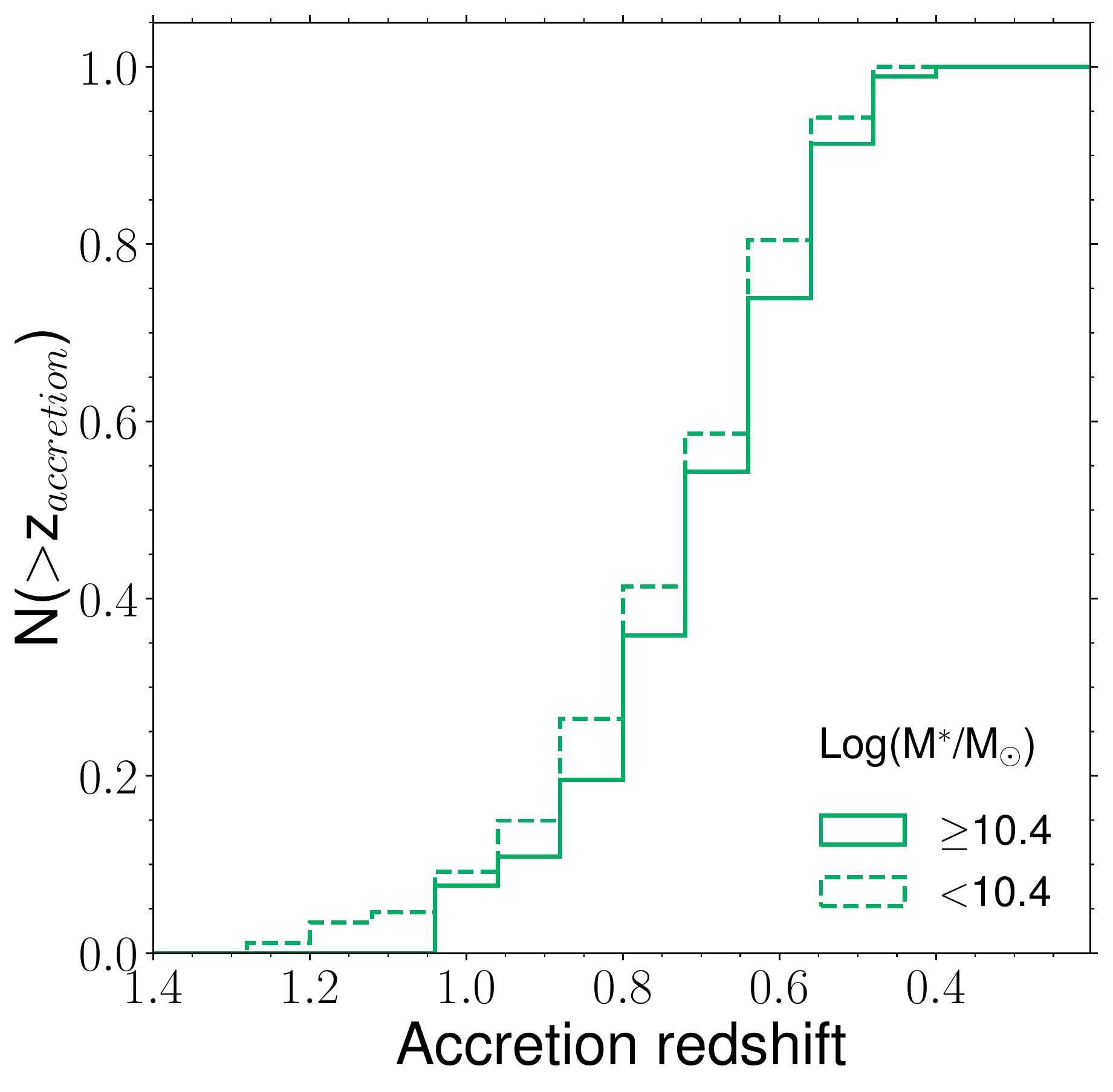}
{\includegraphics[width=0.5\textwidth]{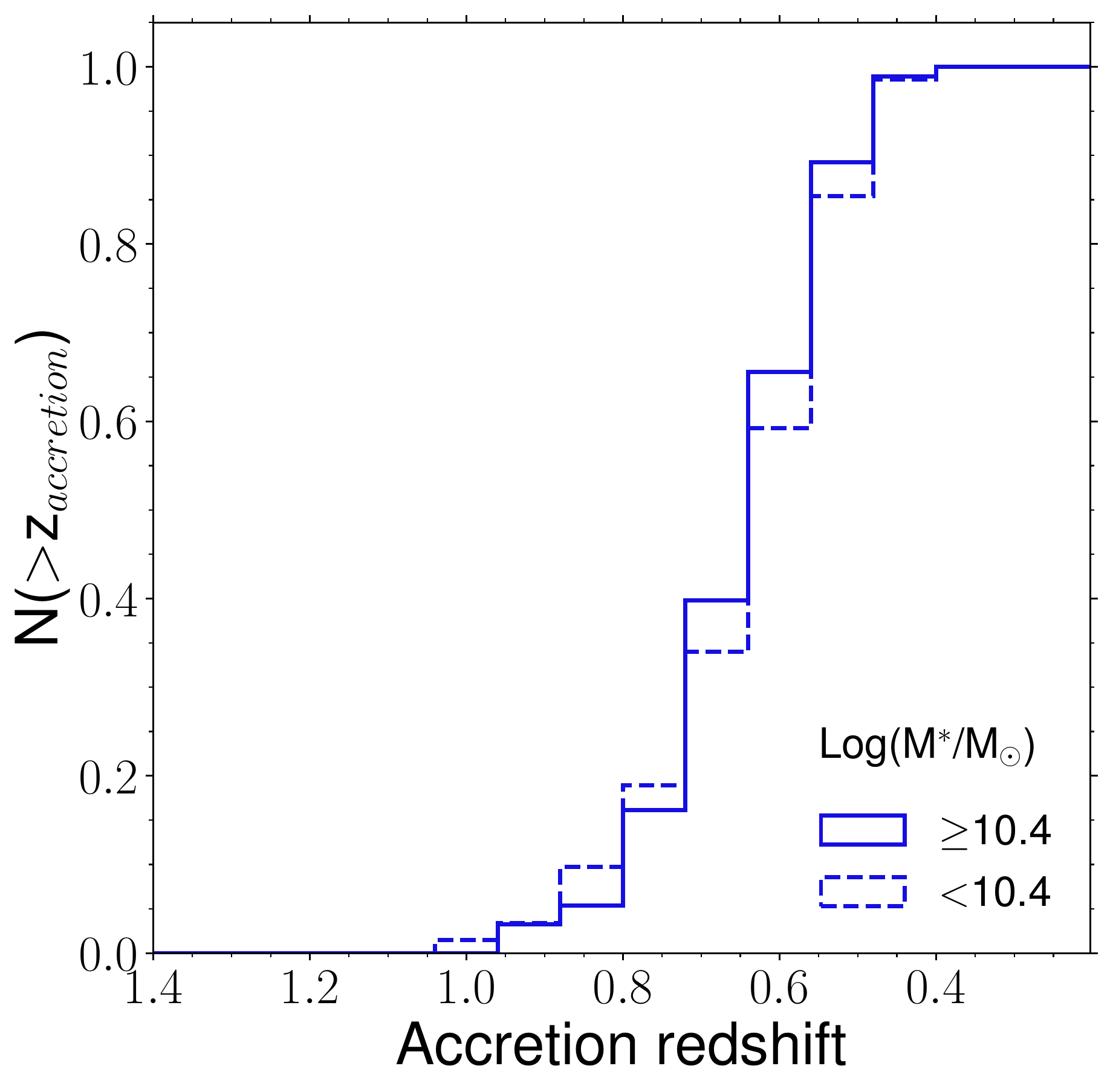}}
}
\caption{{\it Upper left panel}: cumulative distribution of the accretion redshift, z$_{accretion}$, for galaxies with different spectral types: passive + red post-starbursts (orange), blue post-starbursts + weak emission lines (green) and strong emission lines (blue). The other three panels show the same distribution separated into high and low mass bins (using as a threshold Log(M$^*$/M$_{\odot}$)$=$10.4), for the corresponding (colour coded) galaxy spectra types.}
\label{fig20}
\end{figure*}

Using the sample of simulated galaxies selected as described above, we construct the three phase-space diagrams using three different projections (xy,xz,yz). From these three diagrams and the accretion redshift of each simulated galaxy, $\mathrm{z_i}$, we interpolate three different functions, $\mathrm{z_i(R/r_{200},V_{rf}/\sigma)}$, one for each projection. The final accretion redshift obtained for each position in the projected simulated phase-space diagram results from the mean among the three adopted projections of the simulated cluster.

First, we compare the mean accretion redshift of galaxies belonging to the three populations defined in the previous section. The accretion redshift, $\mathrm{z_i}$, is assigned to each observed galaxy corresponding to its position in the projected phase-space diagram. We show in the upper panel of Fig.~\ref{fig20} the cumulative distribution for the three galaxy classes. It shows that galaxies with increasing star-formation rate are progressively accreted at lower redhift. According to the one-dimensional Kolmogorov-Smirnov test (KS-1D), we find that the distributions of the accretion redshifts of the three different spectroscopic populations are not drawn from the same parent distribution at more than 99\% c.l. Then, we investigate if there is a difference in the accretion redshift distributions of galaxies with different stellar masses (see Appendix~\ref{app:C} for the derivation of galaxies stellar mass). According to the KS-1D, we do not see any significant dependence of HDS+wELG and mELG+ELGs15+ELGs40 samples from the galaxy stellar masses. On the other hand the mean accretion redshift distribution of P+red HDS galaxies with Log(M$^*$/M$_{\odot}$)$\ge$10.4 is different with respect to that of galaxies with Log(M$^*$/M$_{\odot}$)$<$10.4 at probability $>0.99$. As shown in the upper right panel of Fig.~\ref{fig20}, the cumulative distribution of the mean accretion redshift of low-mass (Log(M$^*$/M$_{\odot}$)$<$10.4) galaxies moves towards higher accretion redshifts than the distribution of of high-mass (Log(M$^*$/M$_{\odot}$)$\ge$10.4) galaxies. This result is discussed in Sect.~\ref{sec:8}.

\subsection{Orbits}
\label{sec:73}

   \begin{figure}[ht]
   \centering
   \includegraphics[width=0.5\textwidth]{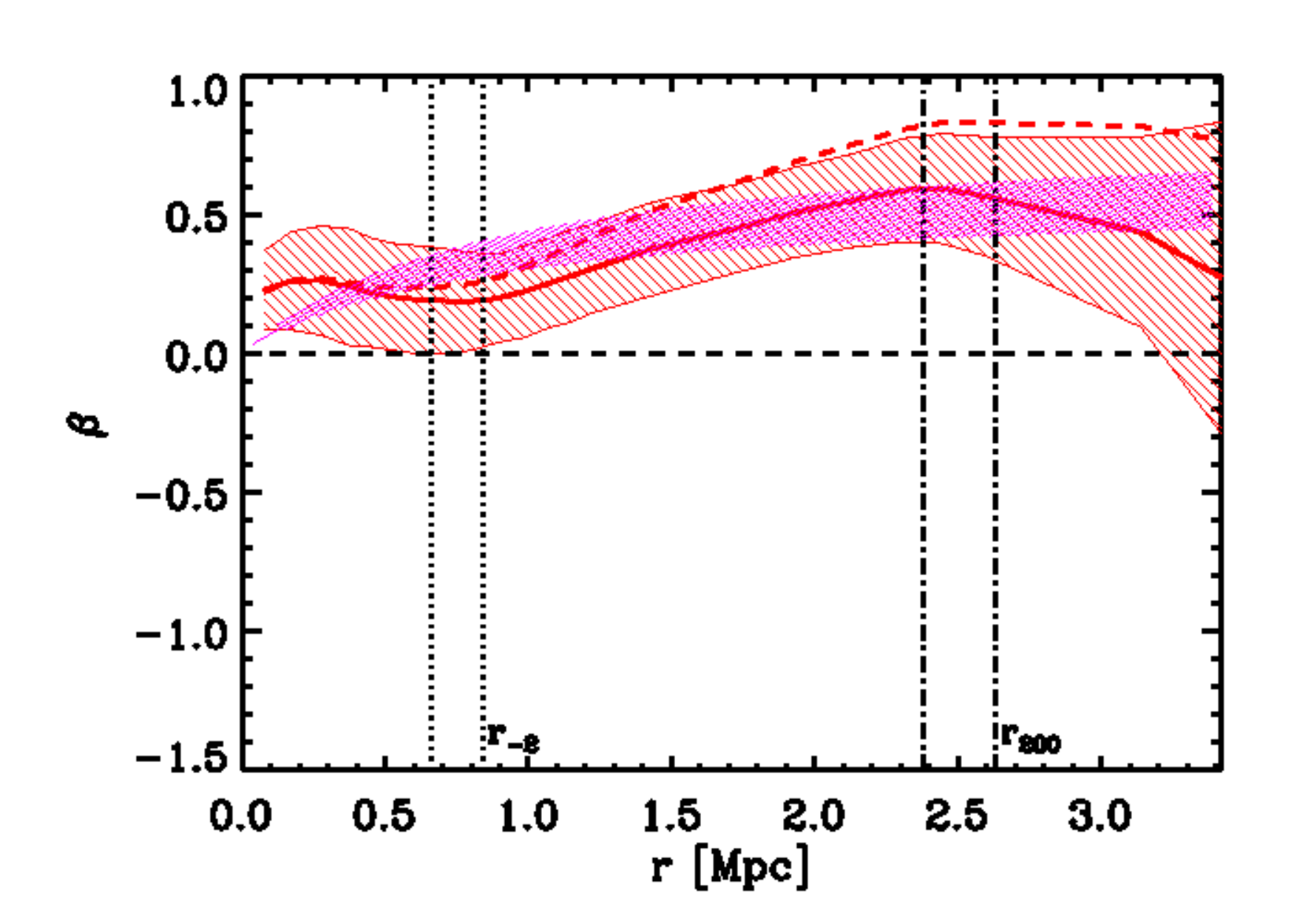}
   \includegraphics[width=0.5\textwidth]{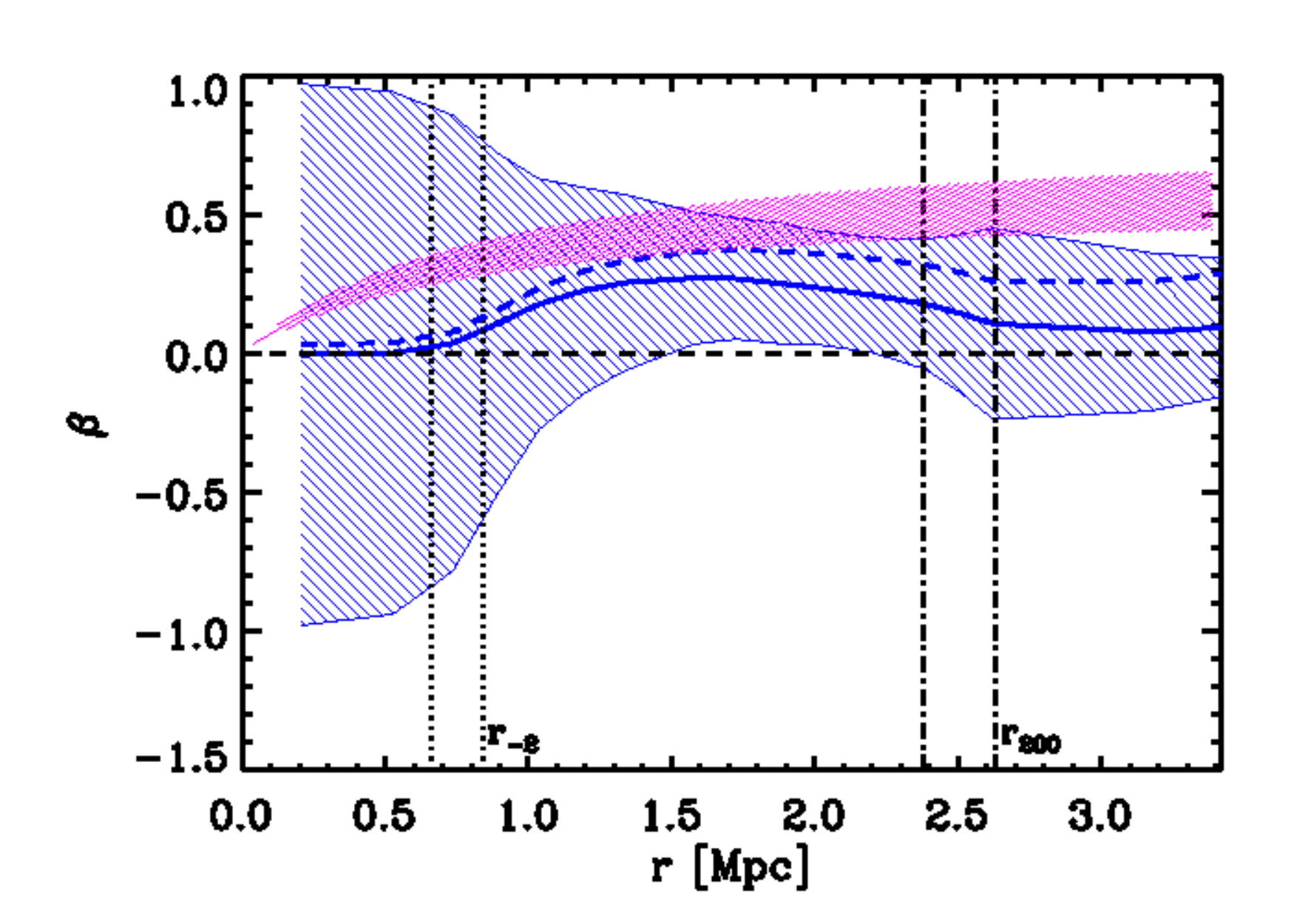}
   \caption{The velocity anisotropy profiles of the red and blue cluster members in the {\it upper} and {\it lower panels}, respectively. The solid (resp. dashed) coloured lines represent the solutions of the inversion of the Jeans equation adopting the mass profile from \citet{sar20} \citep[resp.][]{umetsu2016}. The coloured area indicates the 68\% confidence region around the solution. The magenta shading indicates the 68\% confidence region around the MAMPOSSt best-fit solution for a Tiret model \citep{sar20}. In both panels we use the same Tiret model computed considering the full sample of member galaxies. The vertical black dotted and dash-dotted lines mark r$_{-2}$ (which corresponds to  $\mathrm{r_{s}}$ for a NFW model)
   and r$_{200}$ obtained from \citet{umetsu2016} and the MAMPOSSt analysis of \citet{sar20}, respectively.
   }
          \label{fig21}%
    \end{figure}   

Many of the physical processes that can affect cluster galaxies' properties are likely to be related to their orbits \citep{ton19,jos20}. Thus, to better investigate physical mechanisms responsible for galaxy transformations, we derive orbits for red and blue cluster galaxies separately (see Sect.~\ref{sec:6} for the definition of red and blue galaxies), following the procedure explained in \citet{biv13}. Our procedure is almost fully non-parametric, once the mass profile is specified. We follow \cite{sol90} to invert the Jeans equation, assuming spherical symmetry, and using the projected number density and velocity dispersion profile of red and blue galaxies, separately, and the total mass profile either from \citet{sar20} or from a joint weak and strong lensing analysis of \citet{umetsu2016}. We, then, invert the Jeans equation and obtain a non-parametric form of the 3D velocity anisotropy profile, $\beta(r) \equiv 1-(\sigma_t/\sigma_r)^2$, where $\sigma_t, \sigma_r$ are the tangential and radial components of the velocity dispersion tensor (we adopt the common assumption of two identical tangential components). 
The uncertainties on $\beta(r)$ are evaluated via a bootstrap procedure on the spectroscopic members. At each new bootstrap sampling, the number density and the line-of-sight velocity dispersion profiles are recomputed, and the inversion procedure is carried out. 

In Fig.~\ref{fig21}, we show $\beta(r)$ of the red (top panel) and blue (bottom panel) galaxies separately. We consider only objects with a stellar-mass Log(M$^*$/M$_{\odot}$) $\geq 9.5$ (see Appendix~\ref{app:C} for the derivation of stellar masses) to reduce the completeness correction factor that is applied in the estimate of the number density profiles. We also show the solution obtained by \citet{sar20} using MAMPOSSt \citep{mamon13}, assuming a \citet{tiret2007} model for $\beta(r)$ (magenta shaded region). This solution is in very good agreement with the Jeans inversion solution for the red galaxies, the dominant sample.

Red galaxies in A~S1063 have radial orbits characterised by small pericenters (this result is also discussed in Sect.~\ref{sec:8}). Hence, they may have experienced denser environments than galaxies on more tangential orbits. This can lead to more rapid transformations or a more efficient gas stripping, as suggested by \citet{sol01} by analysing galaxies in the Virgo cluster. We could also expect a difference in the orbits of high-mass and low-mass galaxies, since the more massive galaxies may be better able to survive the passage through small pericenters, i.e. the very hostile and dense environment of the central cluster region. However, dynamical friction acting on the more massive galaxies would have the opposite effect of reducing the radial anisotropy of these galaxies. To this end, we have compared the orbits of red galaxies in different mass ranges, but we do not see any significant difference in the galaxy orbits at varying masses.

The blue population has a systematically lower velocity anisotropy than the red population, at least at $r>0.7 R_{{\rm vir}}$. i.e. red galaxies seem to move on more radially biased orbits than the blue counterparts (see Fig.~\ref{fig21}) although the difference is not statistically significant. 

\section{Discussion and conclusions}
\label{sec:8}

Our analysis of the Frontier Fields cluster A~S1063  is based on an unprecedented sample of 1234 spectroscopic member galaxies out to 1.7$\times$r$_{200}$ (or $R\approx 4.5$ Mpc). Using the whole spectroscopic sample, we estimate the mean cluster redshift $\left<z_{\rm cl}\right>=0.3457\pm0.0001$ and the LOS velocity dispersion of galaxies $\sigma_\mathrm{v}=1380_{-32}^{+26}$. Our estimate of $\sigma_\mathrm{v}$ is significantly smaller than that presented by \citealt{gom12} ($\sigma_\mathrm{v}=1840_{-150}^{+230}$). This difference is likely due to a significantly larger size of our sample ($\simeq 24\times$ larger), which allows us to measure the $\sigma_\mathrm{v}$ radial profile, which is declining with radius (see Fig.\ref{fig15} and \citealt{sar20}), and also to obtain a better rejection of non-member galaxies. 

Based on the spectroscopic classification of 960 galaxies of our sample with high S/N spectra, we find that 50\% of cluster members are passive and red post-starburst galaxies, 18\% blue post-starbursts and weak emission-line galaxies, and 32\% medium, strong and very strong emission lines. To compare our results with the study of \cite{dre13}, who considers five rich clusters at $0.31<z<0.54$, we recompute the fractions per spectral class within 1.5$\times$r$_{200}$ and R$<22.3$.  We find that 59\% of all galaxies are either passive or post-starburst, and, among the members with no emission lines, 15\% are post-starburst. These values are to be compared to those of Dressler et al. (\citeyear{dre13}, see their Fig. 16 and Table 4), who found values in the range 50-80\% for a fraction of P and HDS galaxies among all members and 10-20\% for a fraction of post-starburst galaxies among passive members. We infer that A~S1063 shows a typical distribution of spectral types across its galaxy population.

On the basis of 51 member galaxies, \citet{gom12} detected substructure at the 90\% c.l. according to the DS-test and their Fig.~10 suggested a bimodal galaxy distribution. Our analysis adds several pieces of observational evidence that A~S1063 is far from being dynamically relaxed. We find the presence of two peaks in the velocity distribution, the strong NE-SW elongation in the galaxy distribution,
and the evidence of substructure at the $>99.9\%$ according to the DSv-test (see Sect.~\ref{sec:5} and Figs.\ref{fig9},\ref{fig10} and \ref{fig11}). Although this study is not focused on the cluster structure, we also present a more refined analysis in App.~\ref{app:B} finding five subclumps. We conclude that A~S1063 is far from being dynamically relaxed, and we suggest the merging of two or more subclumps along the NE-SW direction.

To further investigate the merging scenario and related timescales, we used the high-resolution set of adiabatic binary galaxy cluster merger simulations of \citet{ZuHone11}. The qualitative comparison between the X-ray emissivity of simulations and the X-ray contours published in \citet{bon18} suggests that a merger with a mass ratio of 1:3 between the two sub-clusters best represents X-ray observations. This scenario is also in agreement with the presence of a radio halo found by \citet{Xie20}. Radio halos are interpreted as transient components due to the turbulent reacceleration of relativistic electrons generated (and then dissipated) in cluster mergers (e.g., \citealt{bru09}). The acceleration timescale of the emitting electrons is $\sim$0.1-0.2 Gyr. 

Our spectral galaxy classification allows us to obtain new insight into the merging scenario for A~S1063. Figure~\ref{fig15} shows as the NE-SW direction is well traced by passive and red post-starburst galaxies but not by medium/strong/very strong emission-line galaxies. As similar evidence that the cluster elongation is traced by passive and red post-starburst galaxies but not by galaxies with important emission lines has already been found in the massive cluster MACS~J1206.2-0847 (\citealt{girardi2015}). A plausible interpretation is that passive galaxies trace the main accretion filament during the cluster formation and/or the last important cluster merger.  On the other hand, emission line galaxies trace more recent infall through several small groups or eventually randomly backsplash galaxies. 

We then explored the accretion histories of the different sub-populations of galaxies by dividing the projected phase-space diagram into infalling, backsplash, and virialised regions, inside and outside r$_{200}$, guided also by studies based on simulations. We found that 75\% of galaxies within r$_{200}$ are passives or post-starbursts. This result suggests that they passed through the cluster at least once, in agreement with \citet{bakels2020}, who found that roughly 21\% of subhaloes within a host's viral radius are currently on first infall and have not yet reached their first orbital pericentre. 

Moreover, we studied how galaxies with different spectral properties populate these pre-defined phase-space regions: the virialised region is dominated by passive galaxies, i.e. containing galaxies which have passed through the cluster at least once; the region of galaxies with negative large radial velocities ($\Delta{\rm v}/\sigma\lesssim -1.5$) within ${\rm r}\le{\rm r}_{200}$ is mainly dominated by galaxies with medium/strong emission lines, which have not yet passed through the cluster centre; the region with positive radial velocities ($\Delta{\rm v}/\sigma\gtrsim 1.5$), at ${\rm r}\le{\rm r}_{200}$ is populated primarily by weak emission line or post-starburst galaxies, suggesting that these galaxies, after passing the pericenter, may lie along the top edge of the projected phase-space diagram, at positive radial velocities, showing signs of quenching of star-formation. Inside r$_{200}$ the fraction of passive galaxies decreases as a function of the cluster radius, while the fraction of strong emission-line galaxies increases. Outside r$_{200}$ the fraction of strong emission-line galaxies reaches half of the whole population. 

By analysing the position of post-starbursts in the phase-space diagram depending on their colours, we find that blue post-starburst galaxies have negative velocities. We can speculate that blue post-starbursts result from the quenching (or the truncation) of star-formation in star-forming galaxies during the infall (ELG$\Rightarrow$bHDS$\Rightarrow$P). In this case, ram-pressure stripping or strangulation could play an important role in quenching star-formation. On the other hand, red post-starbursts may be the result of an evolutionary path of the type P$\Rightarrow$rHDS$\Rightarrow$P (e.g., \citealt{dre13}). In this case, the infall onto the cluster, or the merging of substructures, may "trigger" the starburst in P galaxies, and we observe these galaxies 1$-$2 Gyr after the burst (e.g., \citealt{girardi2015}). This time scale is comparable with the crossing time of A~S1063 (r$_{200}$/$\sigma$ $\sim$ 1.9 Gyr). The evolutionary scenario, where rHDS  
galaxies could be the remnant of the core of an infalling clump of galaxies that have experienced a merger with the main cluster, and bHDS could be the result of an ICM-related phenomenon, has also been suggested by \citet{mercurio2004}  in the cluster Abell~209. \citet{mercurio2004} also found that bHDS and ELG galaxies are blue disks, while P and rHDS are spheroids. 

By correlating the projected phase-space diagram obtained from the DLB07 simulations of massive cluster assembly with the distribution of different spectral classes in the observed projected phase-space diagram, we are able to estimate the accretion redshift for different galaxy populations (Fig.\ref{fig20}). According to a KS-1D test, the accretion redshifts of the three populations are not drawn from the same parent distribution at c.l.>99\%. Thus, we find a strong correlation between the infall times and the galaxy spectral properties. Galaxies with higher star formation rates are accreted later into the cluster. 

Moreover, high-mass (Log(M$^*$/M$_{\odot}$)$\ge$10.4) passive and red post-starbursts appear to have a different distribution of mean accretion redshift with respect to low-mass (Log(M$^*$/M$_{\odot}$)$<$10.4) passive galaxies (with probability $>0.99$ based a KS-1D test). The accretion redshift distribution of low-mass galaxies moves towards higher redshifts than the distribution of high-mass galaxies. This result may seem at odd with the expectation that more massive galaxies are accreted at earlier epochs as a natural consequence of hierarchical structure growth \citep{del2012}.
However, we should consider that we observe such a difference in the accretion redshift among galaxies that are passive at the time of observation, but we do not know whether they were passive also at the time of the accretion. Mass-quenching processes could explain our result when considering the time needed to quench star-formation as a function of the galaxy stellar mass. If the quenching time is longer for low-mass galaxies than for high-mass ones, we will observe as low-mass passives only those galaxies accreted early in the cluster (when they were star forming), and that had the time to quench their star-formation.

By analysing the orbits of member galaxies we find that red galaxies move on more radial orbits than blue ones. Using a semi-analytical model applied to the Millennium Simulation, \citet{ian2012} found that blue galaxies move on less radial orbits than red galaxies in clusters at any redshift from 0 to 0.7. They predict $\beta \approx 0.4$ (0.2) for red (respectively blue) galaxies in clusters at $z \approx 0.3$, in agreement with our result. This orbital difference is attributed by \citet{ian2012} to the fact that infalling galaxies can remain blue only if they move on tangential orbits.

On the other hand, the cosmological hydrodynamical simulation analysis by \citet{lot19} reaches the opposite conclusion. Blue galaxies move on more radial orbits than red galaxies, a difference that the authors interpret as blue galaxies being more infall dominated than red galaxies. The difference between the results of \citet{ian2012} and \citet{lot19} seems to point towards a different quenching timescale in the two simulations. If blue galaxies are more infall dominated than red galaxies, they would appear on more radial orbits than red galaxies, only if they can survive quenching for at least an orbital time.

Another possible explanation for the more radial orbits of red galaxies, is that at least part of them have been pre-processed before their  infall, and are in fact recent infallers. Thus, they show radial orbits since they still retain part of their original infalling trajectories before reaching the cluster centre for the first time. However, in this case, it is not clear why should we expect that red pre-processed galaxies infall on more radial orbits than blue galaxies that were not pre-processed.

From the observational point of view, our finding that red galaxies move on more radial orbits than blue galaxies is not generally confirmed in other clusters. Although error bars are large, the opposite trend is seen in clusters at any redshift \citep{biv04,biv13,mun14,biv16,mam19}. However, most of the existing results are based on stack samples so that a large spread may exist in the $\beta(r)$ of red and blue galaxies in different clusters. \citet{AADDV17} find that in the nearby cluster Abell~85, red galaxies have more radially elongated orbits than blue galaxies, as we find here. The spread in the $\beta(r)$ of red and blue galaxies among different clusters may be due to different accretion histories, different current accretion rates, or a different quenching efficiency. Thus, we plan to extend the present analysis to quantify the spread in the $\beta(r)$ of red and blue galaxies by analysing the full sample of CLASH-VLT clusters in the near future.

Our analysis shows that extensive spectroscopic information on large samples of cluster members, extending well beyond the virial radius, allows new insights to be obtained on the assembly history of cluster galaxies in relation to their star formation histories via a direct comparison of observed (projected) phase-space diagrams with those derived from cosmological simulations. 

In order to further investigate the accretion history of cluster galaxies, as well as the mechanisms and time scales of star-formation quenching, we plan to extend the present analysis to the full sample of CLASH-VLT clusters in the near future.

\begin{acknowledgements}
The authors thank the anonymous referee for the very useful comments, that improved the manuscript. We acknowledge financial contributions by PRIN-MIUR 2017WSCC32 "Zooming into dark matter and proto-galaxies with massive lensing clusters" (P.I.: P.Rosati), INAF ``main-stream'' 1.05.01.86.20: "Deep and wide view of galaxy clusters (P.I.: M. Nonino)" and INAF ``main-stream'' 1.05.01.86.31 "The deepest view of high-redshift galaxies and globular cluster precursors in the early Universe" (P.I.: E. Vanzella). MB acknowledges financial contributions from the agreement \textit{ASI/INAF 2018-23-HH.0, Euclid ESA mission - Phase D} and with AM the \textit{INAF PRIN-SKA 2017 program 1.05.01.88.04}. C.G. acknowledges support through grant no.~10123 of the \textit{VILLUM FONDEN Young Investigator Programme}. GBC acknowledge the Max Planck Society for financial support through
the Max Planck Research Group for S. H. Suyu and the academic support
from the German Centre for Cosmological Lensing. PB acknowledges financial support from ASI through the agreement ASI-INAF n. 2018-29-HH.0. R.D. gratefully acknowledges support from the Chilean Centro de Excelencia en Astrof\'isica y Tecnolog\'ias Afines (CATA) BASAL grant AFB-170002".
\end{acknowledgements}

\appendix

\section{Photometry}
\label{app:A}

The photometric data includes ground-based wide-field observations in B-, V-, R-, i- and z-band carried out with the wide-field imager (WFI) at the MPG/ESO 2.2-meter telescope at the La Silla Observatory 2.2, giving a field of view of 34$^\prime \times$ 33 $^\prime$,  and HST data in 16 broadband filters, from the UV to the near-IR, as part of the CLASH multi-cycle treasury program (see \citealt{pos12}). Moreover, as part of the FF program, A~S1063 is also imaged with HST, for a total of 140 orbits, divided over seven optical/near-infrared bands (F435W, F606W, F814W, F105W, F125W, F140W, F160W).

\subsection{WFI data}
\label{app:A1}

The ground-based photometric observations were carried out with the wide field imager (WFI\footnote{\footnotesize{MPG/ESO 2.2-metre telescope at La Silla Observatory 2.2, a mosaic of 8 2k $\times$ 4k pixels CCDs giving a field of view of 34$^\prime \times$ 33$^\prime$}}) under the Programme 085.A-9002(A), P.I. S. Seitz. In this paper we analysed B, V, R, i and z band images, having total exposure times of 20297.6137\,s, 23696.685\,s, 34195.3103\,s, 29391.937\,s and 11537.7814\,s, and FWHMs $\sim$1.25, $\sim$0.95, $\sim$0.76, $\sim$1.18, $\sim$1.04, respectively, sampled at 0.20 arcsec per pixel. 

The photometric catalogues are produced using the software SExtractor \citep{bertin96} in conjunction with PSFEx\footnote{Available at \url{http://www.astromatic.net/software/psfex}\ .} (\citealp{bertin11}), which performs PSF fitting photometry. We extract independent catalogues in each band that are then matched across the four wavebands using STILTS \citep{tai06}.
We use a two-step approach in the catalogue extraction as in \cite{mer15}. First, we run SExtractor in a so-called {\it cold mode} where the brightest and extended sources are properly deblended; then, in a second step, we set configuration parameters in the so-called {\it hot mode}, in order to detect fainter objects and to split close sources properly (e.g. \citealt{rix04,cal08}). Finally, we combine the two catalogues by replacing extended objects, properly deblended in cold mode, in the catalogue of sources detected in the hot mode, and by deleting multiple detections of these extended sources.
Among the photometric quantities, we measure aperture magnitudes ({\tt{MAG\_APER}}) in 9 circular apertures with diameters of 1.5, 3.0, 4.0, 5.0, 8.0, 16.0, 30.0, 3*FWHM, 8*FWHM, isophotal magnitudes ({\tt{MAG\_ISO}}), computed by considering the threshold value as the lowest isophote and the Kron magnitude ({\tt{MAG\_AUTO}}, which is estimated through an adaptive elliptical aperture \citep{kron80}.

Using PSFEx it is possible to model the PSF of the images (see \citealt{armstrong2010,bertin11,mohr2012,bou13} for details). Thus, it is possible to run SExtractor taking the PSF models as input to measure the PSF-corrected model fitting photometry for all sources in the image. In this case we extracted magnitudes from: {\it (i)} the PSF fitting ({\tt{MAG\_PSF}}) and {\it(ii)} the sum of the bulge and the disc components, centred on the same position, convolved with the local PSF model ({\tt{MAG\_MODEL}}).

To separate galaxies and point-like sources, we adopt a progressive approach analogous to those described in \cite{ann13} and \cite{mer15}, using: i) the \textit{stellarity index} ({\tt{CLASS\_STAR}}); ii) the half-light radius ({\tt{FLUX\_RADIUS}}); iii) the new SExtractor classifier {\tt{SPREAD\_MODEL}}; iv) the peak of the surface brightness above the background (\texttt{$\mathrm{\mu_{max}}$}); v) a final visual inspection for objects classified as galaxies but with edge values of the stellarity index ({\tt{CLASS\_STAR}}$\ge$0.9, see below).

\begin{figure*}
\hbox{\includegraphics[width=0.50\textwidth, bb= 30 0 700 450,clip]{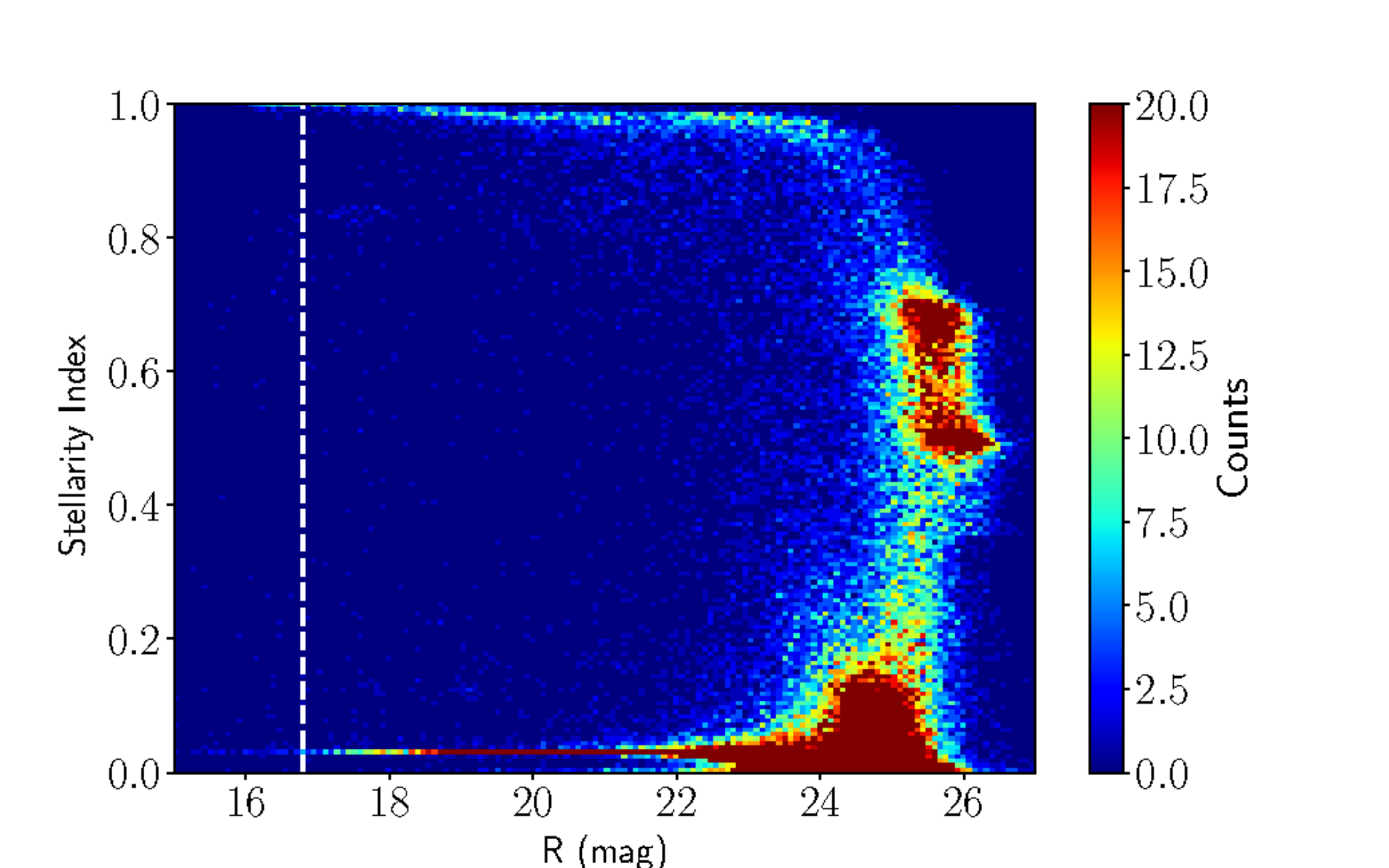}
{\includegraphics[width=0.50\textwidth, bb= 30 0 700 450,clip]{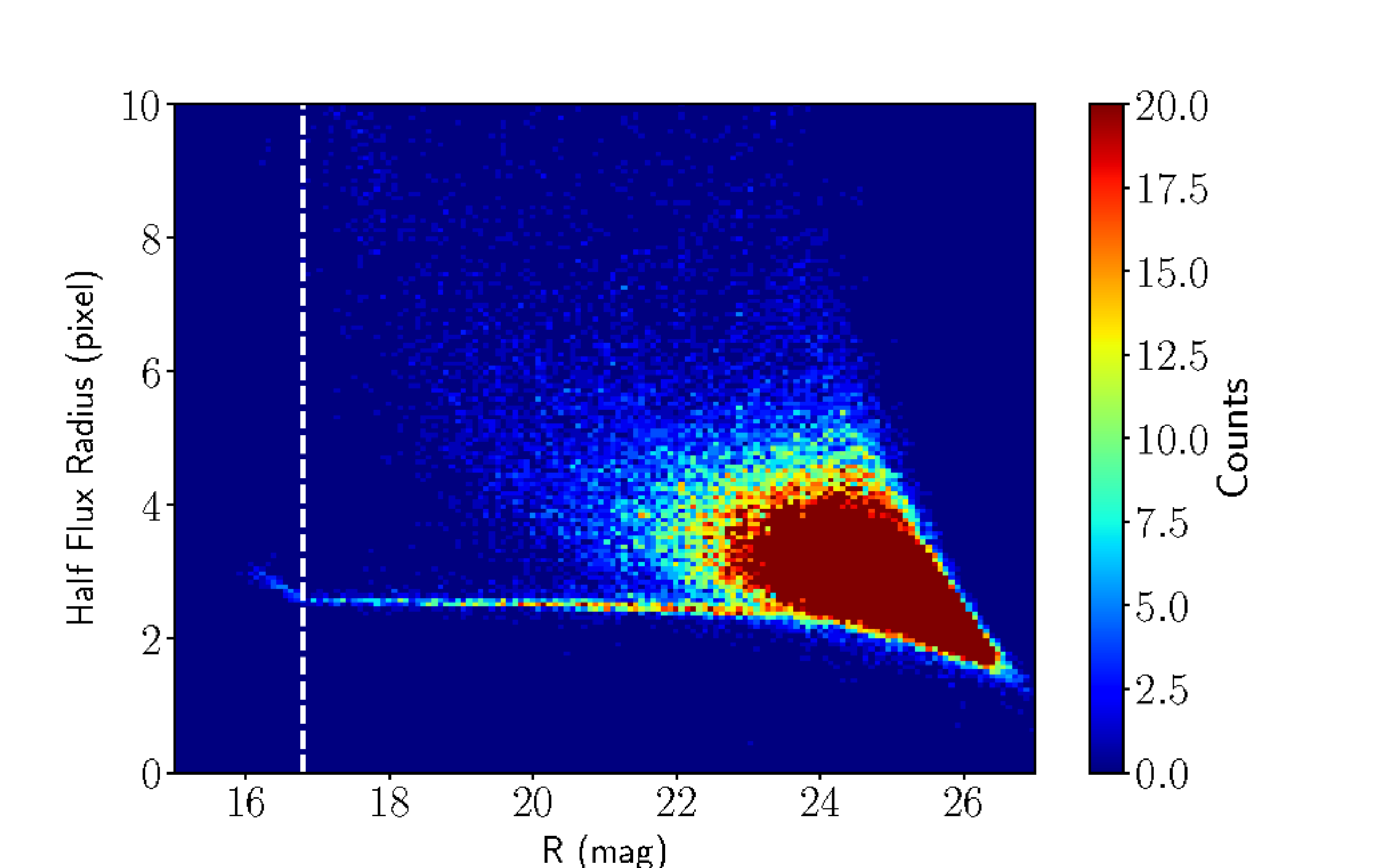}}
}
\hbox{\includegraphics[width=0.5\textwidth, bb= 0 0 700 450,clip]{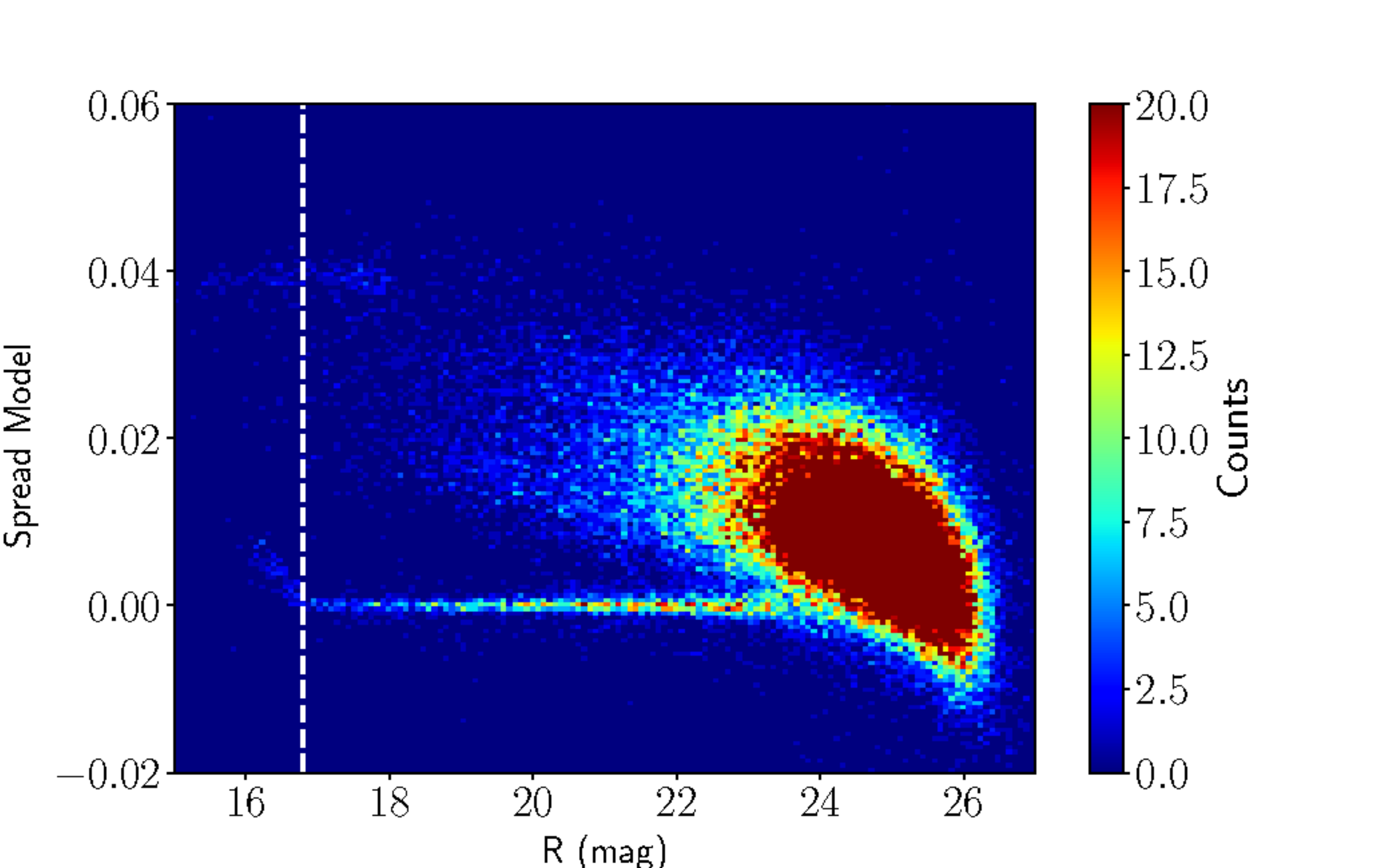}
{\includegraphics[width=0.5\textwidth, bb= 30 0 700 450,clip]{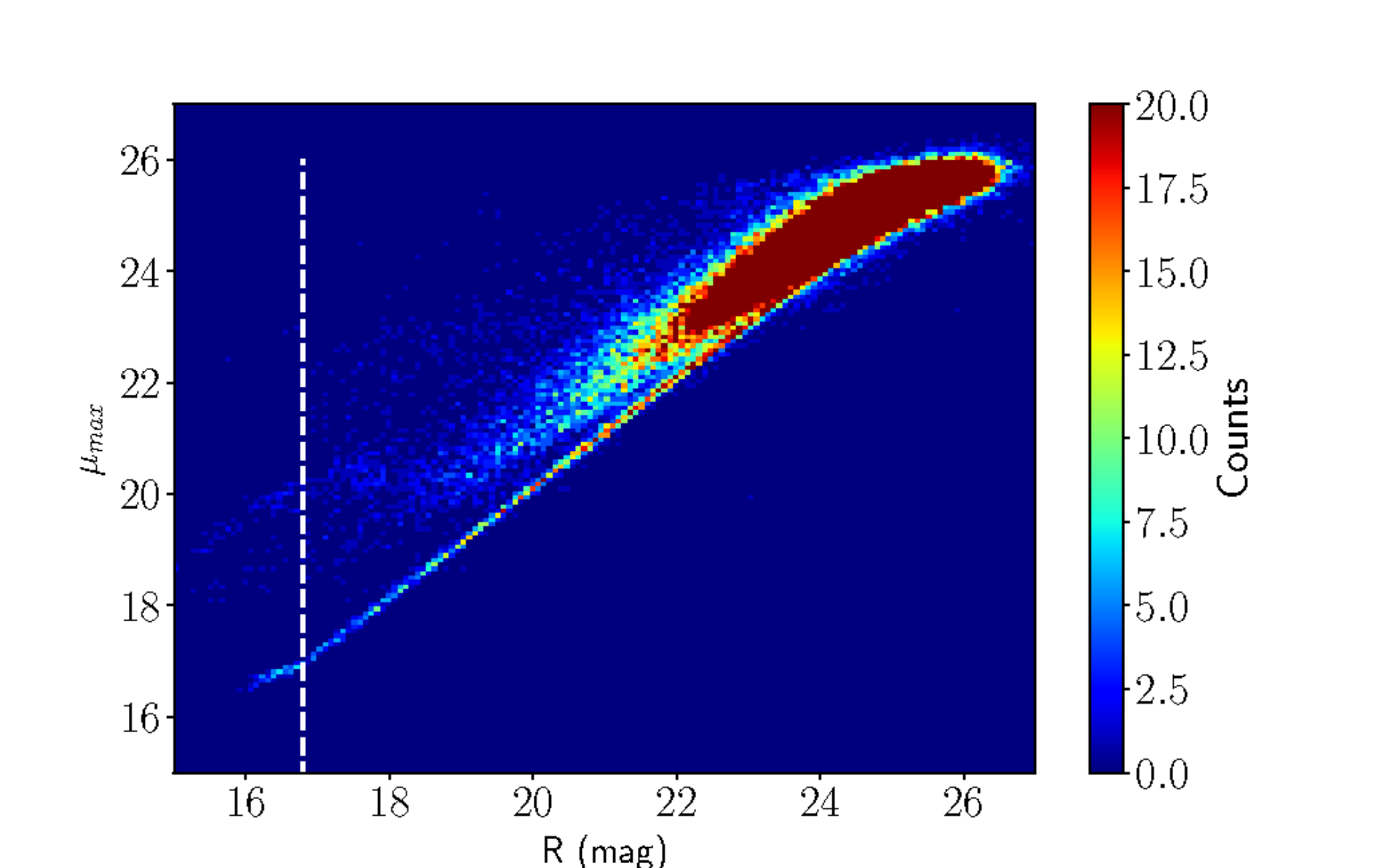}}
}
\caption{Two-dimensional histogram of SExtractor stellarity index ({\it{Top left panel}}), half-light radius ({\it{Top right panel}}), spread model ({\it{Bottom left panel}}), and $\mathrm{\mu_{max}}$ ({\it{Bottom right panel}}) as a function of the Kron magnitudes for sources in the $R$-band image. The vertical dashed line in all panels is the magnitude limit adopted for saturated objects. The points are colour-coded according to their number counts as reported in the vertical colour bars.}
\label{figA1}
\end{figure*}

In Fig.~\ref{figA1}, top left panel, {\tt{CLASS\_STAR}} is plotted as a function of the Kron magnitude for R band. The sequence of unsaturated stars (R$>$16.8 mag) is separated from galaxies by selecting a {\tt{CLASS\_STAR}} value above 0.98 only down to R=21.5 mag. For magnitudes fainter than this value, lowering the established limit to separate stars and galaxies causes an increase in the star subsample contamination from galaxies. We classified sources fainter than R=21.5 mag by using {\tt{FLUX\_RADIUS}} as a measure of source concentration. Figure~\ref{figA1}, top right panel shows that the locus of stars, defined according to the relation between half-light radius and Kron magnitude is reliable down to R=23.0 mag. Finally, to obtain a reliable star/galaxy classification for sources fainter than this limit, we used the new SExtractor classifier, {\tt{SPREAD\_MODEL}}, which takes into account the difference between the model of the source and the model of the local PSF \citep{mohr2012}. By construction, this parameter is close to zero for point sources, positive for extended sources (galaxies), and negative for detections smaller than the PSF, such as cosmic rays. Figure~\ref{figA1}, bottom left panel, shows that stars and galaxies tend to arrange themselves in two different loci in the distribution of the {\tt{SPREAD\_MODEL}} as a function of Kron magnitude. Based on this diagram, we classified as galaxies all sources with {\tt{SPREAD\_MODEL}} $>$ 0.003 and 23.0$<$R$<$25.0 mag. The plot of $\mu_{max}$ as a function of the Kron magnitude is used in order to select saturated stars (vertical dashed line Fig.~\ref{figA1}), as shown in the bottom right panel. We also performed a visual inspection of those objects classified as galaxies but with {\tt{CLASS\_STAR}}$>$0.9. Since the R band is the deepest band of the survey and the one conducted in the best seeing conditions, it is used to classify sources in the five optical bands' cross-correlated catalogue. Finally, the catalogues in all bands have been visually inspected on the images to check the residual presence of spurious or misclassified objects, like traces of satellites, effects of bad columns.

\begin{figure*}
\centering
\includegraphics[scale=0.70]{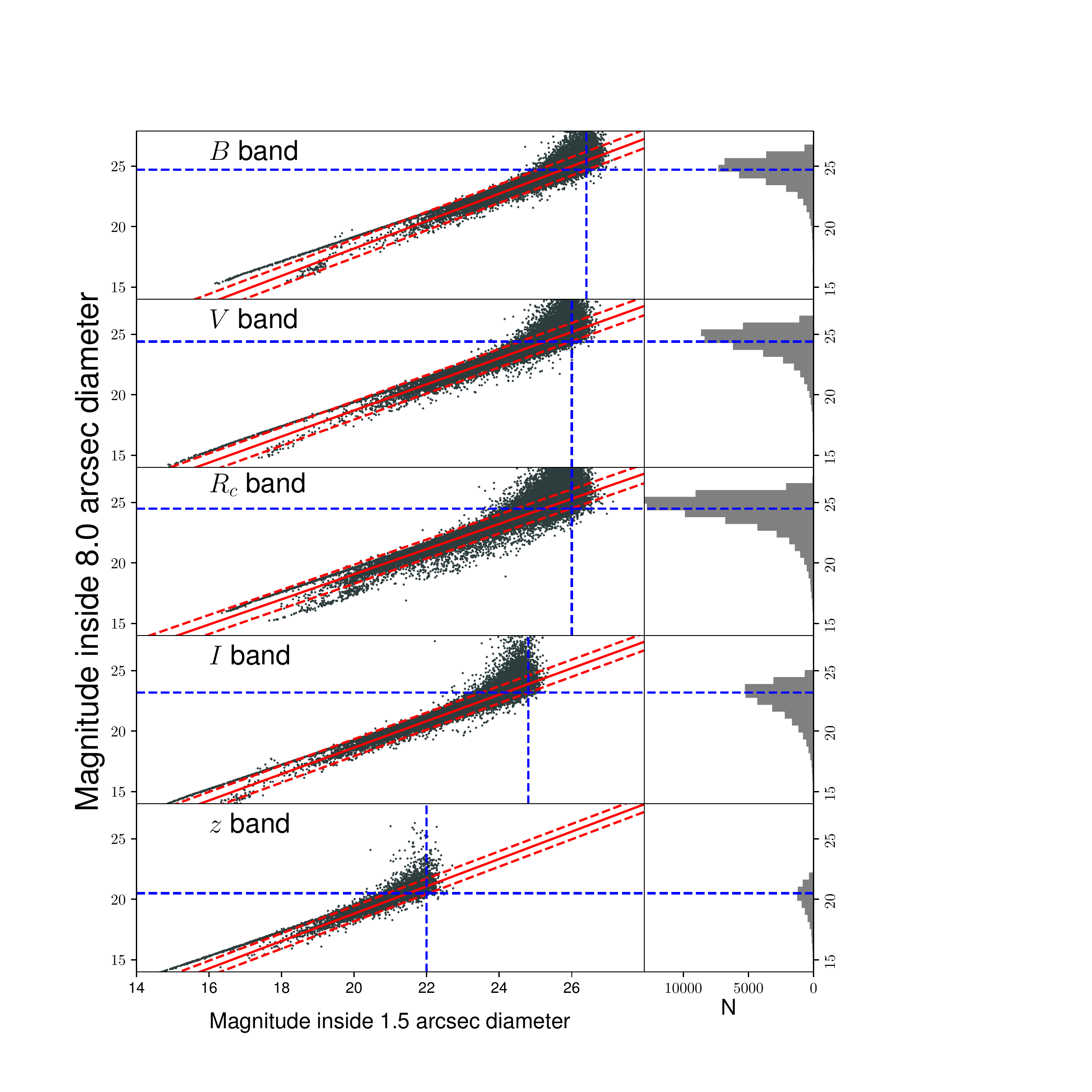}
\caption{{\it Left panels}: Distribution of the SExtractor magnitude inside an 8.0$^{\prime\prime}$ diameter as a function of the magnitude inside a detection aperture of 1.5$^{\prime\prime}$ diameter for BVRiz bands. The horizontal and vertical blue dashed lines indicate the detection and completeness limits, respectively. The red continuous lines are the linear relation between the magnitude within the 8.0$^{\prime\prime}$ diameter aperture and the magnitude within the detection aperture, minus/plus 1$\sigma$ (red dashed lines). {\it Right panels}: Number counts of galaxies for each band. Dashed blue lines mark the completeness magnitudes.}
\label{figA2}
\end{figure*}

The photometric limiting magnitude of the extracted catalogue for galaxies is defined as the magnitude limit below which the completeness drops from 95\%. Following the method of \citet{gar99} and adopted also in \citet{mer15}, we estimate the completeness magnitude limit as the magnitude at which we begin to lose galaxies because they are fainter than the brightness threshold inside a detection aperture of 1.5$^{\prime\prime}$ diameter\footnote{This aperture was adopted as being suitable for all images and comparable to the larger seeing value.}, for all the optical bands. In all panels of Fig.~\ref{figA2}, the vertical blue dashed lines represent the detection limit, while the continuous red lines are the linear empirical relation between the magnitude within a 8.0${\arcsec}$ diameter aperture and the magnitude within the detection aperture. The relation between the two magnitudes shows a scatter, depending essentially on the galaxy profiles. Taking into account this scatter (see dashed red lines in Fig.~\ref{figA2}), we fixed as a completeness magnitude limit (blue dashed horizontal line) the intersection between the lower 2.6*$\sigma$ limit of the relation and the detection limit. 

The catalogues turned out to be 95\% complete at the total magnitudes of 24.7, 24.4, 24.5, 23.2, 20.5 in the BVRiz bands, respectively. The final photometric catalogue contains $\sim$ 34 000 (33 879) objects down to R$_{AB}=$24.6 mag.

\subsection{HST data}
\label{app:A2}

A~S1063 is one of the six clusters imaged in seven optical/near-infrared bands: F435W, F606W, F814W, F105W, F125W, F140W, F160W  with the FF program (P.I.: J. Lotz). This program combines the power of HST with the natural gravitational telescopes of high-magnification clusters of galaxies, producing the deepest observations of clusters and their lensed galaxies ever obtained. These images allow: (1) to study distant galaxy populations $\sim$10-100 times fainter than any previously observed, (2) to improve the statistical understanding of galaxies during the epoch of reionisation, and (3) to provide unprecedented measurements of the dark matter within massive clusters. We analysed the public HST photometric data available at the STScI MAST Archive\footnote{\footnotesize{https://archive.stsci.edu/prepds/frontier/}}. We used the v1.0 release of Epoch 1 images, processed with the new 'self-calibration approach \citep{and14} to reduce low-level dark current artefacts across the detector \citep{tor18}.
A~S1063 is also one of the 25 massive (virial mass $M_{vir}$  5-30 $\times$ 10$^{14}$ M$_{\odot}$) galaxy clusters observed with The Cluster Lensing And Supernova survey with Hubble (CLASH, P.I.: M. Postman). This survey was awarded 524 orbits of HST for observing a sample of clusters, spanning the redshift range z = 0.18-0.90 in 16 broadband filters: four filters from WFC3/UVIS, five from WFC3/IR, and seven from ACS/WFC, ranging from approximately 2000 to 17000 \AA. The sample was carefully chosen to be largely free of lensing bias and representative of relaxed clusters, based on their symmetric and smooth X-ray emission profiles (for a thorough overview, see \citealt{pos12}). CLASH has four main scientific goals: (1) measure the cluster total mass profiles over a wide radial range, through strong and weak lensing analyses; (2) detect new type Ia supernovae out to redshift z $\sim$ 2.5 to improve the constraints on the dark energy equation of state; (3) discover and study some of the first galaxies that formed after the Big Bang; and (4) perform galaxy evolution analyses on cluster members and background galaxies. Ancillary science that can indeed be carried out with CLASH's superb data set is the analysis of several new SL systems on the galaxy scale.
With an averaged exposure time of $\sim$2500 s (one to two orbits) per image (or 20 orbits per cluster if all filters are included), the CLASH observations reach a typical photometric depth of F814W=28.0 or F160W=26.5 (S/N$>$3) (see \citealt{mol17} for details). Image reduction, alignment, and co-adding were made using the MosaicDrizzle pipeline \citep{koe03,koe11}, where a final scale of 0.065 arcsec pixel$^{1}$ was chosen for all the fields. The reduced images and weight maps are available at the Mikulski Archive for Space Telescopes (MAST)\footnote{\footnotesize{https://archive.stsci.edu/prepds/clash/}}.

\begin{figure}[ht]
   \centering
   \includegraphics[width=0.5\textwidth]{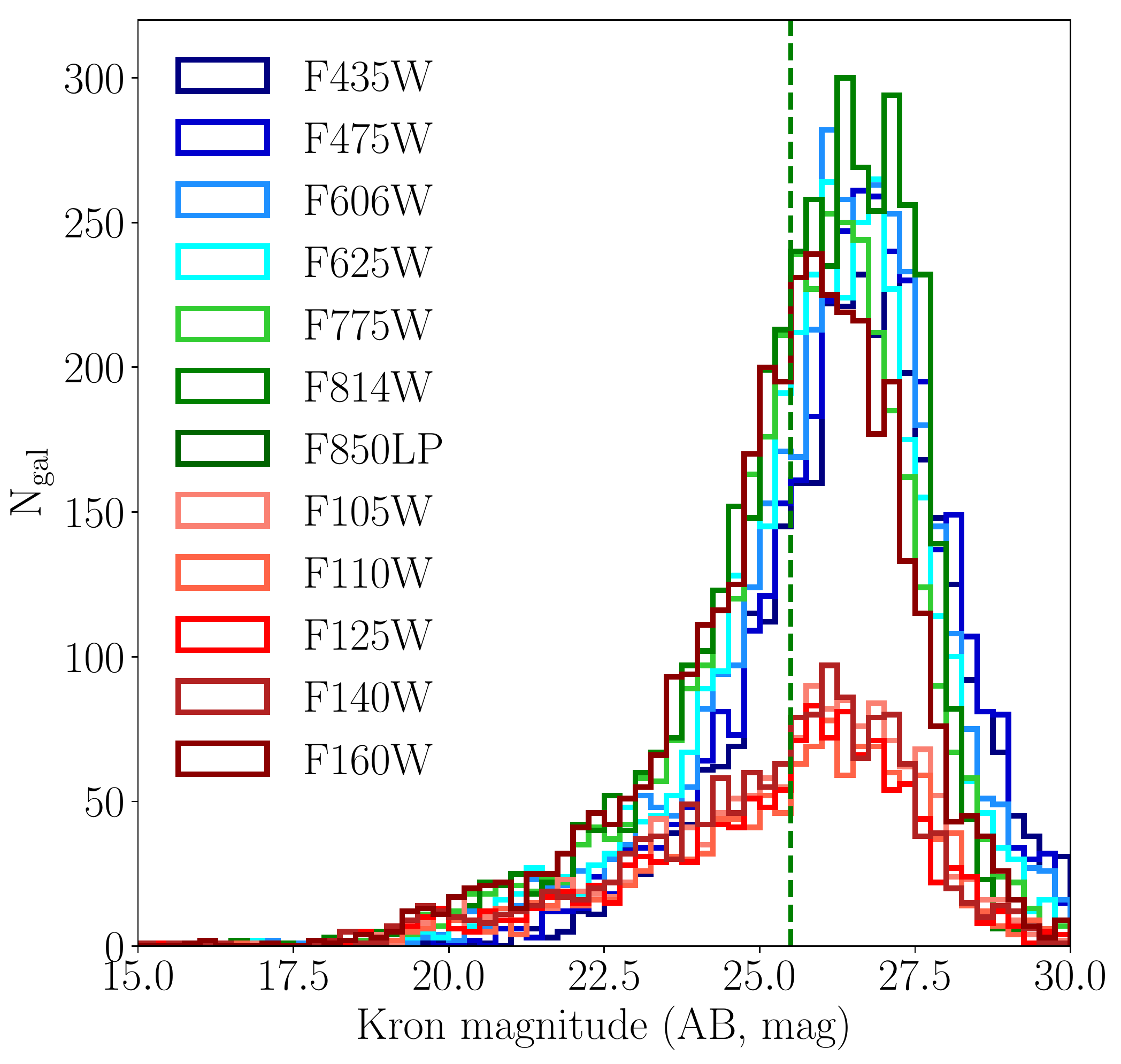}
   \caption{Magnitude distribution of HST sources for 12 wavebands from F435W to F160W in the field of A~S1063. The dashed line indicates the 95\% limiting magnitude in F814W band.}
          \label{figA3}%
    \end{figure}

We extract the catalogues for each of the 16 CLASH bands, in dual-mode, by using as detection image, the deep image obtained combining all the Optical plus NIR images. As for ground-based catalogues, we use a two-step approach in the catalogue extraction using {\it cold} and {\it hot mode} (see above), in order to detect fainter objects and to properly split close sources (e.g. \citealt{rix04,cal08}). We combine, finally, the two catalogues by replacing extended objects, properly deblended in {\it cold mode}, in the catalogue of sources detected in the {\it hot mode}, and by deleting multiple detections of these extended sources. Finally, we visually inspect catalogues to clean them from spurious detections, and we put magnitudes eq. to -99.0 (i.e. no magnitude measurement) in the bands where objects are detected in the combined image but have a value equal to zero in the weight map of those bands. Finally, we obtain a multi-band catalogue of $\sim$ 4300 sources, having $\sim$ 1500 sources down to the 95\% limiting magnitude of F814W=25.5 mag (see the green vertical line in Fig.~\ref{figA3}).

\section{3D dynamical analysis with DEDICA}
\label{app:B}
We apply the 3D-DEDICA method \citep{pisani1993, pisani1996} to the full spectroscopic sample. As in \citet{bal16}, we use a simplified version of the 3D-DEDICA method, based on the same definition of the adaptive kernel estimate of Pisani \citep{pisani1993, pisani1996}, with the same computation of the local bandwidth factor $\lambda_i$ (see Equations (26) and (27) in \citealt{pisani1993}), but using the size parameter of the kernel proposed by \citet{silverman1986}. This simplified procedure is optimised to trace mainly the large-scale structure of the cluster. This procedure identifies five sub-clumps with significance $>99.9\%$, whose properties are described in Table~\ref{tab:2}. 

\begin{table}[]
    \caption{\label{tab:2} 3D substructures.}
    \centering
    \begin{tabular}{l c c c}
\hline
\hline
 Sub-clump  & N$_S$ & velocity peak & RA(J2000) DEC(J2000)\\
 (1) & (2) & \ks (3) & (4) \\
 \hline
   1-red  & 531  & 104566 & 22:48:44.5 -44:31:59 \\
   2-purple  & 336  & 102879 & 22:48:48.0 -44:31:26 \\
   3-blue  & 181  & 101479 & 22:48:47.6 -44:31:17 \\
   4-orange  & 107  & 104103 & 22:47:55.7 -44:40:22 \\
   5-green  & 78   & 103168 & 22:48:01.8 -44:26:32 \\	
\hline
\hline
\end{tabular}
\tablefoot{Results on the detection of 3D substructures with significance $>99.9\%$ in the whole sample of cluster members of A~S1063. Columns list the following information: (1) id of sub-clump, (2) number of assigned members, (3) peak velocity in \ks, (4) coordinates.}
\end{table}

\begin{figure*}
\centering
\includegraphics[width=0.50\textwidth]{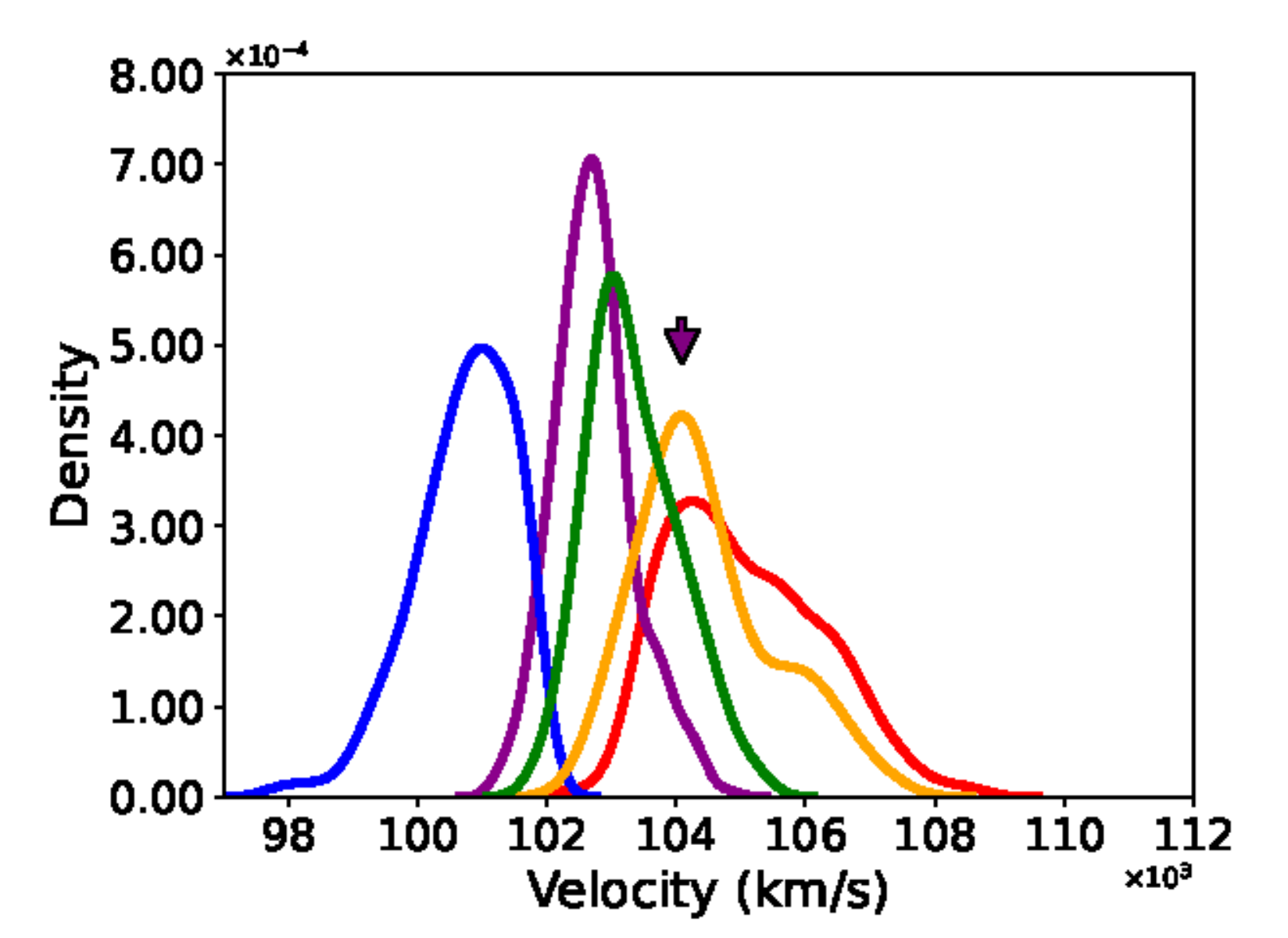}
\includegraphics[width=0.45\textwidth]{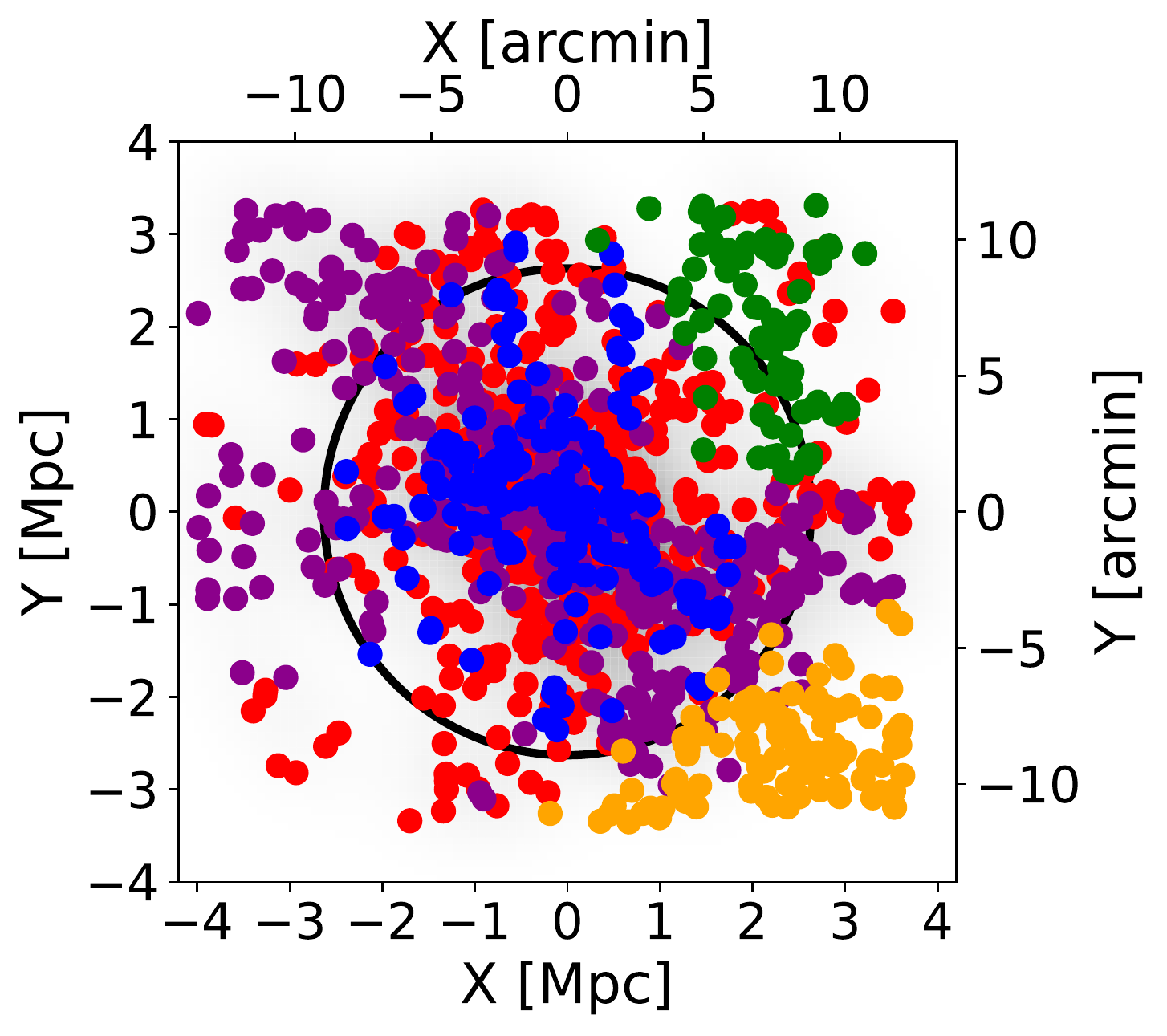}
\caption
{Velocity ({\it Left panel}) and 2D ({\it Right panel}) distributions of the 5 substructures detected with 3D-DEDICA. The arrow indicates the velocity of the BCG. The colour-code, velocity peak, and the centre for each substructure are reported in Tab.~\ref{tab:2}. In the right panel, the circle has a radius equal to r$_{200}$ = 2.63 Mpc.}
\label{figSubs}
\end{figure*}

Figure~\ref{figSubs} shows the velocity (left panel) and 2D (right panel) distributions of the 5 sub-clumps detected with 3D-DEDICA. Sub-clumps 1 (red) and 2 (purple) follow the NE-SW elongation of the cluster (right panel of Fig.~\ref{figSubs}) and show different velocity distributions (left panel of Fig.~\ref{figSubs}), with sub-clump 1 having a broader distribution and a higher velocity peak with respect to sub-clump 2, also producing the two peaks visible in the right panel of fig~\ref{fig9}. Sub-clump 3 (blue) is concentrated inside r$_{200}$, while sub-clumps 4 (orange) and 5 (green) lie outside r$_{200}$. Galaxies in the small high-velocity group detected by the DSv-test at X=1.0~Mpc and Y=$-3.2$~Mpc in Fig.~\ref{fig10} belong to substructure 4. On the other hand, the SW low-velocity region (X=1.0~Mpc and Y=$-1.0$~Mpc in Figs.~\ref{fig10} and~\ref{fig11}), detected by the DSv-test, is composed mainly of galaxies belonging to sub-clump 3. 
The BCG, whose velocity is indicated by the arrow in the left panel of Fig.~\ref{figSubs}, belongs to sub-clump 1.

\begin{figure}[ht]
   \centering
   \includegraphics[width=9.0cm]{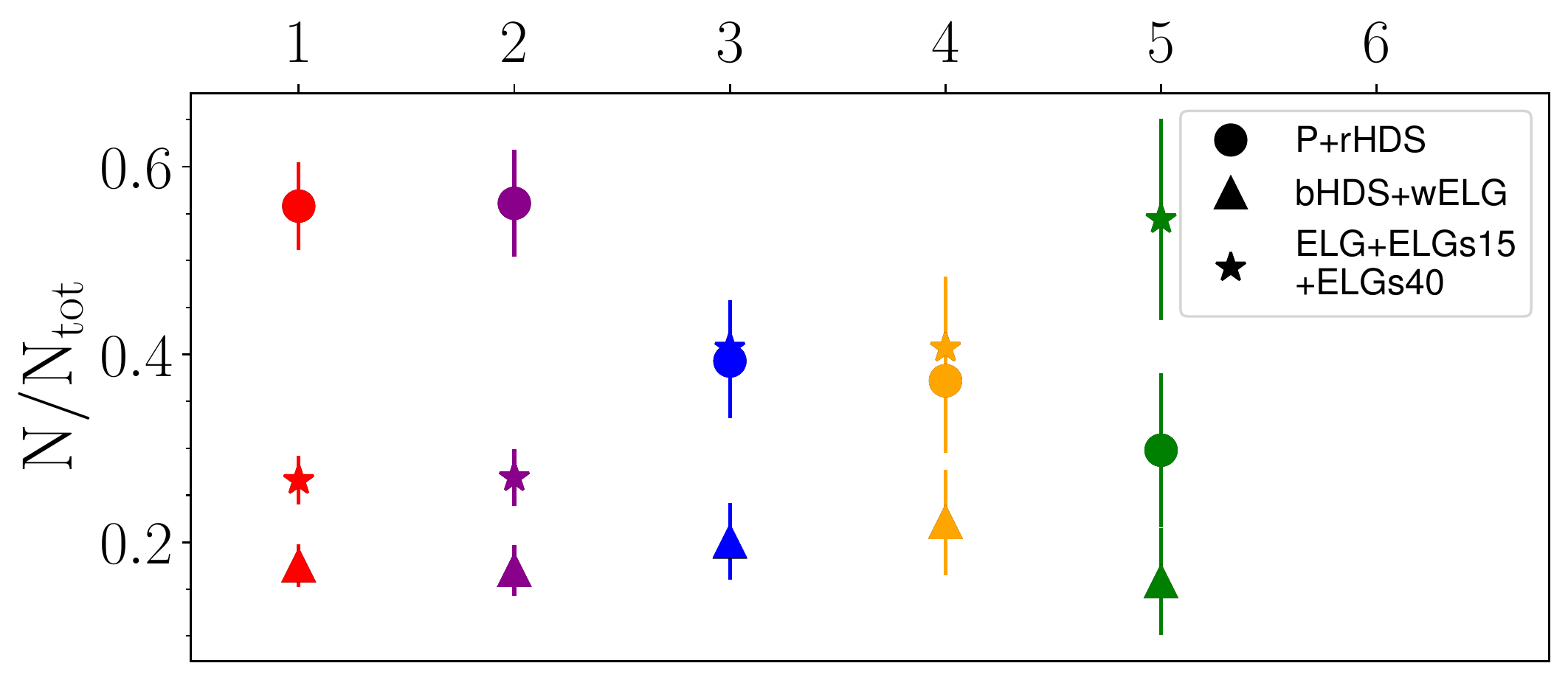}
   \caption{Fraction of galaxies with different spectral types inside each sub-clump.}  
          \label{phasespace_substr}%
    \end{figure}

We also examine the fraction of different spectral types in each identified sub-clump (Fig.~\ref{phasespace_substr}, and Tab.~\ref{tab:3}). Sub-clumps 1 and 2 are composed mainly of P+red HDS, which belong to the virialised region in the projected phase-space diagram. Sub-clump 3 shows similar fractions of P+red and mELG+ELGs15+ELGs40, which belong to the virialised region and region inside r$_{200}$ with negative large velocity dispersion ($\Delta$v/$\sigma<$-1.5) in the projected phase-space diagram. Sub-clump 5 is located outside r$_{200}$ (i.e. belonging to the infalling region) and is composed mainly by galaxies with velocities lower than the mean cluster LOS velocity ($\left<V\right>=103\,640\pm39$ \kss, see Fig.~\ref{figSubs}). It shows, as expected, a high fraction of mELG+ELGs15+ELGs40. On the other hand, sub-clump 4 is located outside r$_{200}$, and is composed of galaxies with velocities higher than the mean cluster LOS velocity (see Fig.~\ref{figSubs}). We found a lower fraction of mELG+ELGs15+ELGs40 and the higher fraction of P+red HDS in sub-clump 4 with respect to sub-clump 5, indicating a different spectral composition of galaxies outside r$_{200}$ but velocities higher and lower than the mean cluster LOS velocity, respectively.

\begin{table}[]
    \caption{\label{tab:3} Fraction of different spectral classes inside 3D Substructures.}
    \centering
    \begin{tabular}{l c c c}
\hline
\hline
 Sub-clump & P+      & blue HDS & mELG+\\
           & red HDS & wELG     & ELGs15+ELGs40\\
 (1) & (2) & (3) & (4) \\
 \hline
   1-red  & 55$\pm$5\% & 18$\pm$2\% & 27$\pm$3\% \\
   2-purple  & 57$\pm$6\% & 16$\pm$3\% & 27$\pm$3\% \\
   3-blue  & 40$\pm$6\%  & 20$\pm$4\% & 40$\pm$3\% \\
   4-orange  & 37$\pm$8\% & 21$\pm$6\% & 42$\pm$7\% \\
   5-green  & 30$\pm$8\% & 16$\pm$6\% & 54$\pm$11\% \\	
\hline
\hline
\end{tabular}
\tablefoot{Fraction of different spectral classes inside each of the 3D Substructures detected in Sect.~\ref{sec:5}. Columns list the following information: (1) id of sub-clump, (2)-(3)-(4) fractions of [P+red HDS], [blue HDS+wELG] and [mELG+ELGs15+ELGs40] for each substructure, respectively.}
\end{table}

\begin{figure}[ht]
\centering
\includegraphics[width=0.45\textwidth]{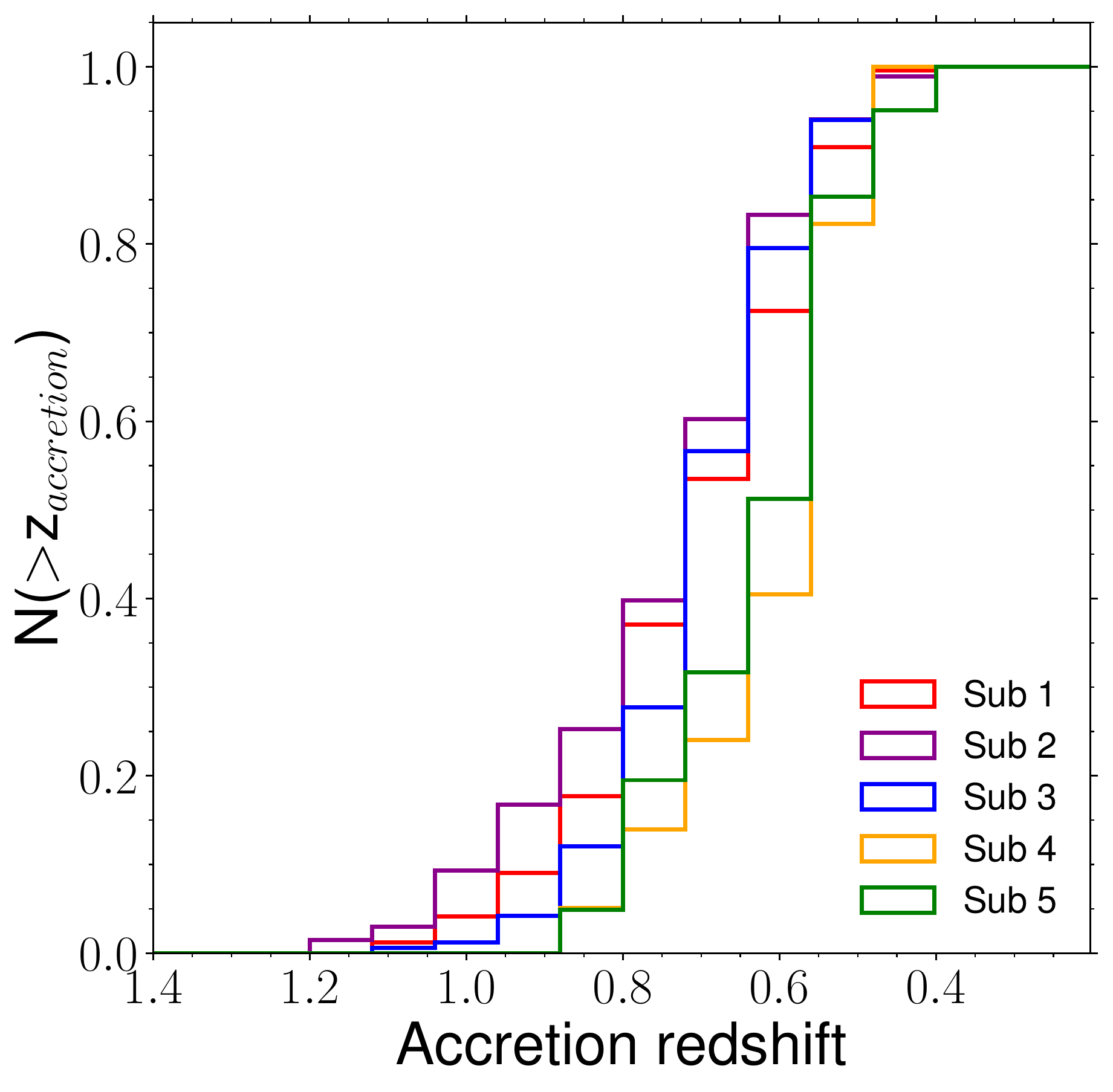}
\caption
{Cumulative distribution of accretion redshift for galaxies belonging to the five sub-clumps.}
\label{accretionreds_subclumps}
\end{figure}

Finally, we compared the accretion redshift distributions of the five sub-clumps (see Fig.~\ref{accretionreds_subclumps}). According to the KS-1D, the accretion redshift distributions of sub-clumps 1, 2, and 3 differ from those of sub-clumps 4 and 5, with the latter being accreted, on average, later as expected, because sub-clumps 4 and 5 are more external and composed mainly by SF galaxies. 

\section{Stellar Masses}
\label{app:C}

\begin{figure*}
\hbox{\includegraphics[width=0.50\textwidth, bb= 0 0 500 450,clip]{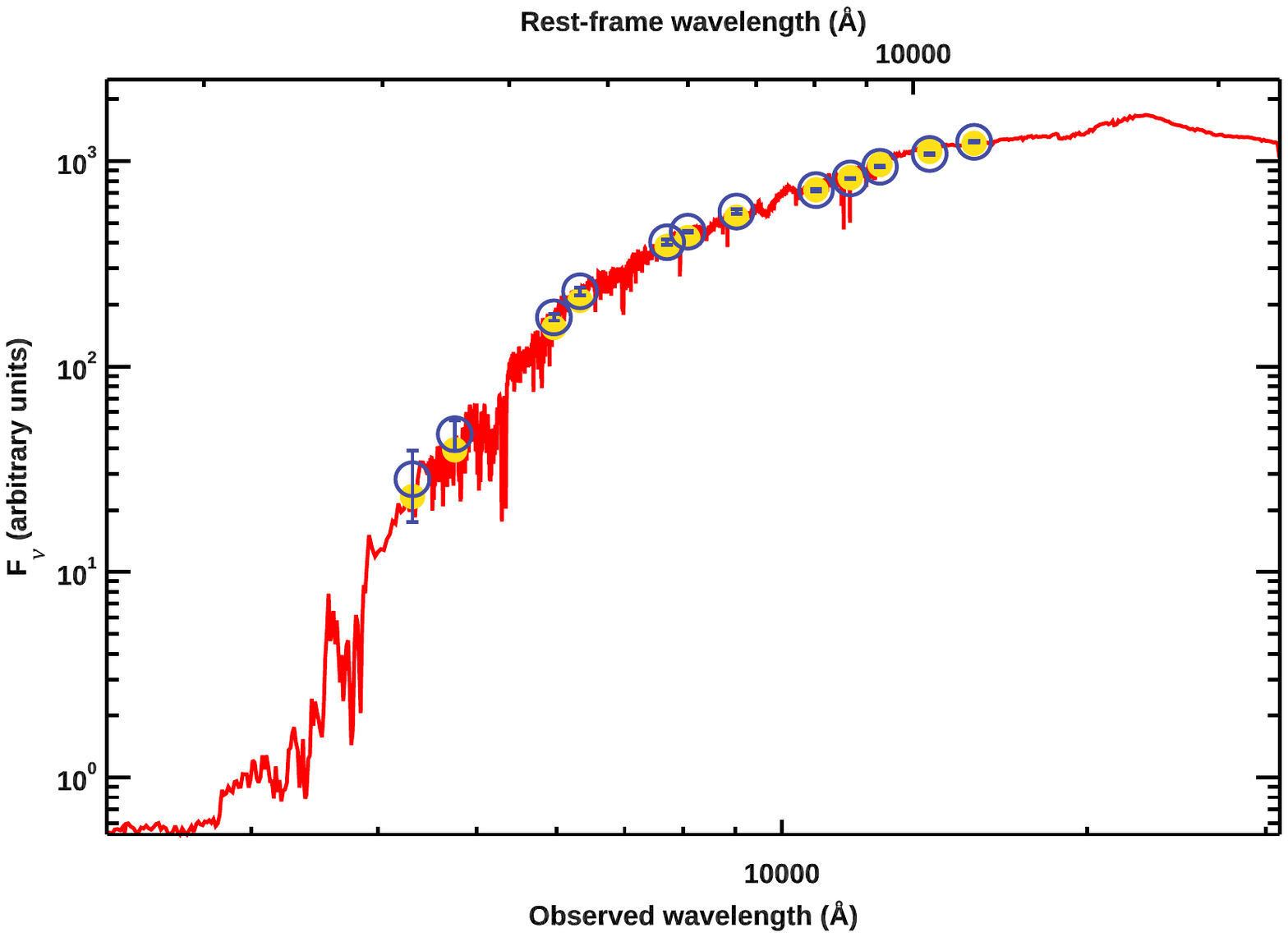}
{\includegraphics[width=0.50\textwidth, bb= 0 0 500 450,clip]{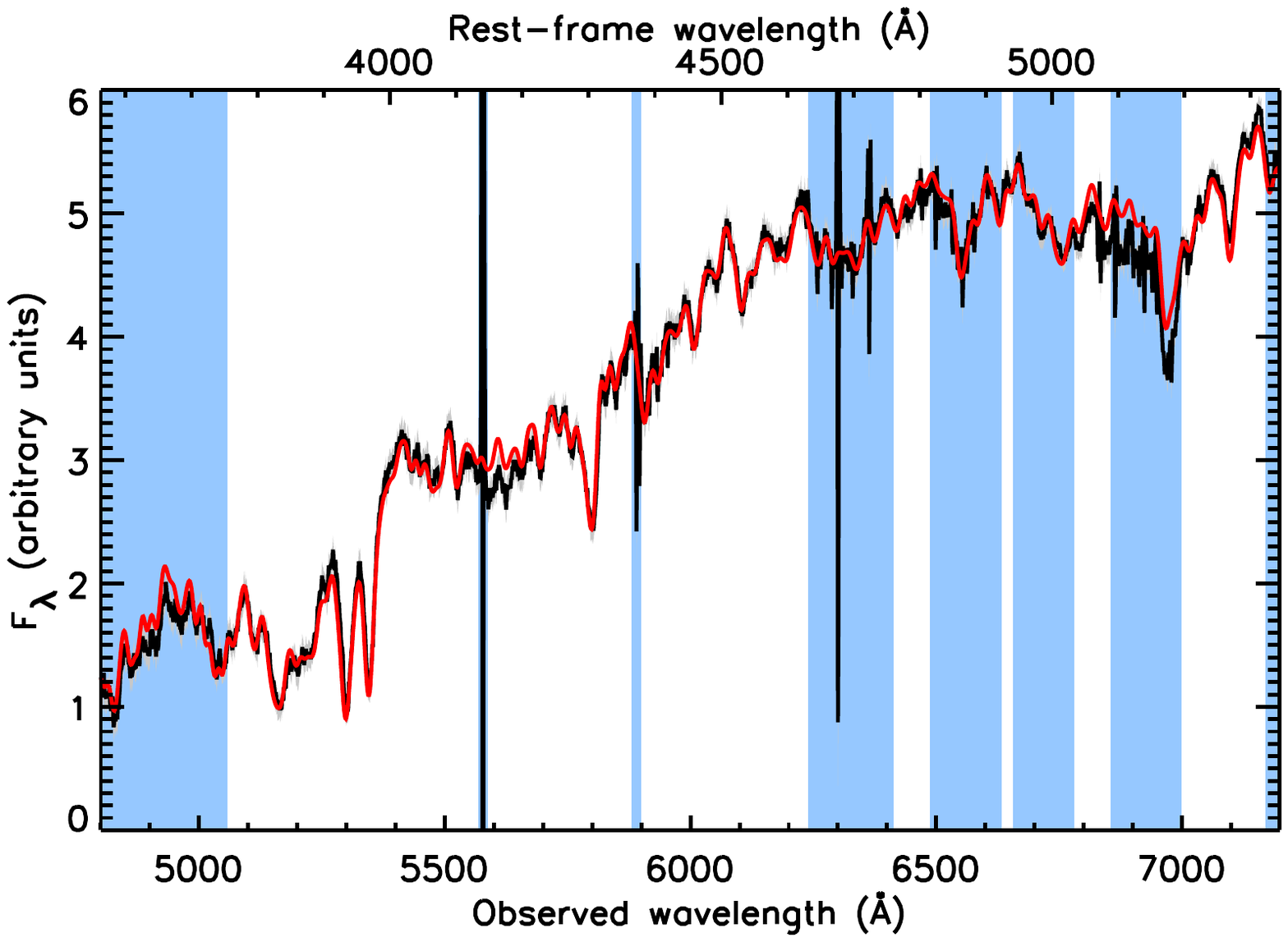}}
}
\caption{Example of the composite stellar population modeling of the 12 OPT/NIR HST bands ({\it left panel}, open blue symbols) plus MUSE spectra ({\it right panel}, black line)  of the BCG. The best-fit is shown in red in both panels. In the {\it left panel} blue empty circles and bars are the observed fluxes with 1$\sigma$ errors and filled yellow circles are model-predicted HST fluxes. In the {\it right panel}, blue areas indicate the masked regions.}
\label{figC1}
\end{figure*}

The available HST photometry from the CLASH survey plus the MUSE spectra were used to determine the stellar mass values of a subsample of 81 spectroscopic members in the centre of the cluster. MUSE spectra plus HST photometry were used to fit these galaxies' spectral energy distributions (SEDs). The SED fitting was performed using composite stellar population models, based on \cite{br03} templates, with delayed exponential star formation histories, solar metallicity, and a Salpeter \citep{sal55} stellar initial mass function (IMF). The presence of dust was taken into account following \cite{calzetti2000}.
For each galaxy, the best-fit (M$_{\mathrm{best}}$) and 1$\sigma$ lower (M$_{\mathrm{low}}$) and upper limit (M$_{\mathrm{high}}$) values of the stellar mass were measured. An example of a SED is shown in Fig.~\ref{figC1}. 

Then we used the stellar masses obtained thanks to the HST+MUSE data to cross-calibrate each member galaxy's stellar mass, which has only the four optical WFI photometric magnitudes available for the majority of members. We obtained that the Salpeter stellar masses follow the relations: 

\begin{equation}
\log M^*=20.12-0.463\times I
\end{equation}
or, if the I$_{\mathrm{c}}$-band is not available,  
\begin{equation}
\log M^*=20.83-0.484\times R \ \ , 
\end{equation}

\noindent with an average scatter of 0.155 and  0.175, respectively.

%
%

\bibliography{biblio}

\begin{thebibliography}{145}
\expandafter\ifx\csname natexlab\endcsname\relax\def\natexlab#1{#1}\fi

\bibitem[{{Abell} {et~al.}(1989){Abell}, {Corwin}, \& {Olowin}}]{abe89}
{Abell}, G.~O., {Corwin}, Harold~G., J., \& {Olowin}, R.~P. 1989, \apjs, 70, 1

\bibitem[{{Aguerri} {et~al.}(2017){Aguerri}, {Agulli}, {Diaferio}, \& {Dalla
  Vecchia}}]{AADDV17}
{Aguerri}, J.~A.~L., {Agulli}, I., {Diaferio}, A., \& {Dalla Vecchia}, C. 2017,
  \mnras, 468, 364

\bibitem[{{Anderson} \& {Ogaz}(2014)}]{and14}
{Anderson}, J. \& {Ogaz}, S. 2014, {hst2galign: an Automated Galaxy-based
  Alignment Routine}, Instrument Science Report ACS 2014-03; Instrument Science
  Report WFC/UVIS 2014-23

\bibitem[{{Annunziatella} {et~al.}(2014){Annunziatella}, {Biviano}, {Mercurio},
  {Nonino}, {Rosati}, {Balestra}, {Presotto}, {Girardi}, {Gobat}, {Grillo},
  {Kelson}, {Medezinski}, {Postman}, {Scodeggio}, {Brescia}, {Demarco},
  {Fritz}, {Koekemoer}, {Lemze}, {Lombardi}, {Sartoris}, {Umetsu}, {Vanzella},
  {Bradley}, {Coe}, {Donahue}, {Infante}, {Kuchner}, {Maier}, {Reg{\H{o}}s},
  {Verdugo}, \& {Ziegler}}]{ann14}
{Annunziatella}, M., {Biviano}, A., {Mercurio}, A., {et~al.} 2014, \aap, 571,
  A80

\bibitem[{{Annunziatella} {et~al.}(2016){Annunziatella}, {Mercurio}, {Biviano},
  {Girardi}, {Nonino}, {Balestra}, {Rosati}, {Bartosch Caminha}, {Brescia},
  {Gobat}, {Grillo}, {Lombardi}, {Sartoris}, {De Lucia}, {Demarco}, {Frye},
  {Fritz}, {Moustakas}, {Scodeggio}, {Kuchner}, {Maier}, \& {Ziegler}}]{ann16}
{Annunziatella}, M., {Mercurio}, A., {Biviano}, A., {et~al.} 2016, \aap, 585,
  A160

\bibitem[{{Annunziatella} {et~al.}(2013){Annunziatella}, {Mercurio}, {Brescia},
  {Cavuoti}, \& {Longo}}]{ann13}
{Annunziatella}, M., {Mercurio}, A., {Brescia}, M., {Cavuoti}, S., \& {Longo},
  G. 2013, \pasp, 125, 68

\bibitem[{{Armstrong} {et~al.}(2010){Armstrong}, {Mohr}, {Adams}, {Beldica},
  {Cai}, {Darnell}, {Daues}, {Desai}, {Gower}, {Mossessian}, {Ngeow}, {Lin},
  {Neilson}, {Tucker}, {Bertin}, \& {BCS Collaboration}}]{armstrong2010}
{Armstrong}, B., {Mohr}, J., {Adams}, D., {et~al.} 2010, in American
  Astronomical Society Meeting Abstracts, Vol. 215, American Astronomical
  Society Meeting Abstracts \#215, 438.07

\bibitem[{{Ashman} {et~al.}(1994){Ashman}, {Bird}, \& {Zepf}}]{ashman1994}
{Ashman}, K.~M., {Bird}, C.~M., \& {Zepf}, S.~E. 1994, \aj, 108, 2348

\bibitem[{{Bakels} {et~al.}(2020){Bakels}, {Ludlow}, \& {Power}}]{bakels2020}
{Bakels}, L., {Ludlow}, A.~D., \& {Power}, C. 2020, arXiv e-prints,
  arXiv:2008.05475

\bibitem[{{Balestra} {et~al.}(2016){Balestra}, {Mercurio}, {Sartoris},
  {Girardi}, {Grillo}, {Nonino}, {Rosati}, {Biviano}, {Ettori}, {Forman},
  {Jones}, {Koekemoer}, {Medezinski}, {Merten}, {Ogrean}, {Tozzi}, {Umetsu},
  {Vanzella}, {van Weeren}, {Zitrin}, {Annunziatella}, {Caminha}, {Broadhurst},
  {Coe}, {Donahue}, {Fritz}, {Frye}, {Kelson}, {Lombardi}, {Maier},
  {Meneghetti}, {Monna}, {Postman}, {Scodeggio}, {Seitz}, \& {Ziegler}}]{bal16}
{Balestra}, I., {Mercurio}, A., {Sartoris}, B., {et~al.} 2016, \apjs, 224, 33

\bibitem[{{Balogh} {et~al.}(2004){Balogh}, {Baldry}, {Nichol}, {Miller},
  {Bower}, \& {Glazebrook}}]{bal04}
{Balogh}, M.~L., {Baldry}, I.~K., {Nichol}, R., {et~al.} 2004, \apjl, 615, L101

\bibitem[{{Balogh} {et~al.}(1999){Balogh}, {Morris}, {Yee}, {Carlberg}, \&
  {Ellingson}}]{bal99}
{Balogh}, M.~L., {Morris}, S.~L., {Yee}, H.~K.~C., {Carlberg}, R.~G., \&
  {Ellingson}, E. 1999, \apj, 527, 54

\bibitem[{{Barger} {et~al.}(1996){Barger}, {Aragon-Salamanca}, {Ellis},
  {Couch}, {Smail}, \& {Sharples}}]{bar96}
{Barger}, A.~J., {Aragon-Salamanca}, A., {Ellis}, R.~S., {et~al.} 1996, \mnras,
  279, 1

\bibitem[{{Beers} {et~al.}(1990){Beers}, {Flynn}, \& {Gebhardt}}]{bee90}
{Beers}, T.~C., {Flynn}, K., \& {Gebhardt}, K. 1990, \aj, 100, 32

\bibitem[{{Behroozi} {et~al.}(2013){Behroozi}, {Wechsler}, {Wu}, {Busha},
  {Klypin}, \& {Primack}}]{behroozi2013}
{Behroozi}, P.~S., {Wechsler}, R.~H., {Wu}, H.-Y., {et~al.} 2013, \apj, 763, 18

\bibitem[{{Bertin}(2011)}]{bertin11}
{Bertin}, E. 2011, Astronomical Society of the Pacific Conference Series, Vol.
  442, {Automated Morphometry with SExtractor and PSFEx}, ed. I.~N. {Evans},
  A.~{Accomazzi}, D.~J. {Mink}, \& A.~H. {Rots}, 435

\bibitem[{{Bertin} \& {Arnouts}(1996)}]{bertin96}
{Bertin}, E. \& {Arnouts}, S. 1996, \aaps, 117, 393

\bibitem[{{Biviano} \& {Katgert}(2004)}]{biv04}
{Biviano}, A. \& {Katgert}, P. 2004, \aap, 424, 779

\bibitem[{{Biviano} {et~al.}(2013){Biviano}, {Rosati}, {Balestra}, {Mercurio},
  {Girardi}, {Nonino}, {Grillo}, {Scodeggio}, {Lemze}, {Kelson}, {Umetsu},
  {Postman}, {Zitrin}, {Czoske}, {Ettori}, {Fritz}, {Lombardi}, {Maier},
  {Medezinski}, {Mei}, {Presotto}, {Strazzullo}, {Tozzi}, {Ziegler},
  {Annunziatella}, {Bartelmann}, {Benitez}, {Bradley}, {Brescia}, {Broadhurst},
  {Coe}, {Demarco}, {Donahue}, {Ford}, {Gobat}, {Graves}, {Koekemoer},
  {Kuchner}, {Melchior}, {Meneghetti}, {Merten}, {Moustakas}, {Munari},
  {Reg{\H{o}}s}, {Sartoris}, {Seitz}, \& {Zheng}}]{biv13}
{Biviano}, A., {Rosati}, P., {Balestra}, I., {et~al.} 2013, \aap, 558, A1

\bibitem[{{Biviano} {et~al.}(2016){Biviano}, {van der Burg}, {Muzzin},
  {Sartoris}, {Wilson}, \& {Yee}}]{biv16}
{Biviano}, A., {van der Burg}, R.~F.~J., {Muzzin}, A., {et~al.} 2016, \aap,
  594, A51

\bibitem[{{Blanton} {et~al.}(2005){Blanton}, {Eisenstein}, {Hogg}, {Schlegel},
  \& {Brinkmann}}]{bla05}
{Blanton}, M.~R., {Eisenstein}, D., {Hogg}, D.~W., {Schlegel}, D.~J., \&
  {Brinkmann}, J. 2005, \apj, 629, 143

\bibitem[{{Bonamigo} {et~al.}(2018){Bonamigo}, {Grillo}, {Ettori}, {Caminha},
  {Rosati}, {Mercurio}, {Munari}, {Annunziatella}, {Balestra}, \&
  {Lombardi}}]{bon18}
{Bonamigo}, M., {Grillo}, C., {Ettori}, S., {et~al.} 2018, \apj, 864, 98

\bibitem[{{Boselli} \& {Gavazzi}(2006)}]{bos06}
{Boselli}, A. \& {Gavazzi}, G. 2006, \pasp, 118, 517

\bibitem[{{Boselli} \& {Gavazzi}(2014)}]{bos14}
{Boselli}, A. \& {Gavazzi}, G. 2014, \aapr, 22, 74

\bibitem[{{Boselli} {et~al.}(2016){Boselli}, {Roehlly}, {Fossati}, {Buat},
  {Boissier}, {Boquien}, {Burgarella}, {Ciesla}, {Gavazzi}, \& {Serra}}]{bos16}
{Boselli}, A., {Roehlly}, Y., {Fossati}, M., {et~al.} 2016, \aap, 596, A11

\bibitem[{{Bouy} {et~al.}(2013){Bouy}, {Bertin}, {Moraux}, {Cuillandre},
  {Bouvier}, {Barrado}, {Solano}, \& {Bayo}}]{bou13}
{Bouy}, H., {Bertin}, E., {Moraux}, E., {et~al.} 2013, \aap, 554, A101

\bibitem[{{Brunetti}(2009)}]{bru09}
{Brunetti}, G. 2009, in Revista Mexicana de Astronomia y Astrofisica Conference
  Series, Vol.~36, Revista Mexicana de Astronomia y Astrofisica Conference
  Series, 201--208

\bibitem[{{Bruzual} \& {Charlot}(2003)}]{br03}
{Bruzual}, G. \& {Charlot}, S. 2003, \mnras, 344, 1000

\bibitem[{{Butcher} \& {Oemler}(1978)}]{but78}
{Butcher}, H. \& {Oemler}, A., J. 1978, \apj, 226, 559

\bibitem[{{Butcher} \& {Oemler}(1984)}]{but84}
{Butcher}, H. \& {Oemler}, A., J. 1984, \apj, 285, 426

\bibitem[{{Caldwell} {et~al.}(2008){Caldwell}, {McIntosh}, {Rix}, {Barden},
  {Beckwith}, {Bell}, {Borch}, {Heymans}, {H{\"a}u{\ss}ler}, {Jahnke}, {Jogee},
  {Meisenheimer}, {Peng}, {S{\'a}nchez}, {Somerville}, {Wisotzki}, \&
  {Wolf}}]{cal08}
{Caldwell}, J. A.~R., {McIntosh}, D.~H., {Rix}, H.-W., {et~al.} 2008, \apjs,
  174, 136

\bibitem[{{Calzetti} {et~al.}(2000){Calzetti}, {Tremonti}, {Heckman}, \&
  {Leitherer}}]{calzetti2000}
{Calzetti}, D., {Tremonti}, C.~A., {Heckman}, T.~M., \& {Leitherer}, C. 2000,
  Astronomical Society of the Pacific Conference Series, Vol. 211, {The
  Evolution of the Interstellar Medium Around Young Stellar Clusters.}, ed.
  A.~{Lan{\c{c}}on} \& C.~M. {Boily}, 25

\bibitem[{{Caminha} {et~al.}(2016){Caminha}, {Grillo}, {Rosati}, {Balestra},
  {Karman}, {Lombardi}, {Mercurio}, {Nonino}, {Tozzi}, {Zitrin}, {Biviano},
  {Girardi}, {Koekemoer}, {Melchior}, {Meneghetti}, {Munari}, {Suyu}, {Umetsu},
  {Annunziatella}, {Borgani}, {Broadhurst}, {Caputi}, {Coe}, {Delgado-Correal},
  {Ettori}, {Fritz}, {Frye}, {Gobat}, {Maier}, {Monna}, {Postman}, {Sartoris},
  {Seitz}, {Vanzella}, \& {Ziegler}}]{cam16}
{Caminha}, G.~B., {Grillo}, C., {Rosati}, P., {et~al.} 2016, \aap, 587, A80

\bibitem[{{Caminha} {et~al.}(2017){Caminha}, {Grillo}, {Rosati}, {Balestra},
  {Mercurio}, {Vanzella}, {Biviano}, {Caputi}, {Delgado-Correal}, {Karman},
  {Lombardi}, {Meneghetti}, {Sartoris}, \& {Tozzi}}]{cam17}
{Caminha}, G.~B., {Grillo}, C., {Rosati}, P., {et~al.} 2017, \aap, 600, A90

\bibitem[{{Ciocan} {et~al.}(2020){Ciocan}, {Maier}, {Ziegler}, \&
  {Verdugo}}]{ciocan2020}
{Ciocan}, B.~I., {Maier}, C., {Ziegler}, B.~L., \& {Verdugo}, M. 2020, \aap,
  633, A139

\bibitem[{{Couch} \& {Sharples}(1987)}]{cou87}
{Couch}, W.~J. \& {Sharples}, R.~M. 1987, \mnras, 229, 423

\bibitem[{{Danese} {et~al.}(1980){Danese}, {de Zotti}, \& {di Tullio}}]{dan80}
{Danese}, L., {de Zotti}, G., \& {di Tullio}, G. 1980, \aap, 82, 322

\bibitem[{{De Grandi} {et~al.}(1999){De Grandi}, {B{\"o}hringer}, {Guzzo},
  {Molendi}, {Chincarini}, {Collins}, {Cruddace}, {Neumann}, {Schindler},
  {Schuecker}, \& {Voges}}]{deg99}
{De Grandi}, S., {B{\"o}hringer}, H., {Guzzo}, L., {et~al.} 1999, \apj, 514,
  148

\bibitem[{{De Lucia} \& {Blaizot}(2007)}]{del07}
{De Lucia}, G. \& {Blaizot}, J. 2007, \mnras, 375, 2

\bibitem[{{De Lucia} {et~al.}(2004){De Lucia}, {Kauffmann}, {Springel},
  {White}, {Lanzoni}, {Stoehr}, {Tormen}, \& {Yoshida}}]{del04}
{De Lucia}, G., {Kauffmann}, G., {Springel}, V., {et~al.} 2004, \mnras, 348,
  333

\bibitem[{{De Lucia} {et~al.}(2012){De Lucia}, {Weinmann}, {Poggianti},
  {Arag{\'o}n-Salamanca}, \& {Zaritsky}}]{del2012}
{De Lucia}, G., {Weinmann}, S., {Poggianti}, B.~M., {Arag{\'o}n-Salamanca}, A.,
  \& {Zaritsky}, D. 2012, \mnras, 423, 1277

\bibitem[{{Diaferio}(1999)}]{diaferio99}
{Diaferio}, A. 1999, \mnras, 309, 610

\bibitem[{{Diaferio} \& {Geller}(1997)}]{diaferio97}
{Diaferio}, A. \& {Geller}, M.~J. 1997, \apj, 481, 633

\bibitem[{{Dressler}(1980)}]{dre80}
{Dressler}, A. 1980, \apj, 236, 351

\bibitem[{{Dressler} {et~al.}(1997){Dressler}, {Oemler}, {Couch}, {Smail},
  {Ellis}, {Barger}, {Butcher}, {Poggianti}, \& {Sharples}}]{dre97}
{Dressler}, A., {Oemler}, Augustus, J., {Couch}, W.~J., {et~al.} 1997, \apj,
  490, 577

\bibitem[{{Dressler} {et~al.}(2013){Dressler}, {Oemler}, {Poggianti},
  {Gladders}, {Abramson}, \& {Vulcani}}]{dre13}
{Dressler}, A., {Oemler}, Augustus, J., {Poggianti}, B.~M., {et~al.} 2013,
  \apj, 770, 62

\bibitem[{{Dressler} \& {Shectman}(1988)}]{dressler1988}
{Dressler}, A. \& {Shectman}, S.~A. 1988, \aj, 95, 985

\bibitem[{{Dressler} {et~al.}(1999){Dressler}, {Smail}, {Poggianti}, {Butcher},
  {Couch}, {Ellis}, \& {Oemler}}]{dre99}
{Dressler}, A., {Smail}, I., {Poggianti}, B.~M., {et~al.} 1999, \apjs, 122, 51

\bibitem[{{D{\"u}nner} {et~al.}(2007){D{\"u}nner}, {Reisenegger}, {Meza},
  {Araya}, \& {Quintana}}]{dunner07}
{D{\"u}nner}, R., {Reisenegger}, A., {Meza}, A., {Araya}, P.~A., \& {Quintana},
  H. 2007, \mnras, 376, 1577

\bibitem[{{Fadda} {et~al.}(1996){Fadda}, {Girardi}, {Giuricin}, {Mardirossian},
  \& {Mezzetti}}]{fadda1996}
{Fadda}, D., {Girardi}, M., {Giuricin}, G., {Mardirossian}, F., \& {Mezzetti},
  M. 1996, \apj, 473, 670

\bibitem[{{Fasano} \& {Franceschini}(1987)}]{FF87}
{Fasano}, G. \& {Franceschini}, A. 1987, \mnras, 225, 155

\bibitem[{{Foltz} {et~al.}(2018){Foltz}, {Wilson}, {Muzzin}, {Cooper},
  {Nantais}, {van der Burg}, {Cerulo}, {Chan}, {Fillingham}, {Surace}, {Webb},
  {Noble}, {Lacy}, {McDonald}, {Rudnick}, {Lidman}, {Demarco},
  {Hlavacek-Larrondo}, {Yee}, {Perlmutter}, \& {Hayden}}]{Foltz18}
{Foltz}, R., {Wilson}, G., {Muzzin}, A., {et~al.} 2018, \apj, 866, 136

\bibitem[{{Fritz} {et~al.}(2009){Fritz}, {B{\"o}hm}, \& {Ziegler}}]{Fritz2009}
{Fritz}, A., {B{\"o}hm}, A., \& {Ziegler}, B.~L. 2009, \mnras, 393, 1467

\bibitem[{{Fritz} {et~al.}(2014){Fritz}, {Scodeggio}, {Ilbert}, {Bolzonella},
  {Davidzon}, {Coupon}, {Garilli}, {Guzzo}, {Zamorani}, {Abbas}, {Adami},
  {Arnouts}, {Bel}, {Bottini}, {Branchini}, {Cappi}, {Cucciati}, {De Lucia},
  {de la Torre}, {Franzetti}, {Fumana}, {Granett}, {Iovino}, {Krywult}, {Le
  Brun}, {Le F{\`e}vre}, {Maccagni}, {Ma{\l}ek}, {Marulli}, {McCracken},
  {Paioro}, {Polletta}, {Pollo}, {Schlagenhaufer}, {Tasca}, {Tojeiro},
  {Vergani}, {Zanichelli}, {Burden}, {Di Porto}, {Marchetti}, {Marinoni},
  {Mellier}, {Moscardini}, {Nichol}, {Peacock}, {Percival}, {Phleps}, \&
  {Wolk}}]{Fritz2014}
{Fritz}, A., {Scodeggio}, M., {Ilbert}, O., {et~al.} 2014, \aap, 563, A92

\bibitem[{{Garilli} {et~al.}(2010){Garilli}, {Fumana}, {Franzetti}, {Paioro},
  {Scodeggio}, {Le F{\`e}vre}, {Paltani}, \& {Scaramella}}]{gar10}
{Garilli}, B., {Fumana}, M., {Franzetti}, P., {et~al.} 2010, \pasp, 122, 827

\bibitem[{{Garilli} {et~al.}(1999){Garilli}, {Maccagni}, \& {Andreon}}]{gar99}
{Garilli}, B., {Maccagni}, D., \& {Andreon}, S. 1999, \aap, 342, 408

\bibitem[{{Gebhardt} \& {Beers}(1991)}]{geb91}
{Gebhardt}, K. \& {Beers}, T.~C. 1991, \apj, 383, 72

\bibitem[{{Gehrels}(1986)}]{geh86}
{Gehrels}, N. 1986, \apj, 303, 336

\bibitem[{{Girardi} {et~al.}(2008){Girardi}, {Barrena}, {Boschin}, \&
  {Ellingson}}]{girardi2008}
{Girardi}, M., {Barrena}, R., {Boschin}, W., \& {Ellingson}, E. 2008, \aap,
  491, 379

\bibitem[{{Girardi} {et~al.}(2010){Girardi}, {Boschin}, \&
  {Barrena}}]{girardi2010}
{Girardi}, M., {Boschin}, W., \& {Barrena}, R. 2010, \aap, 517, A65

\bibitem[{{Girardi} {et~al.}(1997){Girardi}, {Escalera}, {Fadda}, {Giuricin},
  {Mardirossian}, \& {Mezzetti}}]{girardi1997a}
{Girardi}, M., {Escalera}, E., {Fadda}, D., {et~al.} 1997, \apj, 482, 41

\bibitem[{{Girardi} {et~al.}(1996){Girardi}, {Fadda}, {Giuricin},
  {Mardirossian}, {Mezzetti}, \& {Biviano}}]{girardi1996}
{Girardi}, M., {Fadda}, D., {Giuricin}, G., {et~al.} 1996, Astronomical Society
  of the Pacific Conference Series, Vol.~94, {The Galaxy Velocity Dispersion -
  X-ray Temperature Helation in Galaxy Clusters}, ed. P.~{Coles},
  V.~{Martinez}, \& M.-J. {Pons-Borderia}, 221

\bibitem[{{Girardi} {et~al.}(2015){Girardi}, {Mercurio}, {Balestra}, {Nonino},
  {Biviano}, {Grillo}, {Rosati}, {Annunziatella}, {Demarco}, {Fritz}, {Gobat},
  {Lemze}, {Presotto}, {Scodeggio}, {Tozzi}, {Bartosch Caminha}, {Brescia},
  {Coe}, {Kelson}, {Koekemoer}, {Lombardi}, {Medezinski}, {Postman},
  {Sartoris}, {Umetsu}, {Zitrin}, {Boschin}, {Czoske}, {De Lucia}, {Kuchner},
  {Maier}, {Meneghetti}, {Monaco}, {Monna}, {Munari}, {Seitz}, {Verdugo}, \&
  {Ziegler}}]{girardi2015}
{Girardi}, M., {Mercurio}, A., {Balestra}, I., {et~al.} 2015, \aap, 579, A4

\bibitem[{{G{\'o}mez} {et~al.}(2003){G{\'o}mez}, {Nichol}, {Miller}, {Balogh},
  {Goto}, {Zabludoff}, {Romer}, {Bernardi}, {Sheth}, {Hopkins}, {Castander},
  {Connolly}, {Schneider}, {Brinkmann}, {Lamb}, {SubbaRao}, \& {York}}]{gom03}
{G{\'o}mez}, P.~L., {Nichol}, R.~C., {Miller}, C.~J., {et~al.} 2003, \apj, 584,
  210

\bibitem[{{G{\'o}mez} {et~al.}(2012){G{\'o}mez}, {Valkonen}, {Romer},
  {Lloyd-Davies}, {Verdugo}, {Cantalupo}, {Daub}, {Goldstein}, {Kuo}, {Lange},
  {Lueker}, {Holzapfel}, {Peterson}, {Ruhl}, {Runyan}, {Reichardt}, \&
  {Sabirli}}]{gom12}
{G{\'o}mez}, P.~L., {Valkonen}, L.~E., {Romer}, A.~K., {et~al.} 2012, \aj, 144,
  79

\bibitem[{{Gruen} {et~al.}(2013){Gruen}, {Brimioulle}, {Seitz}, {Lee}, {Young},
  {Koppenhoefer}, {Eichner}, {Riffeser}, {Vikram}, {Weidinger}, \&
  {Zenteno}}]{gru13}
{Gruen}, D., {Brimioulle}, F., {Seitz}, S., {et~al.} 2013, \mnras, 432, 1455

\bibitem[{{Guzzo} {et~al.}(1999){Guzzo}, {B{\"o}hringer}, {Schuecker},
  {Collins}, {Schindler}, {Neumann}, {de Grand i}, {Cruddace}, {Chincarini},
  {Edge}, {Shaver}, \& {Voges}}]{guz99}
{Guzzo}, L., {B{\"o}hringer}, H., {Schuecker}, P., {et~al.} 1999, The
  Messenger, 95, 27

\bibitem[{{Haines} {et~al.}(2007){Haines}, {Gargiulo}, {La Barbera},
  {Mercurio}, {Merluzzi}, \& {Busarello}}]{hai07}
{Haines}, C.~P., {Gargiulo}, A., {La Barbera}, F., {et~al.} 2007, \mnras, 381,
  7

\bibitem[{{Haines} {et~al.}(2006){Haines}, {La Barbera}, {Mercurio},
  {Merluzzi}, \& {Busarello}}]{hai06}
{Haines}, C.~P., {La Barbera}, F., {Mercurio}, A., {Merluzzi}, P., \&
  {Busarello}, G. 2006, \apjl, 647, L21

\bibitem[{Haines {et~al.}(2012)Haines, Pereira, Sanderson, Smith, Egami, Babul,
  Edge, Finoguenov, Moran, \& Okabe}]{haines2012}
Haines, C.~P., Pereira, M.~J., Sanderson, A. J.~R., {et~al.} 2012, The
  Astrophysical Journal, 754, 97

\bibitem[{{Haines} {et~al.}(2015){Haines}, {Pereira}, {Smith}, {Egami},
  {Babul}, {Finoguenov}, {Ziparo}, {McGee}, {Rawle}, {Okabe}, \&
  {Moran}}]{haines2015}
{Haines}, C.~P., {Pereira}, M.~J., {Smith}, G.~P., {et~al.} 2015, \apj, 806,
  101

\bibitem[{{Iannuzzi} \& {Dolag}(2012)}]{ian2012}
{Iannuzzi}, F. \& {Dolag}, K. 2012, \mnras, 427, 1024

\bibitem[{{Jaff{\'e}} {et~al.}(2015){Jaff{\'e}}, {Smith}, {Candlish},
  {Poggianti}, {Sheen}, \& {Verheijen}}]{Jaffe15}
{Jaff{\'e}}, Y.~L., {Smith}, R., {Candlish}, G.~N., {et~al.} 2015, \mnras, 448,
  1715

\bibitem[{{Joshi} {et~al.}(2020){Joshi}, {Pillepich}, {Nelson}, {Marinacci},
  {Springel}, {Rodriguez-Gomez}, {Vogelsberger}, \& {Hernquist}}]{jos20}
{Joshi}, G.~D., {Pillepich}, A., {Nelson}, D., {et~al.} 2020, \mnras, 496, 2673

\bibitem[{{Just} {et~al.}(2015){Just}, {Fuchs}, {Jahrei{\ss}}, {Flynn},
  {Dettbarn}, \& {Rybizki}}]{jus15}
{Just}, A., {Fuchs}, B., {Jahrei{\ss}}, H., {et~al.} 2015, \mnras, 451, 149

\bibitem[{{Karman} {et~al.}(2017){Karman}, {Caputi}, {Caminha}, {Gronke},
  {Grillo}, {Balestra}, {Rosati}, {Vanzella}, {Coe}, {Dijkstra}, {Koekemoer},
  {McLeod}, {Mercurio}, \& {Nonino}}]{kar17}
{Karman}, W., {Caputi}, K.~I., {Caminha}, G.~B., {et~al.} 2017, \aap, 599, A28

\bibitem[{{Karman} {et~al.}(2015){Karman}, {Caputi}, {Grillo}, {Balestra},
  {Rosati}, {Vanzella}, {Coe}, {Christensen}, {Koekemoer}, {Kr{\"u}hler},
  {Lombardi}, {Mercurio}, {Nonino}, \& {van der Wel}}]{kar15}
{Karman}, W., {Caputi}, K.~I., {Grillo}, C., {et~al.} 2015, \aap, 574, A11

\bibitem[{{Kauffmann} {et~al.}(2004){Kauffmann}, {White}, {Heckman},
  {M{\'e}nard}, {Brinchmann}, {Charlot}, {Tremonti}, \& {Brinkmann}}]{kau04}
{Kauffmann}, G., {White}, S. D.~M., {Heckman}, T.~M., {et~al.} 2004, \mnras,
  353, 713

\bibitem[{{Koekemoer} {et~al.}(2003){Koekemoer}, {Fruchter}, \& {Hack}}]{koe03}
{Koekemoer}, A., {Fruchter}, A., \& {Hack}, W. 2003, Space Telescope European
  Coordinating Facility Newsletter, 33, 10

\bibitem[{{Koekemoer} {et~al.}(2011){Koekemoer}, {Faber}, {Ferguson}, {Grogin},
  {Kocevski}, {Koo}, {Lai}, {Lotz}, {Lucas}, {McGrath}, {Ogaz}, {Rajan},
  {Riess}, {Rodney}, {Strolger}, {Casertano}, {Castellano}, {Dahlen},
  {Dickinson}, {Dolch}, {Fontana}, {Giavalisco}, {Grazian}, {Guo}, {Hathi},
  {Huang}, {van der Wel}, {Yan}, {Acquaviva}, {Alexander}, {Almaini}, {Ashby},
  {Barden}, {Bell}, {Bournaud}, {Brown}, {Caputi}, {Cassata}, {Challis},
  {Chary}, {Cheung}, {Cirasuolo}, {Conselice}, {Roshan Cooray}, {Croton},
  {Daddi}, {Dav{\'e}}, {de Mello}, {de Ravel}, {Dekel}, {Donley}, {Dunlop},
  {Dutton}, {Elbaz}, {Fazio}, {Filippenko}, {Finkelstein}, {Frazer}, {Gardner},
  {Garnavich}, {Gawiser}, {Gruetzbauch}, {Hartley}, {H{\"a}ussler},
  {Herrington}, {Hopkins}, {Huang}, {Jha}, {Johnson}, {Kartaltepe},
  {Khostovan}, {Kirshner}, {Lani}, {Lee}, {Li}, {Madau}, {McCarthy},
  {McIntosh}, {McLure}, {McPartland}, {Mobasher}, {Moreira}, {Mortlock},
  {Moustakas}, {Mozena}, {Nandra}, {Newman}, {Nielsen}, {Niemi}, {Noeske},
  {Papovich}, {Pentericci}, {Pope}, {Primack}, {Ravindranath}, {Reddy},
  {Renzini}, {Rix}, {Robaina}, {Rosario}, {Rosati}, {Salimbeni}, {Scarlata},
  {Siana}, {Simard}, {Smidt}, {Snyder}, {Somerville}, {Spinrad}, {Straughn},
  {Telford}, {Teplitz}, {Trump}, {Vargas}, {Villforth}, {Wagner}, {Wand ro},
  {Wechsler}, {Weiner}, {Wiklind}, {Wild}, {Wilson}, {Wuyts}, \& {Yun}}]{koe11}
{Koekemoer}, A.~M., {Faber}, S.~M., {Ferguson}, H.~C., {et~al.} 2011, \apjs,
  197, 36

\bibitem[{{Kova{\v{c}}} {et~al.}(2014){Kova{\v{c}}}, {Lilly}, {Knobel},
  {Bschorr}, {Peng}, {Carollo}, {Contini}, {Kneib}, {Le F{\'e}vre}, {Mainieri},
  {Renzini}, {Scodeggio}, {Zamorani}, {Bardelli}, {Bolzonella}, {Bongiorno},
  {Caputi}, {Cucciati}, {de la Torre}, {de Ravel}, {Franzetti}, {Garilli},
  {Iovino}, {Kampczyk}, {Lamareille}, {Le Borgne}, {Le Brun}, {Maier},
  {Mignoli}, {Oesch}, {Pello}, {Montero}, {Presotto}, {Silverman}, {Tanaka},
  {Tasca}, {Tresse}, {Vergani}, {Zucca}, {Aussel}, {Koekemoer}, {Le Floc'h},
  {Moresco}, \& {Pozzetti}}]{kov14}
{Kova{\v{c}}}, K., {Lilly}, S.~J., {Knobel}, C., {et~al.} 2014, \mnras, 438,
  717

\bibitem[{{Kron}(1980)}]{kron80}
{Kron}, R.~G. 1980, \apjs, 43, 305

\bibitem[{{Lauer} {et~al.}(2014){Lauer}, {Postman}, {Strauss}, {Graves}, \&
  {Chisari}}]{lauer2014}
{Lauer}, T.~R., {Postman}, M., {Strauss}, M.~A., {Graves}, G.~J., \& {Chisari},
  N.~E. 2014, \apj, 797, 82

\bibitem[{{Lee} {et~al.}(2017){Lee}, {Kang}, {Lee}, \& {Jang}}]{lee17}
{Lee}, M.~G., {Kang}, J., {Lee}, J.~H., \& {Jang}, I.~S. 2017, \apj, 844, 157

\bibitem[{{Lemaux} {et~al.}(2019){Lemaux}, {Tomczak}, {Lubin}, {Gal}, {Shen},
  {Pelliccia}, {Wu}, {Hung}, {Mei}, {Le F{\`e}vre}, {Rumbaugh}, {Kocevski}, \&
  {Squires}}]{lem19}
{Lemaux}, B.~C., {Tomczak}, A.~R., {Lubin}, L.~M., {et~al.} 2019, \mnras, 490,
  1231

\bibitem[{{{\L}okas} \& {Mamon}(2001)}]{lokas01}
{{\L}okas}, E.~L. \& {Mamon}, G.~A. 2001, \mnras, 321, 155

\bibitem[{{Lotz} {et~al.}(2017){Lotz}, {Koekemoer}, {Coe}, {Grogin}, {Capak},
  {Mack}, {Anderson}, {Avila}, {Barker}, {Borncamp}, {Brammer}, {Durbin},
  {Gunning}, {Hilbert}, {Jenkner}, {Khandrika}, {Levay}, {Lucas}, {MacKenty},
  {Ogaz}, {Porterfield}, {Reid}, {Robberto}, {Royle}, {Smith},
  {Storrie-Lombardi}, {Sunnquist}, {Surace}, {Taylor}, {Williams}, {Bullock},
  {Dickinson}, {Finkelstein}, {Natarajan}, {Richard}, {Robertson}, {Tumlinson},
  {Zitrin}, {Flanagan}, {Sembach}, {Soifer}, \& {Mountain}}]{lot17}
{Lotz}, J.~M., {Koekemoer}, A., {Coe}, D., {et~al.} 2017, \apj, 837, 97

\bibitem[{{Lotz} {et~al.}(2019){Lotz}, {Remus}, {Dolag}, {Biviano}, \&
  {Burkert}}]{lot19}
{Lotz}, M., {Remus}, R.-S., {Dolag}, K., {Biviano}, A., \& {Burkert}, A. 2019,
  \mnras, 488, 5370

\bibitem[{{Mahajan} {et~al.}(2011){Mahajan}, {Mamon}, \&
  {Raychaudhury}}]{mahajan2011}
{Mahajan}, S., {Mamon}, G.~A., \& {Raychaudhury}, S. 2011, \mnras, 416, 2882

\bibitem[{{Mamon} {et~al.}(2013){Mamon}, {Biviano}, \& {Bou{\'e}}}]{mamon13}
{Mamon}, G.~A., {Biviano}, A., \& {Bou{\'e}}, G. 2013, \mnras, 429, 3079

\bibitem[{{Mamon} {et~al.}(2019){Mamon}, {Cava}, {Biviano}, {Moretti},
  {Poggianti}, \& {Bettoni}}]{mam19}
{Mamon}, G.~A., {Cava}, A., {Biviano}, A., {et~al.} 2019, \aap, 631, A131

\bibitem[{{Mamon} {et~al.}(2004){Mamon}, {Sanchis}, {Salvador-Sol{\'e}}, \&
  {Solanes}}]{mamon04}
{Mamon}, G.~A., {Sanchis}, T., {Salvador-Sol{\'e}}, E., \& {Solanes}, J.~M.
  2004, \aap, 414, 445

\bibitem[{{Mercurio} {et~al.}(2004){Mercurio}, {Busarello}, {Merluzzi}, {La
  Barbera}, {Girardi}, \& {Haines}}]{mercurio2004}
{Mercurio}, A., {Busarello}, G., {Merluzzi}, P., {et~al.} 2004, \aap, 424, 79

\bibitem[{{Mercurio} {et~al.}(2010){Mercurio}, {Haines}, {Gargiulo}, {La
  Barbera}, {Merluzzi}, \& {Busarello}}]{mer10}
{Mercurio}, A., {Haines}, C.~P., {Gargiulo}, A., {et~al.} 2010, arXiv e-prints,
  arXiv:1006.5001

\bibitem[{{Mercurio} {et~al.}(2015){Mercurio}, {Merluzzi}, {Busarello},
  {Grado}, {Limatola}, {Haines}, {Brescia}, {Cavuoti}, {Dopita}, {Dall'Ora},
  {Capaccioli}, {Napolitano}, \& {Pimbblet}}]{mer15}
{Mercurio}, A., {Merluzzi}, P., {Busarello}, G., {et~al.} 2015, \mnras, 453,
  3685

\bibitem[{{Mohr} {et~al.}(2012){Mohr}, {Armstrong}, {Bertin}, {Daues}, {Desai},
  {Gower}, {Gruendl}, {Hanlon}, {Kuropatkin}, {Lin}, {Marriner}, {Petravic},
  {Sevilla}, {Swanson}, {Tomashek}, {Tucker}, \& {Yanny}}]{mohr2012}
{Mohr}, J.~J., {Armstrong}, R., {Bertin}, E., {et~al.} 2012, Society of
  Photo-Optical Instrumentation Engineers (SPIE) Conference Series, Vol. 8451,
  {The Dark Energy Survey data processing and calibration system}, 84510D

\bibitem[{{Molino} {et~al.}(2017){Molino}, {Ben{\'\i}tez}, {Ascaso}, {Coe},
  {Postman}, {Jouvel}, {Host}, {Lahav}, {Seitz}, {Medezinski}, {Rosati},
  {Schoenell}, {Koekemoer}, {Jimenez-Teja}, {Broadhurst}, {Melchior},
  {Balestra}, {Bartelmann}, {Bouwens}, {Bradley}, {Czakon}, {Donahue}, {Ford},
  {Graur}, {Graves}, {Grillo}, {Infante}, {Jha}, {Kelson}, {Lazkoz}, {Lemze},
  {Maoz}, {Mercurio}, {Meneghetti}, {Merten}, {Moustakas}, {Nonino}, {Orgaz},
  {Riess}, {Rodney}, {Sayers}, {Umetsu}, {Zheng}, \& {Zitrin}}]{mol17}
{Molino}, A., {Ben{\'\i}tez}, N., {Ascaso}, B., {et~al.} 2017, \mnras, 470, 95

\bibitem[{{Moresco} {et~al.}(2010){Moresco}, {Pozzetti}, {Cimatti}, {Zamorani},
  {Mignoli}, {di Cesare}, {Bolzonella}, {Zucca}, {Lilly}, {Kova{\v{c}}},
  {Scodeggio}, {Cassata}, {Tasca}, {Vergani}, {Halliday}, {Carollo}, {Contini},
  {Kneib}, {Le F{\'e}vre}, {Mainieri}, {Renzini}, {Bardelli}, {Bongiorno},
  {Caputi}, {Coppa}, {Cucciati}, {de la Torre}, {de Ravel}, {Franzetti},
  {Garilli}, {Iovino}, {Kampczyk}, {Knobel}, {Lamareille}, {Le Borgne}, {Le
  Brun}, {Maier}, {Pell{\`o}}, {Peng}, {Perez Montero}, {Ricciardelli},
  {Silverman}, {Tanaka}, {Tresse}, {Abbas}, {Bottini}, {Cappi}, {Guzzo},
  {Koekemoer}, {Leauthaud}, {Maccagni}, {Marinoni}, {McCracken}, {Memeo},
  {Meneux}, {Nair}, {Oesch}, {Porciani}, {Scaramella}, {Scarlata}, \&
  {Scoville}}]{mor10}
{Moresco}, M., {Pozzetti}, L., {Cimatti}, A., {et~al.} 2010, \aap, 524, A67

\bibitem[{{Munari} {et~al.}(2014){Munari}, {Biviano}, \& {Mamon}}]{mun14}
{Munari}, E., {Biviano}, A., \& {Mamon}, G.~A. 2014, \aap, 566, A68

\bibitem[{{Muzzin} {et~al.}(2014){Muzzin}, {van der Burg}, {McGee}, {Balogh},
  {Franx}, {Hoekstra}, {Hudson}, {Noble}, {Taranu}, {Webb}, {Wilson}, \&
  {Yee}}]{muzzin2014}
{Muzzin}, A., {van der Burg}, R.~F.~J., {McGee}, S.~L., {et~al.} 2014, \apj,
  796, 65

\bibitem[{{Nandra} {et~al.}(2012){Nandra}, {Lasenby}, \& {Hobson}}]{nandra12}
{Nandra}, R., {Lasenby}, A.~N., \& {Hobson}, M.~P. 2012, \mnras, 422, 2931

\bibitem[{{Nantais} {et~al.}(2016){Nantais}, {van der Burg}, {Lidman},
  {Demarco}, {Noble}, {Wilson}, {Muzzin}, {Foltz}, {DeGroot}, \&
  {Cooper}}]{nan16}
{Nantais}, J.~B., {van der Burg}, R. F.~J., {Lidman}, C., {et~al.} 2016, \aap,
  592, A161

\bibitem[{{Navarro} {et~al.}(1997){Navarro}, {Frenk}, \& {White}}]{navarro1997}
{Navarro}, J.~F., {Frenk}, C.~S., \& {White}, S. D.~M. 1997, \apj, 490, 493

\bibitem[{{Noble} {et~al.}(2013){Noble}, {Webb}, {Muzzin}, {Wilson}, {Yee}, \&
  {van der Burg}}]{Noble2013}
{Noble}, A.~G., {Webb}, T.~M.~A., {Muzzin}, A., {et~al.} 2013, \apj, 768, 118

\bibitem[{{Oemler} {et~al.}(2017){Oemler}, {Abramson}, {Gladders}, {Dressler},
  {Poggianti}, \& {Vulcani}}]{oem17}
{Oemler}, Augustus, J., {Abramson}, L.~E., {Gladders}, M.~D., {et~al.} 2017,
  \apj, 844, 45

\bibitem[{{Oemler} {et~al.}(2009){Oemler}, {Dressler}, {Kelson}, {Rigby},
  {Poggianti}, {Fritz}, {Morrison}, \& {Smail}}]{oem09}
{Oemler}, Augustus, J., {Dressler}, A., {Kelson}, D., {et~al.} 2009, \apj, 693,
  152

\bibitem[{{Owers} {et~al.}(2019){Owers}, {Hudson}, {Oman}, {Bland -Hawthorn},
  {Brough}, {Bryant}, {Cortese}, {Couch}, {Croom}, {van de Sande}, {Federrath},
  {Groves}, {Hopkins}, {Lawrence}, {Lorente}, {McDermid}, {Medling},
  {Richards}, {Scott}, {Taranu}, {Welker}, \& {Yi}}]{owe19}
{Owers}, M.~S., {Hudson}, M.~J., {Oman}, K.~A., {et~al.} 2019, \apj, 873, 52

\bibitem[{{Pasquali} {et~al.}(2019){Pasquali}, {Smith}, {Gallazzi}, {De Lucia},
  {Zibetti}, {Hirschmann}, \& {Yi}}]{Pasquali19}
{Pasquali}, A., {Smith}, R., {Gallazzi}, A., {et~al.} 2019, \mnras, 484, 1702

\bibitem[{{Peng} {et~al.}(2010){Peng}, {Lilly}, {Kova{\v{c}}}, {Bolzonella},
  {Pozzetti}, {Renzini}, {Zamorani}, {Ilbert}, {Knobel}, {Iovino}, {Maier},
  {Cucciati}, {Tasca}, {Carollo}, {Silverman}, {Kampczyk}, {de Ravel},
  {Sanders}, {Scoville}, {Contini}, {Mainieri}, {Scodeggio}, {Kneib}, {Le
  F{\`e}vre}, {Bardelli}, {Bongiorno}, {Caputi}, {Coppa}, {de la Torre},
  {Franzetti}, {Garilli}, {Lamareille}, {Le Borgne}, {Le Brun}, {Mignoli},
  {Perez Montero}, {Pello}, {Ricciardelli}, {Tanaka}, {Tresse}, {Vergani},
  {Welikala}, {Zucca}, {Oesch}, {Abbas}, {Barnes}, {Bordoloi}, {Bottini},
  {Cappi}, {Cassata}, {Cimatti}, {Fumana}, {Hasinger}, {Koekemoer},
  {Leauthaud}, {Maccagni}, {Marinoni}, {McCracken}, {Memeo}, {Meneux}, {Nair},
  {Porciani}, {Presotto}, \& {Scaramella}}]{pen10}
{Peng}, Y.-j., {Lilly}, S.~J., {Kova{\v{c}}}, K., {et~al.} 2010, \apj, 721, 193

\bibitem[{{Pisani}(1993)}]{pisani1993}
{Pisani}, A. 1993, \mnras, 265, 706

\bibitem[{{Pisani}(1996)}]{pisani1996}
{Pisani}, A. 1996, \mnras, 278, 697

\bibitem[{{Plagge} {et~al.}(2010){Plagge}, {Bonamente}, \& {South Pole
  Telescope Collaboration}}]{pla10}
{Plagge}, T.~J., {Bonamente}, M., \& {South Pole Telescope Collaboration}.
  2010, in AAS/High Energy Astrophysics Division \#11, AAS/High Energy
  Astrophysics Division, 31.06

\bibitem[{{Planck Collaboration} {et~al.}(2011){Planck Collaboration}, {Ade},
  {Aghanim}, {Arnaud}, {Ashdown}, {Aumont}, {Baccigalupi}, {Balbi}, {Banday},
  {Barreiro}, {Bartelmann}, {Bartlett}, {Battaner}, {Battye}, {Benabed},
  {Beno{\^\i}t}, {Bernard}, {Bersanelli}, {Bhatia}, {Bock}, {Bonaldi}, {Bond},
  {Borrill}, {Bouchet}, {Brown}, {Bucher}, {Burigana}, {Cabella}, {Cantalupo},
  {Cardoso}, {Carvalho}, {Catalano}, {Cay{\'o}n}, {Challinor}, {Chamballu},
  {Chary}, {Chiang}, {Chiang}, {Chon}, {Christensen}, {Churazov}, {Clements},
  {Colafrancesco}, {Colombi}, {Couchot}, {Coulais}, {Crill}, {Cuttaia}, {da
  Silva}, {Dahle}, {Danese}, {Davis}, {de Bernardis}, {de Gasperis}, {de Rosa},
  {de Zotti}, {Delabrouille}, {Delouis}, {D{\'e}sert}, {Dickinson}, {Diego},
  {Dolag}, {Dole}, {Donzelli}, {Dor{\'e}}, {D{\"o}rl}, {Douspis}, {Dupac},
  {Efstathiou}, {Eisenhardt}, {En{\ss}lin}, {Feroz}, {Finelli}, {Flores-Cacho},
  {Forni}, {Fosalba}, {Frailis}, {Franceschi}, {Fromenteau}, {Galeotta},
  {Ganga}, {G{\'e}nova-Santos}, {Giard}, {Giardino}, {Giraud-H{\'e}raud},
  {Gonz{\'a}lez-Nuevo}, {Gonz{\'a}lez-Riestra}, {G{\'o}rski}, {Grainge},
  {Gratton}, {Gregorio}, {Gruppuso}, {Harrison}, {Hein{\"a}m{\"a}ki},
  {Henrot-Versill{\'e}}, {Hern{\'a}ndez-Monteagudo}, {Herranz}, {Hildebrand t},
  {Hivon}, {Hobson}, {Holmes}, {Hovest}, {Hoyland}, {Huffenberger}, {Hurier},
  {Hurley-Walker}, {Jaffe}, {Jones}, {Juvela}, {Keih{\"a}nen}, {Keskitalo},
  {Kisner}, {Kneissl}, {Knox}, {Kurki-Suonio}, {Lagache}, {Lamarre}, {Lasenby},
  {Laureijs}, {Lawrence}, {Le Jeune}, {Leach}, {Leonardi}, {Li}, {Liddle},
  {Lilje}, {Linden-V{\o}rnle}, {L{\'o}pez-Caniego}, {Lubin},
  {Mac{\'\i}as-P{\'e}rez}, {MacTavish}, {Maffei}, {Maino}, {Mandolesi}, {Mann},
  {Maris}, {Marleau}, {Mart{\'\i}nez-Gonz{\'a}lez}, {Masi}, {Matarrese},
  {Matthai}, {Mazzotta}, {Mei}, {Meinhold}, {Melchiorri}, {Melin}, {Mendes},
  {Mennella}, {Mitra}, {Miville-Desch{\^e}nes}, {Moneti}, {Montier},
  {Morgante}, {Mortlock}, {Munshi}, {Murphy}, {Naselsky}, {Nati}, {Natoli},
  {Netterfield}, {N{\o}rgaard-Nielsen}, {Noviello}, {Novikov}, {Novikov},
  {Olamaie}, {Osborne}, {Pajot}, {Pasian}, {Patanchon}, {Pearson}, {Perdereau},
  {Perotto}, {Perrotta}, {Piacentini}, {Piat}, {Pierpaoli}, {Piffaretti},
  {Plaszczynski}, {Pointecouteau}, {Polenta}, {Ponthieu}, {Poutanen}, {Pratt},
  {Pr{\'e}zeau}, {Prunet}, {Puget}, {Rachen}, {Reach}, {Rebolo}, {Reinecke},
  {Renault}, {Ricciardi}, {Riller}, {Ristorcelli}, {Rocha}, {Rosset},
  {Rubi{\~n}o-Mart{\'\i}n}, {Rusholme}, {Saar}, {Sandri}, {Santos}, {Saunders},
  {Savini}, {Schaefer}, {Scott}, {Seiffert}, {Shellard}, {Smoot}, {Stanford},
  {Starck}, {Stivoli}, {Stolyarov}, {Stompor}, {Sudiwala}, {Sunyaev}, {Sutton},
  {Sygnet}, {Taburet}, {Tauber}, {Terenzi}, {Toffolatti}, {Tomasi}, {Torre},
  {Tristram}, {Tuovinen}, {Valenziano}, {Vibert}, {Vielva}, {Villa},
  {Vittorio}, {Wade}, {Wandelt}, {Weller}, {White}, {White}, {Yvon}, {Zacchei},
  \& {Zonca}}]{plk11}
{Planck Collaboration}, {Ade}, P.~A.~R., {Aghanim}, N., {et~al.} 2011, \aap,
  536, A8

\bibitem[{{Postman} {et~al.}(2012){Postman}, {Coe}, {Ben{\'\i}tez}, {Bradley},
  {Broadhurst}, {Donahue}, {Ford}, {Graur}, {Graves}, {Jouvel}, {Koekemoer},
  {Lemze}, {Medezinski}, {Molino}, {Moustakas}, {Ogaz}, {Riess}, {Rodney},
  {Rosati}, {Umetsu}, {Zheng}, {Zitrin}, {Bartelmann}, {Bouwens}, {Czakon},
  {Golwala}, {Host}, {Infante}, {Jha}, {Jimenez-Teja}, {Kelson}, {Lahav},
  {Lazkoz}, {Maoz}, {McCully}, {Melchior}, {Meneghetti}, {Merten}, {Moustakas},
  {Nonino}, {Patel}, {Reg{\"o}s}, {Sayers}, {Seitz}, \& {Van der Wel}}]{pos12}
{Postman}, M., {Coe}, D., {Ben{\'\i}tez}, N., {et~al.} 2012, \apjs, 199, 25

\bibitem[{{Postman} {et~al.}(2005){Postman}, {Franx}, {Cross}, {Holden},
  {Ford}, {Illingworth}, {Goto}, {Demarco}, {Rosati}, {Blakeslee}, {Tran},
  {Ben{\'\i}tez}, {Clampin}, {Hartig}, {Homeier}, {Ardila}, {Bartko},
  {Bouwens}, {Bradley}, {Broadhurst}, {Brown}, {Burrows}, {Cheng}, {Feldman},
  {Golimowski}, {Gronwall}, {Infante}, {Kimble}, {Krist}, {Lesser}, {Martel},
  {Mei}, {Menanteau}, {Meurer}, {Miley}, {Motta}, {Sirianni}, {Sparks}, {Tran},
  {Tsvetanov}, {White}, \& {Zheng}}]{pos05}
{Postman}, M., {Franx}, M., {Cross}, N.~J.~G., {et~al.} 2005, \apj, 623, 721

\bibitem[{{Rahaman} {et~al.}(2021){Rahaman}, {Raja}, {Datta}, {Burns}, {Alden},
  \& {Rapetti}}]{ram21}
{Rahaman}, M., {Raja}, R., {Datta}, A., {et~al.} 2021, \mnras, 505, 480

\bibitem[{{Rhee} {et~al.}(2020){Rhee}, {Smith}, {Choi}, {Contini}, {Jung},
  {Han}, \& {Yi}}]{rhee20}
{Rhee}, J., {Smith}, R., {Choi}, H., {et~al.} 2020, \apjs, 247, 45

\bibitem[{{Rhee} {et~al.}(2017){Rhee}, {Smith}, {Choi}, {Yi}, {Jaff{\'e}},
  {Candlish}, \& {S{\'a}nchez-J{\'a}nssen}}]{rhee2017}
{Rhee}, J., {Smith}, R., {Choi}, H., {et~al.} 2017, \apj, 843, 128

\bibitem[{{Rix} {et~al.}(2004){Rix}, {Barden}, {Beckwith}, {Bell}, {Borch},
  {Caldwell}, {H{\"a}ussler}, {Jahnke}, {Jogee}, {McIntosh}, {Meisenheimer},
  {Peng}, {Sanchez}, {Somerville}, {Wisotzki}, \& {Wolf}}]{rix04}
{Rix}, H.-W., {Barden}, M., {Beckwith}, S. V.~W., {et~al.} 2004, \apjs, 152,
  163

\bibitem[{{Roberts} {et~al.}(2019){Roberts}, {Parker}, {Brown}, {Joshi},
  {Hlavacek-Larrondo}, \& {Wadsley}}]{rob19}
{Roberts}, I.~D., {Parker}, L.~C., {Brown}, T., {et~al.} 2019, \apj, 873, 42

\bibitem[{{Rosati} {et~al.}(2014){Rosati}, {Balestra}, {Grillo}, {Mercurio},
  {Nonino}, {Biviano}, {Girardi}, {Vanzella}, \& {Clash-VLT Team}}]{ros14}
{Rosati}, P., {Balestra}, I., {Grillo}, C., {et~al.} 2014, The Messenger, 158,
  48

\bibitem[{{Salpeter}(1955)}]{sal55}
{Salpeter}, E.~E. 1955, \apj, 121, 161

\bibitem[{{Sartoris} {et~al.}(2020){Sartoris}, {Biviano}, {Rosati}, {Mercurio},
  {Grillo}, {Ettori}, {Nonino}, {Umetsu}, {Bergamini}, {Caminha}, \&
  {Girardi}}]{sar20}
{Sartoris}, B., {Biviano}, A., {Rosati}, P., {et~al.} 2020, \aap, 637, A34

\bibitem[{{Scodeggio} {et~al.}(2005){Scodeggio}, {Franzetti}, {Garilli},
  {Zanichelli}, {Paltani}, {Maccagni}, {Bottini}, {Le Brun}, {Contini},
  {Scaramella}, {Adami}, {Bardelli}, {Zucca}, {Tresse}, {Ilbert}, {Foucaud},
  {Iovino}, {Merighi}, {Zamorani}, {Gavignaud}, {Rizzo}, {McCracken}, {Le
  F{\`e}vre}, {Picat}, {Vettolani}, {Arnaboldi}, {Arnouts}, {Bolzonella},
  {Cappi}, {Charlot}, {Ciliegi}, {Guzzo}, {Marano}, {Marinoni}, {Mathez},
  {Mazure}, {Meneux}, {Pell{\`o}}, {Pollo}, {Pozzetti}, \& {Radovich}}]{sco05}
{Scodeggio}, M., {Franzetti}, P., {Garilli}, B., {et~al.} 2005, \pasp, 117,
  1284

\bibitem[{{Sheen} {et~al.}(2017){Sheen}, {Smith}, {Jaff{\'e}}, {Kim}, {Yi},
  {Duc}, {Nantais}, {Candlish}, {Demarco}, \& {Treister}}]{sheen17}
{Sheen}, Y.-K., {Smith}, R., {Jaff{\'e}}, Y., {et~al.} 2017, \apjl, 840, L7

\bibitem[{{Silverman}(1986)}]{silverman1986}
{Silverman}, B.~W. 1986, {Density estimation for statistics and data analysis}

\bibitem[{{Smith} {et~al.}(2005){Smith}, {Treu}, {Ellis}, {Moran}, \&
  {Dressler}}]{smi05}
{Smith}, G.~P., {Treu}, T., {Ellis}, R.~S., {Moran}, S.~M., \& {Dressler}, A.
  2005, \apj, 620, 78

\bibitem[{{Solanes} {et~al.}(2001){Solanes}, {Manrique},
  {Garc{\'\i}a-G{\'o}mez}, {Gonz{\'a}lez-Casado}, {Giovanelli}, \&
  {Haynes}}]{sol01}
{Solanes}, J.~M., {Manrique}, A., {Garc{\'\i}a-G{\'o}mez}, C., {et~al.} 2001,
  \apj, 548, 97

\bibitem[{{Solanes} \& {Salvador-Sole}(1990)}]{sol90}
{Solanes}, J.~M. \& {Salvador-Sole}, E. 1990, \aap, 234, 93

\bibitem[{{Springel}(2005)}]{springel2005}
{Springel}, V. 2005, \mnras, 364, 1105

\bibitem[{{Springel} {et~al.}(2001){Springel}, {White}, {Tormen}, \&
  {Kauffmann}}]{springel2001}
{Springel}, V., {White}, S. D.~M., {Tormen}, G., \& {Kauffmann}, G. 2001,
  \mnras, 328, 726

\bibitem[{{Stark} {et~al.}(2016){Stark}, {Miller}, \& {Gifford}}]{stark16}
{Stark}, A., {Miller}, C.~J., \& {Gifford}, D. 2016, \apj, 830, 109

\bibitem[{{Tanaka} {et~al.}(2004){Tanaka}, {Goto}, {Okamura}, {Shimasaku}, \&
  {Brinkmann}}]{tan04}
{Tanaka}, M., {Goto}, T., {Okamura}, S., {Shimasaku}, K., \& {Brinkmann}, J.
  2004, \aj, 128, 2677

\bibitem[{{Tanaka} {et~al.}(2005){Tanaka}, {Kodama}, {Arimoto}, {Okamura},
  {Umetsu}, {Shimasaku}, {Tanaka}, \& {Yamada}}]{tan05}
{Tanaka}, M., {Kodama}, T., {Arimoto}, N., {et~al.} 2005, \mnras, 362, 268

\bibitem[{{Taylor}(2006)}]{tai06}
{Taylor}, M.~B. 2006, Astronomical Society of the Pacific Conference Series,
  Vol. 351, {STILTS - A Package for Command-Line Processing of Tabular Data},
  ed. C.~{Gabriel}, C.~{Arviset}, D.~{Ponz}, \& S.~{Enrique}, 666

\bibitem[{{Tiret} {et~al.}(2007){Tiret}, {Combes}, {Angus}, {Famaey}, \&
  {Zhao}}]{tiret2007}
{Tiret}, O., {Combes}, F., {Angus}, G.~W., {Famaey}, B., \& {Zhao}, H.~S. 2007,
  \aap, 476, L1

\bibitem[{{Tonnesen}(2019)}]{ton19}
{Tonnesen}, S. 2019, \apj, 874, 161

\bibitem[{{Tortorelli} {et~al.}(2018){Tortorelli}, {Mercurio}, {Paolillo},
  {Rosati}, {Gargiulo}, {Gobat}, {Balestra}, {Caminha}, {Annunziatella},
  {Grillo}, {Lombardi}, {Nonino}, {Rettura}, {Sartoris}, \&
  {Strazzullo}}]{tor18}
{Tortorelli}, L., {Mercurio}, A., {Paolillo}, M., {et~al.} 2018, \mnras, 477,
  648

\bibitem[{{Treu} {et~al.}(2003){Treu}, {Ellis}, {Kneib}, {Dressler}, {Smail},
  {Czoske}, {Oemler}, \& {Natarajan}}]{tre03}
{Treu}, T., {Ellis}, R.~S., {Kneib}, J.-P., {et~al.} 2003, \apj, 591, 53

\bibitem[{{Treu} \& {GLASS Team}(2016)}]{tre16}
{Treu}, T. \& {GLASS Team}. 2016, in American Astronomical Society Meeting
  Abstracts, Vol. 227, American Astronomical Society Meeting Abstracts \#227,
  324.04

\bibitem[{{Umetsu} {et~al.}(2016){Umetsu}, {Zitrin}, {Gruen}, {Merten},
  {Donahue}, \& {Postman}}]{umetsu2016}
{Umetsu}, K., {Zitrin}, A., {Gruen}, D., {et~al.} 2016, \apj, 821, 116

\bibitem[{{van der Burg} {et~al.}(2020){van der Burg}, {Rudnick}, {Balogh},
  {Muzzin}, {Lidman}, {Old}, {Shipley}, {Gilbank}, {McGee}, {Biviano},
  {Cerulo}, {Chan}, {Cooper}, {De Lucia}, {Demarco}, {Forrest}, {Gwyn},
  {Jablonka}, {Kukstas}, {Marchesini}, {Nantais}, {Noble}, {Pintos-Castro},
  {Poggianti}, {Reeves}, {Stefanon}, {Vulcani}, {Webb}, {Wilson}, {Yee}, \&
  {Zaritsky}}]{vdb20}
{van der Burg}, R. F.~J., {Rudnick}, G., {Balogh}, M.~L., {et~al.} 2020, \aap,
  638, A112

\bibitem[{{White} \& {Frenk}(1991)}]{white1991}
{White}, S. D.~M. \& {Frenk}, C.~S. 1991, \apj, 379, 52

\bibitem[{{Xie} {et~al.}(2020){Xie}, {van Weeren}, {Lovisari},
  {Andrade-Santos}, {Botteon}, {Br{\"u}ggen}, {Bulbul}, {Churazov}, {Clarke},
  {Forman}, {Intema}, {Jones}, {Kraft}, {Lal}, {Mroczkowski}, \&
  {Zitrin}}]{Xie20}
{Xie}, C., {van Weeren}, R.~J., {Lovisari}, L., {et~al.} 2020, \aap, 636, A3

\bibitem[{{ZuHone}(2011)}]{ZuHone11}
{ZuHone}, J.~A. 2011, \apj, 728, 54

\end{thebibliography}
\bibliographystyle{aa}

\end{document}